\documentclass[aps,prd,superscriptaddress,nofootinbib,preprintnumbers,floatfix,notitlepage]{revtex4-1}

\usepackage{amsmath,amsthm,amssymb}
\usepackage{graphicx}
\usepackage{hyperref}
\usepackage{datetime}

\begin{document}

\title{CONSTRAINTS ON PRIMORDIAL BLACK HOLES}

\author{Bernard Carr}\email[]{B.J.Carr@qmul.ac.uk}
\affiliation{
School of Physics and Astronomy, Queen Mary University of London,
Mile End Road, London E1 4NS, UK
}
\affiliation{
Research Center for the Early Universe (RESCEU),
Graduate School of Science, The University of Tokyo,
Tokyo 113-0033, Japan
}
\author{Kazunori Kohri}\email[]{kohri@post.kek.jp}
\affiliation{
KEK Theory Center, IPNS, KEK, Tsukuba, Ibaraki 305-0801, Japan
}
\affiliation{
The Graduate University for Advanced Studies (SOKENDAI),
Tsukuba, Ibaraki 305-0801, Japan
}
\affiliation{
Kavli Institute for the Physics and Mathematics of the Universe,
The University of Tokyo,
Kashiwa, Chiba 277-8568, Japan
}
\author{Yuuiti Sendouda}\email[]{sendouda@hirosaki-u.ac.jp}
\affiliation{
Graduate School of Science and Technology, Hirosaki University,
Hirosaki, Aomori 036-8561, Japan
}
\author{Jun'ichi Yokoyama}\email[]{yokoyama@resceu.s.u-tokyo.ac.jp}
\affiliation{
Research Center for the Early Universe (RESCEU),
Graduate School of Science, The University of Tokyo,
Tokyo 113-0033, Japan
}
\affiliation{
Kavli Institute for the Physics and Mathematics of the Universe,
The University of Tokyo,
Kashiwa, Chiba 277-8568, Japan
}
\affiliation{
Department of Physics,
Graduate School of Science, The University of Tokyo,
Tokyo 113-0033, Japan
}
\affiliation{
Trans-scale Quantum Science Institute, The University of Tokyo,
Tokyo 113-0033, Japan
}
\date{\formatdate{\day}{\month}{\year}, \currenttime}

\begin{abstract}
We update the constraints on the fraction of the Universe that may have gone into primordial black holes (PBHs) over the mass range $10^{-5}\text{--}10^{50}\,\mathrm g$.
Those smaller than $\sim 10^{15}\,\mathrm g$ would have evaporated by now due to Hawking radiation, so their abundance at formation is constrained by the effects of evaporated particles on big bang nucleosynthesis, the cosmic microwave background (CMB), the Galactic and extragalactic $\gamma$-ray and cosmic ray backgrounds and the possible generation of stable Planck mass relics.
PBHs larger than $\sim 10^{15}\,\mathrm g$ are subject to a variety of constraints associated with gravitational lensing, dynamical effects, influence on large-scale structure, accretion and gravitational waves.
We discuss the constraints on both the initial collapse fraction and the current fraction of the CDM in PBHs at each mass scale but stress that many of the constraints are associated with observational or theoretical uncertainties.
We also consider indirect constraints associated with the amplitude of the primordial density fluctuations, such as second-order tensor perturbations and $\mu$-distortions arising from the effect of acoustic reheating on the CMB, if PBHs are created from the high-$\sigma$ peaks of nearly Gaussian fluctuations.
Finally we discuss how the constraints are modified if the PBHs have an extended mass function, this being relevant if PBHs provide some combination of the dark matter, the LIGO/Virgo coalescences and the seeds for cosmic structure.
Even if PBHs make a small contribution to the dark matter, they could play an important cosmological role and provide a unique probe of the early Universe.
\end{abstract}

\maketitle

\tableofcontents


\section{Introduction}

\subsection{Overview}

Primordial black holes (PBHs) have been a source of intense interest for more than 50 years \cite{1967SvA....10..602Z}, despite the fact that there is still no evidence for them.
One reason for this interest is that only PBHs could be small enough for Hawking radiation to be important \cite{Hawking:1974rv}.
This has not yet been confirmed experimentally but this discovery is generally recognised as one of the key developments in 20th century physics and Hawking was only led to it through contemplating the properties of PBHs.
Indeed, those smaller than about $10^{15}\,\mathrm g$ would have evaporated by now but could still have many interesting cosmological consequences.

PBHs much larger than $10^{15}\,\mathrm g$ are unaffected by Hawking radiation but have also attracted interest because of the possibility that they provide the dark matter (DM) which comprises 25\,\% of the critical density \cite{Ade:2015xua}, an idea that goes back to the earliest days of PBH research \cite{1975Natur.253..251C}.
If PBHs formed at all, they would probably have done so before the end of the radiation-dominated era and are not subject to the well-known big bang nucleosynthesis (BBN) constraint that baryons can have at most 5\,\% of the critical density \cite{Cyburt:2003fe}.
They should therefore be classed as non-baryonic and from a dynamical perspective they behave like any other cold dark matter (CDM) candidate.
There is still no compelling evidence that PBHs provide the DM, but nor is there for any of the more traditional CDM candidates.

Despite the lack of evidence for them, PBHs have been invoked to explain numerous cosmological features.
For example, evaporating PBHs have been invoked to explain the extragalactic \cite{Page:1976wx} and Galactic \cite{Lehoucq:2009ge} $\gamma$-ray backgrounds, antimatter in cosmic rays \cite{Barrau:1999sk}, the annihilation line radiation from the Galactic centre \cite{1980A&A....81..263O}, the reionisation of the pregalactic medium \cite{Belotsky:2014twa} and some short-period gamma-ray bursts \cite{Cline:1996zg}.
Non-evaporating PBHs -- even if they do not provide the DM -- have been invoked to explain lensing effects \cite{1993Natur.366..242H}, the heating of the stars in our Galactic disc \cite{1985ApJ...299..633L}, the origin of massive compact halo objects (MACHOs) \cite{Yokoyama:1995ex}, the seeds for the supermassive black holes in galactic nuclei \cite{Bean:2002kx}, the generation of large-scale structure through Poisson fluctuations \cite{Afshordi:2003zb}, effects on the thermal and ionisation history of the Universe \cite{Ricotti:2007au} and the LIGO/Virgo gravitational wave events \cite{Bird:2016dcv}.
Here we only cite the original paper on each topic, even if this has been superseded by later work.

There are usually other possible explanations for these features, so there is still no definitive evidence for PBHs.
This should be born in mind throughout this review since -- in discussing the possible consequences of PBHs -- we will not repeatedly stress that they may not exist at all.
Nevertheless, studying each of these effects allows one to place interesting constraints on the number of PBHs of mass $M$ and this in turn places constraints on the cosmological models which would generate them.
We will be discussing all these constraints in this review.
For evaporating PBHs these are usually expressed as constraints on the fraction of the Universe collapsing into PBHs at their formation epoch, denoted as $\beta(M)$.
Indeed, this must be less than $10^{-18}$ over the entire mass range $10^9\text{--}10^{15}\,\mathrm g$.
For non-evaporating ones they are most usefully expressed as constraints on the fraction of the DM in PBHs, denoted as $f(M)$.
It was already clear a decade ago that there were only a few mass windows where PBHs could provide all the dark matter \cite{Barrau:2003xp}: the asteroid range ($10^{16}\text{--}10^{17}\,\mathrm g$), the sublunar range ($10^{20}\text{--}10^{26}\,\mathrm g$) and the intermediate-mass range ($10\text{--}10^{3}\,M_{\odot}$).
Since then, interest in PBHs as dark matter has increased, primarily because of the LIGO/Virgo events.
The lowest and highest mass windows have now narrowed and it is sometimes argued that they are excluded, while the middle mass window has shifted to $10^{17}\text{--}10^{23}\,\mathrm g$, with the both mass limits having decreased.
However, we stress that most of the limits assume that PBH mass spectrum is quasi-monochromatic (i.e.\ having a width $\Delta M \lesssim M$) and it could well be extended \cite{Carr:2016drx}.
As discussed later, whether this makes it easier or more difficult for PBHs to provide the DM in the various mass windows has been the subject of some dispute.

In this review, we discuss all the PBH constraints, summarizing our results as excluded regions in the $\beta(M)$ and $f(M)$ planes.
Similar diagrams have been produced by numerous authors -- indeed almost every paper on PBHs now includes such a diagram.
However, we hope this review will be more comprehensive and up-to-date than previous ones.
We should stress at the outset that the limits are constantly changing as a result of both observational and theoretical developments, so our claim for comprehensiveness may be short-lived.
We will also discuss some limits which are no longer believed because this is historically illuminating.
Even wrong calculations are worth recording because this may avoid their being repeated in the future.
In fact, all of the limits come with caveats and few can be regarded as 100\,\% secure.
(For example, the evaporation constraints assume the validity of Hawking radiation, even though there is still no direct observational evidence for this.)
While we emphasize which limits have gone away or are questionable, we are not always able to specify the confidence levels precisely.
Since our primary purpose is to discuss the constraints, there will be little attempt to describe the more positive aspects of PBHs (i.e.\ the ways in which they can probe the early Universe or explain various cosmological conundra \cite{Carr:2019kxo}).
However, such positive aspects are covered in another recent PBH review~\cite{Carr:2020xqk} and this complements the present work.

\subsection{PBH formation}

Black holes with a wide range of masses could have formed in the early Universe as a result of the great compression associated with the big bang \cite{Hawking:1971ei,Carr:1974nx}.
A comparison of the cosmological density at a time $ t $ after the big bang with the density associated with a black hole of mass $ M $ suggests that such PBHs would have a mass of order
\begin{equation}
M
\sim
  \frac{c^3\,t}{G}
\sim
  10^{15}\,\left(\frac{t}{10^{-23}\,\mathrm s}\right)\,\mathrm g\,.
\label{eq:Moft}
\end{equation}
This roughly corresponds to the Hubble mass at time $t$.
PBHs could thus span an enormous mass range:
those formed at the Planck time ($ 10^{-43}\,\mathrm s $) would have the Planck mass ($ 10^{-5}\,\mathrm g $), whereas those formed at $ 1\,\mathrm s $ would be as large as $ 10^5\,M_\odot $\,, comparable to the mass of the holes thought to reside in galactic nuclei.
By contrast, black holes forming at the present epoch could never be smaller than about $ 1\,M_\odot $\,.
The high density of the early Universe is a necessary but not sufficient condition for PBH formation.
One possibility is that they formed from large inhomogeneities -- either fed into the initial conditions of the Universe or arising spontaneously in an initially smooth universe (e.g.\ through quantum effects during inflation).
Another possibility is that some sort of phase transition may have enhanced PBH formation from primordial inhomogeneities or triggered it even if there were none.
We now briefly discuss the various formation scenarios, although the references are incomplete and papers are cited without the authors being named.

\subsubsection{Collapse from inhomogeneities during radiation-dominated era}
\label{sec:collapse}

Many PBH formation scenarios depend on the development of inhomogeneities of some kind.
Overdense regions could then stop expanding and recollapse \cite{Carr:1974nx}.
Whatever the source of the fluctuations, they would need to be larger than the Jeans length at maximum expansion in order to collapse against the pressure and this is $ \sqrt{w}\,ct$ for equation of state $p = w\,\rho\,c^2$.
A simple analytic argument then implies that the overdensity $\delta$ when the region enters the horizon must exceed a critical value $\delta_\mathrm c \approx w$~\cite{Carr:1975qj}.
This gives $1/3$ in the radiation era and early numerical calculations \cite{1978SvA....22..129N,1979ApJ...232..670B} for overdense spherically symmetric regions surrounded by a compensating void roughly confirmed this prediction.
If the fluctuations on a given mass scale have Gaussian distribution, with a root-mean-square amplitude $\sigma$ at the horizon epoch, then the collapse fraction for that scale should be roughly $\exp[-(\delta_\mathrm c/ \sqrt{2} \sigma)^2]$~\cite{Carr:1975qj}, which is expected to be tiny.
However, there is some ambiguity in how one defines fluctuations on scales larger than the horizon, especially when they are non-linear.
These issues were later addressed in Ref.~\cite{Shibata:1999zs}, this leading to estimates for $\delta_\mathrm c$ in the range $0.3\text{--}0.5$~\cite{Green:2004wb}.
A more precise analytical expression for $\delta_\mathrm c$ was later given in Ref.~\cite{ Harada:2013epa} and this corresponded to $0.4$ in the radiation era.
However, this did not allow for pressure gradient effects and an improved calculation, allowing for the sharpness of the transition in the compensated model, gave a larger value~\cite{Harada:2015yda}.
As discussed below, further insights came with the realisation that PBH formation is an application of critical phenomena~\cite{Niemeyer:1997mt,Niemeyer:1999ak}, with subsequent studies indicating that the value of $\delta_\mathrm c$ depends on the form of the density profile in the collapsing region~\cite{Musco:2004ak,Musco:2008hv,Polnarev:2006aa}.
Later numerical calculations~\cite{Nakama:2013ica} identified the radial integral of the curvature profile in the central region and the transition scale as the most important features of the density profile and indicated a value for $\delta_\mathrm c$ in the range $0.37\text{--}0.43$.
Subsequent work considered the relationship between the threshold and curvature profile more precisely and obtained $0.4\text{--}2/3$~\cite{Musco:2018rwt}, with an analytic expression being obtained in Ref.~\cite{Escriva:2019phb}.
In the context of inflationary scenarios, it was also shown that the collapse fraction depends on the shape of the peak in the power spectrum \cite{Germani:2018jgr}.
Recent work~\cite{Musco:2020jjb} uses numerical simulations to derive a simple expression for $\delta_\mathrm c$ which accounts for the non-linear relation between the curvature perturbation and the density contrast and for non-linear effects arising at horizon crossing.
The value of $\delta_\mathrm c$ could also be affected by non-Gaussianity, this slightly changing the relationship between the curvature perturbation and the overdensity~\cite{Saito:2008em,Kehagias:2019eil}.

\subsubsection{Critical collapse}

When the density perturbation approaches the threshold value $\delta_\mathrm c$ required for PBH formation, a critical phenomenon occurs in which the black hole mass scales as $(\delta - \delta_\mathrm c)^{\gamma}$ and therefore extends down to arbitrarily small scales~\cite{Niemeyer:1997mt,Niemeyer:1999ak,Musco:2004ak,Musco:2008hv}.
Here the exponent $\gamma$ is independent of the density profile and just depends on the equation of state ($0.35$ in the radiation case).
Most of the density is in PBHs with the horizon mass but there is also a power-law tail at lower masses \cite{Yokoyama:1998xd,Green:1999xm,Kribs:1999bs}.
The PBHs much smaller than the horizon are expected to have large spins~\cite{Harada:2020pzb}.
Because $\delta_\mathrm c$ is sensitive to the equation of state parameter $w$, even a slight reduction in this can enhance PBH production.
For example, this may happen at the Quantum Chromodynamics (QCD) era \cite{Jedamzik:1996mr,Widerin:1998my,Jedamzik:1999am} and the critical collapse analysis then predicts the mass function very precisely \cite{Byrnes:2018clq}.

\subsubsection{Collapse from single-field inflation}

The most natural source of the fluctuations would be quantum effects during inflation.
Although any PBHs formed before the end of inflation (i.e.\ with mass exceeding about $1\,\mathrm g$) would be diluted exponentially, the inflationary fluctuations themselves could generate PBHs on scales larger than this.
In the simplest (single scalar field) scenario, the inflationary fluctuations would depend upon the form of the inflaton potential $V(\phi)$ and is expected to have a power-law form.
Since the fluctuations are only $10^{-5}$ on the CMB scale, they would then need to be ``blue'' (i.e.\ increasing on smaller scales) for PBH formation, with most PBH production occurring shortly after reheating~\cite{Carr:1993aq}.
However, the observed fluctuations on the CMB scale are ``red'', so one needs a more complicated scenario, with a running of the spectral index or some feature in the power spectrum at the PBH scale.
Since the horizon fluctuation scales as $(dV/d\phi)^{-1}$ in slow-roll inflation, one expects a peak in the power spectrum at an inflection point of $V(\phi)$, so this was a feature of an early model of PBH production~\cite{Ivanov:1994pa}.
Similar models were subsequently studied for a polynomial potential with \cite{Yokoyama:1998pt} or without \cite{Garcia-Bellido:2017mdw} a logarithmic correction, a non-minimally coupled scalar field with a quartic potential~\cite{Ballesteros:2017fsr} and a quintic potential fine-tuned so that a local minimum is followed by a small maximum (so that the field is slowed down but can still exit this region)~\cite{Hertzberg:2017dkh}.
Ultra-slow roll inflation occurs in the limit of a flat potential~\cite{Kinney:2005vj} and curvature perturbations then grow rapidly on superhorizon scales, rather than remaining constant, so the standard expression for the power spectrum is not valid and a more precise numerical calculation is required~\cite{Germani:2017bcs}.
Rapid growth of superhorizon fluctuation is not limited to the case of ultra-slow roll, as it also happens if the first and second Hubble slow-roll parameters satisfy a certain condition \cite{Saito:2008em}.
Whenever the predicted fluctuations are large, one expects quantum corrections to the standard analysis to be important, so this has led to a study of the link between quantum diffusion and PBH production~\cite{Pattison:2017mbe,Ezquiaga:2018gbw,Biagetti:2018pjj}.
Other single-field proposals are the running-mass model~\cite{Stewart:1997wg}, in which the power spectrum grows sufficiently for PBHs to form without violating constraints on cosmological scales~\cite{Kohri:2007qn}, and the hill-top model~\cite{Alabidi:2009bk}, in which inflation occurs as the field evolves away from a local maximum towards a minimum.
However, in both of these cases, an auxiliary mechanism is required to terminate inflation.
Reheating at the end of inflation, where the inflaton oscillates around its minimum and decays, may also generate PBHs~\cite{Green:2000he,Bassett:2000ha}, with perturbations being enhanced by a resonant instability~\cite{Martin:2019nuw}.

\subsubsection{Collapse from multi-field inflation}

A different mechanism for PBH production arises in multi-field scenarios~\cite{Randall:1995dj,GarciaBellido:1996qt}.
In hybrid inflation one of the fields, $\phi$, initially slow-rolls while the accelerated expansion is driven by a second field $\psi$.
At a critical value of $\phi$, $\psi$ undergoes a waterfall transition to a global minimum and inflation ends, with quantum fluctuations at the transition being large.
For some parameter values, however, the transition can be mild, so that there is a second inflationary phase as $\psi$ evolves to the minimum of its potential.
In this case, isocurvature perturbations are generated during the waterfall transition, leading to a broad peak in the power spectrum~\cite{Kawasaki:1997ju,Clesse:2015wea}.
Perturbations on cosmological scales are generated during the first period of inflation, while those on PBH scales are generated during the second period.
Another model invokes a curvaton field, which is dynamically unimportant during inflation but generates curvature perturbations after the inflaton has decayed~\cite{Lyth:2001nq}.
This makes it easier to produce PBHs than in standard single field models.
The first model of PBH formation with such a field was proposed even before the term curvaton was introduced \cite{Yokoyama:1995ex}.
Other examples are the axion-like curvaton~\cite{Kawasaki:2012wr} and the inflating curvaton~\cite{Kohri:2012yw}.

\subsubsection{Collapse from inhomogeneities during matter-dominated era}

Whatever the source of the inhomogeneities, PBH formation would be enhanced if some phase transitions led to a sudden reduction in the pressure -- for example, if the early Universe went through a dustlike phase at early times as a result of being dominated by non-relativistic particles for a period \cite{Khlopov:1980mg,1981SvA....25..406P,1982SvA....26..391P} or undergoing slow reheating after inflation \cite{Khlopov:1985jw,Carr:1994ar,Carr:2018nkm,Martin:2020fgl,Allahverdi:2020bys}.
In such cases, the effect of pressure in stopping collapse is unimportant and the probability of PBH formation just depends upon the fraction of regions which are sufficiently spherical to undergo collapse.
If a perturbation is not sufficiently spherically symmetric, it will collapse to form a pancake or cigar.
The mass of PBHs formed in this case may be much smaller than the horizon mass.
For a given spectrum of primordial fluctuations, the mass range over which the PBHs form is determined by the period of the soft equation of state, extending from the horizon mass at the start of the matter-dominated period to the mass of the regions binding at the end of that period (this being much smaller than the horizon mass then).
The fraction of regions which collapse to PBHs is the product of the fractions of regions which satisfy the inhomogeneity and anisotropy criteria.
The first fraction can be derived from the condition that the fluctuation must collapse within its Schwarzschild radius before a caustic can form at its centre and was originally estimated as $ \sigma^{3/2}$~\cite{1981SvA....25..406P}.
However, taking into account the finite propagation speed of information leads to a larger estimate $ 3.7\,\sigma^{3/2}$~\cite{Kokubu:2018fxy}.
The original expression for second fraction was $0.02\,\sigma^5$~\cite{Khlopov:1980mg} but this was later revised to $0.056\,\sigma^5$~\cite{Harada:2016mhb}.
The total collapse fraction is therefore $0.2\,\sigma^{13/2}$ and PBHs form more abundantly during matter-domination than radiation-domination if $\sigma < 0.05$.
Note that PBHs can form with large spins during matter-domination, with the exact value depending on $\sigma$ and the duration of the matter-dominated period~\cite{Harada:2017fjm}.

\subsubsection{Collapse of cosmic string loops}

In the cosmic string scenario, one expects some strings to self-intersect and form cosmic loops.
A typical loop will be larger than its Schwarzschild radius by the factor $(G\mu)^{-1}$, where $\mu$ is the string mass per unit length.
Observations of the CMB imply that $G\mu$ must be less than about $10^{-7}$ \cite{Ade:2013xla}, while a comparison of the gravitational wave background expected from loop oscillations with pulsar timing array (PTA) data gives a limit of $10^{-11}$~\cite{Blanco-Pillado:2017rnf}.
However, as discussed by many authors \cite{Hawking:1987bn,Polnarev:1988dh,Hansen:1999su,Hogan:1984zb,Nagasawa:2005hv,Honma:1991na,James-Turner:2019ssu}, there is still a small probability that a cosmic loop will get into a configuration in which every dimension lies within its Schwarzschild radius, numerical simulations~\cite{PhysRevD.45.3447} indicating a collapse fraction $10^{4.9}\,(G \mu)^{4.1}$.
Since the PBHs form with equal probability at every epoch, they should have an extended mass spectrum~\cite{MacGibbon:1997pu}.
It has recently been argued that PBHs would form more abundantly from the collapse of cosmic string cusps~\cite{Jenkins:2020ctp}.
Critical phenomena may arise in loop collapse, with the PBH mass scaling as a power of the difference between the loop and Schwarzschild radius~\cite{Helfer:2018qgv}.
PBHs might also form through the collapse of string necklaces \cite{Matsuda:2005ez,Lake:2009nq}.

\subsubsection{Collapse from bubble collisions}

Bubbles of broken symmetry might arise at any spontaneously broken symmetry epoch and many people have suggested that PBHs could form as a result of bubble collisions \cite{Crawford:1982yz,Hawking:1982ga,La:1989st,Moss:1994iq,1998AstL...24..413K,Konoplich:1999qq}.
The production of PBHs from bubble collisions associated with a first-order phase transition at the end of the old inflation model \cite{Sato:1980yn,PhysRevD.23.347} has also been studied \cite{Kodama:1982sf}.
However, this happens only if the bubble formation rate per Hubble volume is finely tuned:
if it is much larger than the Hubble rate, the entire Universe undergoes the phase transition immediately and there is not time to form black holes;
if it is much less than the Hubble rate, the bubbles are very rare and never collide.
The holes should have a mass of order the horizon mass at the phase transition, so PBHs forming at the GUT (Grand Unified Theory) epoch would have a mass of $10^3\,\mathrm g$, those forming at the electroweak unification epoch would have a mass of $10^{28}\,\mathrm g$, and those forming at the QCD phase transition would have mass of around $1\,M_{\odot}$\,.
The production of PBHs through the nucleation of false vacuum bubbles during inflation has recently been reconsidered in Ref.~\cite{Kusenko:2020pcg}.

\subsubsection{Collapse of scalar field}

A scalar condensate can form in the early Universe and collapse into Q-balls before decaying \cite{Cotner:2016cvr}.
If the Q-balls dominate the energy density for some period, the statistical fluctuations in their number density can lead to PBH formation \cite{Cotner:2017tir}.
For a general charged scalar field, this can generate PBHs over the mass range allowed by observational constraints and with sufficient abundance to account for the dark matter and the LIGO observations.
If the scalar field is associated with supersymmetry, the mass range must be below $10^{23}\,\mathrm g$.
The fragmentation of the inflaton into oscillons might also lead to PBH production in the sublunar range \cite{Cotner:2018vug,Cotner:2019ykd}.

\subsubsection{Collapse of domain walls}

The collapse of sufficiently large closed domain walls produced at a 2nd order phase transition in the vacuum state of a scalar field, such as might be associated with inflation, could lead to PBH formation \cite{Rubin:2000dq,Rubin:2001yw,Dokuchaev:2004kr}.
These PBHs would have a small mass for a thermal phase transition but they could be much larger if one invoked a non-equilibrium scenario.
Indeed, they could then span a wide range of masses, with a fractal structure of smaller PBHs clustered around larger ones \cite{Khlopov:2000js}.
PBHs in the sublunar range could form through the collapse of closed domain walls in QCD axion models, together with a string-wall network~\cite{Ge:2019ihf}.
It has been argued that bubbles formed during inflation would (depending on their size) form either black holes or baby universes connected to our Universe by wormholes \cite{Garriga:2015fdk,Deng:2016vzb}.
In this case, the PBH mass function would be very broad and extend to very high masses \cite{Deng:2017uwc,Liu:2019lul}.

\vspace{\baselineskip}

In some of these scenarios, the PBH mass spectrum is expected to be narrow and centred around the mass given by Eq.~\eqref{eq:Moft} with $ t $ corresponding to the time at which the PBH scale re-enters the horizon in the inflationary model or to the time of the relevant cosmological phase transition otherwise.
However, there are some circumstances in which the spectrum would be extended and the constraint on one mass-scale would then imply a constraint on neighbouring scales.
For example, PBHs may be much smaller than the horizon if they form as a result of critical phenomena or during a matter-dominated phase and their spectrum could then extend well below the horizon mass.
It used to be claimed that a PBH could not be much larger than the value given by Eq.~\eqref{eq:Moft} at formation else it would be a separate closed universe rather than part of our Universe \cite{Carr:1974nx}.
This interpretation is misleading because the PBH mass necessarily goes to zero when the size of the overdense region reaches a certain limit~\cite{Kopp:2010sh} but there is still a maximum mass for a PBH forming at a given epoch and this is of order the horizon mass~\cite{Harada:2004pe}.

\subsection{Mass and density of PBHs}
\label{sec:bg}

In the following discussion, we assume that the standard Lambda Cold Dark Matter ($ \Lambda $CDM) model applies, with the age of the Universe being $ t_0 = 13.8\,\mathrm{Gyr} $ and the Hubble parameter being $H_0 \equiv 100\,h\,\mathrm{km}\,\mathrm s^{-1}\,\mathrm{Mpc}^{-1}$ with $ h = 0.67 $ \cite{Akrami:2018odb}.
We also put $ c = \hbar = k_\mathrm B = 1 $.
The Friedmann equation implies that the density $\rho$ and temperature $T$ during the radiation era are given by
\begin{equation}
H^2
= \frac{8\pi\,G}{3}\,\rho
= \frac{4\pi^3\,G}{45}\,g_*\,T^4\,,
\end{equation}
where $ g_* $ counts the number of relativistic degrees of freedom.
This can be integrated to give
\begin{equation}
t
\approx
  0.738\,
  \left(\frac{g_*}{10.75}\right)^{-1/2}\,
  \left(\frac{T}{1\,\mathrm{MeV}}\right)^{-2}\,\mathrm s\,,
\end{equation}
where $ g_* $ and $ T $ are normalised to their values at the start of the BBN epoch.
For PBHs which form during the radiation era (the ones generated before inflation being diluted to negligible density), the initial PBH mass $ M $ is close to the Hubble horizon mass $ M_\mathrm{H} $ and therefore given by
\begin{equation}
M
= \gamma\,M_\mathrm{H}
= \frac{4\pi}{3}\,\gamma\,\rho\,R_\mathrm{H}^{-3}
= \frac{\gamma c^3 t}{G}
\approx
  2.03 \times 10^5\,
  \gamma\,
  \left(\frac{t}{1\,\mathrm s}\right)\,M_\odot\,.
\label{mass}
\end{equation}
Here $ \gamma $ is a numerical factor (somewhat below $1$) which depends on the details of gravitational collapse and the penultimate expression applies \emph{exactly} in a radiation Universe (for which $R_\mathrm{H} = 2ct$).
However, Eq.~\eqref{mass} does not apply for PBHs which form in an early matter-dominated (dust-like) era or for the low mass PBHs forming through critical collapse.
Note that the normalisation in Eq.~\eqref{mass} is sometimes taken to be the particle horizon mass; that is inappropriate if there is an inflationary phase, although the two masses are the same in a non-inflationary radiation-dominated model.
For much of the following discussion, we will assume that the PBHs all have the same mass or at least a mass width $\Delta M$ no larger than $ M $.
This simplifies the analysis considerably and suffices providing we only require limits on the PBH abundance at particular values of $ M $.
If the PBHs have an extended mass function, the analysis of the constraints is more complicated and we discuss this situation later.

Assuming adiabatic cosmic expansion after PBH formation, the ratio of the PBH number density to the entropy density, $ n_\mathrm{PBH}/s $, is conserved.
Using the relation $ \rho = 3sT/4 $, the fraction of the Universe's mass in PBHs at their formation time is then related to their number density $ n_\mathrm{PBH}(t) $ during the radiation era by
\begin{equation}
\beta(M)
\equiv
  \frac{\rho_\mathrm{PBH}(t_\mathrm i)}{\rho(t_\mathrm i)}
= \frac{4 M}{3 T_\mathrm i}\,
  \frac{n_\mathrm{PBH}(t)}{s(t)} \\
\approx
  7.99 \times 10^{-29}\,\gamma^{-1/2}\,
  \left(\frac{g_{*\mathrm i}}{106.75}\right)^{1/4}\,
  \left(\frac{M}{M_\odot}\right)^{3/2}\,
  \left(\frac{n_\mathrm{PBH}(t_0)}{1\,\mathrm{Gpc}^{-3}}\right)\,,
\label{eq:beta}
\end{equation}
where the subscript ``$ \mathrm i $'' indicates values at the epoch of PBH formation and we have assumed $ s = 8.54 \times 10^{85}\,\mathrm{Gpc}^{-3} $ today.
$ g_{*\mathrm i} $ is now normalised to the value of $ g_* $ at around $ 10^{-5}\,\mathrm s $ since it does not increase much before that in the Standard Model and most PBHs are likely to form before then.
We can also express this as
\begin{equation}
\beta(M)
\approx
  7.06 \times 10^{-18}\,
  \gamma^{-1/2}\,
  \left( \frac{h}{0.67}\right)^2\,
  \left(\frac{g_{*\mathrm i}}{106.75}\right)^{1/4}\,
  \Omega_\mathrm{PBH}(M)\,
  \left(\frac{M}{10^{15}\,\mathrm g}\right)^{1/2}\,,
\label{eq:density}
\end{equation}
where $\Omega_\mathrm{PBH} \equiv \rho_\mathrm{PBH}(t_0)/\rho_\mathrm{crit}$ is the current density parameter of the PBHs.
A less precise form of this equation can be obtained from the relation
\begin{equation}
\Omega_\mathrm{PBH}
= \beta\,\Omega_\mathrm{CMB}\,(1+z)
\sim
  10^6\,\beta\,\left(\frac{t}{\mathrm s}\right)^{-1/2}
\sim
  10^{9}\,\beta\,\left(\frac{M}{M_{\odot}}\right)^{-1/2}\,,
\end{equation}
where $\Omega_\mathrm{CMB} \approx 5 \times 10^{-5}$ is the density parameter of the CMB and we have used Eq.~\eqref{eq:Moft}.
The $(1+z)$ factor can be understood as arising because the radiation density scales as $(1+z)^4$, whereas the PBH density scales as $(1+z)^3$.

Since $ \beta $ always appears in combination with $ \gamma^{1/2}\,g_{*\mathrm i}^{-1/4}\,h^{-2}$, it is convenient to define a new parameter
\begin{equation}
\beta'(M)
\equiv
  \gamma^{1/2}\,\left(\frac{g_{*\mathrm i}}{106.75}\right)^{-1/4}\,\left(\frac{h}{0.67}\right)^{-2}\,\beta(M)\,,
\end{equation}
where $ g_{*\mathrm i} $ and $h$ can be specified very precisely but $ \gamma $ is rather uncertain.
Note that the relationship between $ \beta $ and $ \Omega_\mathrm{PBH} $ must be modified if the Universe ever deviates from the standard radiation-dominated behaviour -- for example, if there is a dust-like stage, a second inflationary phase, a change of dimensionality \cite{Sendouda:2006nu,Johnson:2020tiw} or a variation in the gravitational constant at early times \cite{Barrow:1992ay,Barrow:1996jk,Harada:2001kc}.

Any limit on $\Omega_\mathrm{PBH}$ places a constraint on $\beta'(M)$.
For non-evaporating PBHs with $M>10^{15}\,\mathrm g$, one constraint comes from requiring that $\Omega_\mathrm{PBH}$ be less than the CDM density, $ \Omega_\mathrm{CDM} = 0.264 \pm 0.006$ with $ h = 0.67 $ \cite{Akrami:2018odb}, so the $3\sigma$ upper limit is $ \Omega_\mathrm{PBH} < \Omega_\mathrm{CDM} < 0.282 $.
Much stronger constraints are associated with PBHs smaller than $10^{15}\,\mathrm g$ since these would have evaporated by now.
For example, the $ \gamma $-ray background limit implies $ \beta(10^{15}\,\mathrm g) \lesssim 10^{-29} $ and this is the strongest constraint on $ \beta $ over all mass ranges.
Other ones are associated with the generation of entropy, modifications to the cosmological production of light elements and the CMB anisotropies.
There are also constraints below $10^{6}\,\mathrm g$ based on the (uncertain) assumption that evaporating PBHs leave stable Planck mass relics, an issue which is discussed later.

\begin{figure}[ht]
\begin{center}
\includegraphics[scale=0.8]{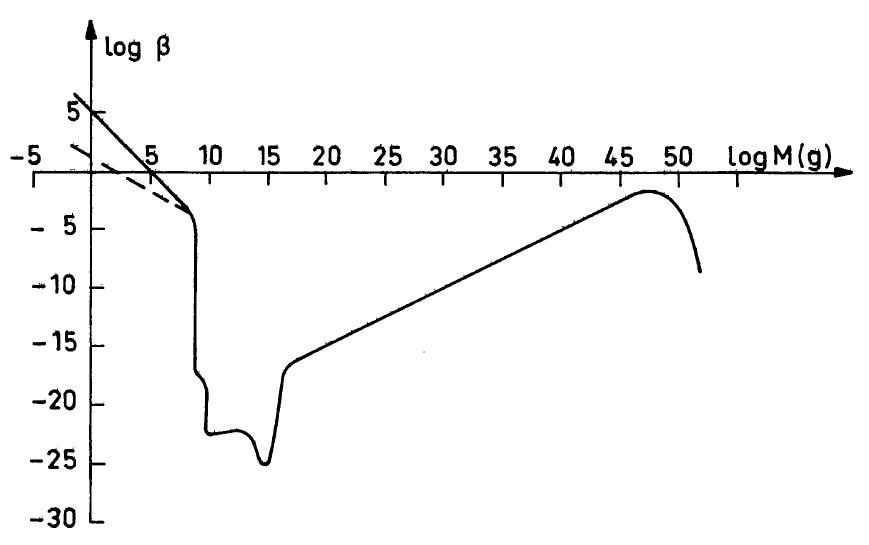}
\includegraphics[scale=0.35]{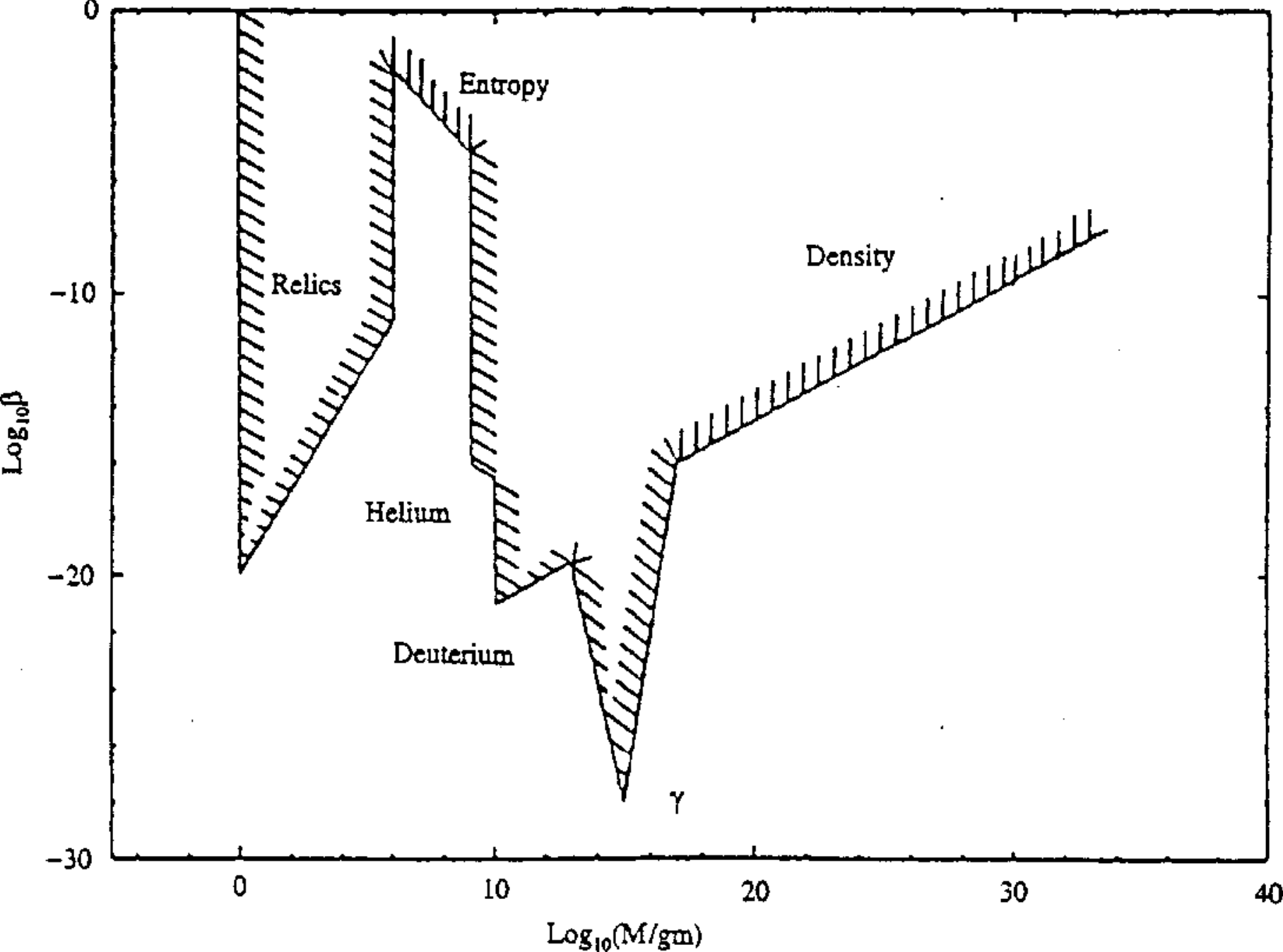}
\caption{Constraints on $\beta(M)$ from Ref.~\cite{1979A&A....80..104N} in 1979 (left) and Ref.~\cite{Carr:1994ar} in 1994 (right).}
\label{fig:bc1}
\end{center}
\end{figure}

The constraints on $ \beta(M) $ were first brought together by Novikov \textit{et al.} \cite{1979A&A....80..104N} 40 years ago.
Besides the entropy, BBN, $\gamma$-ray background and density limit, they included a strong constraint above $10^{15} M_{\odot}$ associated with the upper limit on the CMB dipole anisotropy.
An updated version of this diagram appeared in Ref.~\cite{Carr:1994ar} about 15 years later but was essentially the same, apart from the omission of the dipole constraint (which was outside the considered mass range) and the addition of a `relics' constraint.
Because of their historical interest, both diagrams are shown in Fig.~\ref{fig:bc1}, although the later one is slightly adapted.
Subsequently, this diagram has frequently been revised as the relevant effects have been studied in greater detail.
For example, Josan \textit{et al.} \cite{Josan:2009qn} produced a comprehensive version a decade ago and we produced a version at about the same time in Ref.~\cite{Carr:2009jm} (henceforth CKSY), covering both the evaporating and non-evaporating PBHs.
However, there have been many further developments since then, on both the observational and theoretical fronts.
An updated review of the constraints was given in Ref.~\cite{Carr:2016drx} several years ago and the present discussion might be regarded as another (albeit more comprehensive) update.
The important qualitative point of all such diagrams is that the value of $ \beta(M) $ must be tiny over every mass range, even if the PBH density is large today, so any cosmological model which would entail more than a tiny fraction of the Universe going into PBHs is immediately excluded.

\subsection{Evaporation of PBHs}

The realisation that PBHs might be small prompted Hawking to study their quantum properties.
This led to his famous discovery \cite{Hawking:1974sw} that black holes radiate thermally with a temperature
\begin{equation}
T_\mathrm{BH}
= \frac{\hbar\,c^3}{8\pi\,G\,M\,k_\mathrm B}
\sim
  10^{-7}\,\left(\frac{M}{M_\odot}\right)^{-1}\,\mathrm K\,,
\label{eq:BHtemp}
\end{equation}
so they evaporate completely on a timescale
\begin{equation}
\tau(M)
\sim
  \frac{G^2\,M^3}{\hbar\,c^4}
\sim
  10^{64}\,\left(\frac{M}{M_\odot}\right)^3\,\mathrm{yr}\,.
\end{equation}
More precise expressions are given below but this implies that only PBHs smaller than about $ 10^{15}\,\mathrm g $ would have evaporated by the present epoch, so Eq.~\eqref{eq:Moft} implies that this effect could be important only for ones which formed before $ 10^{-23}\,\mathrm s $.
Since PBHs with a mass of around $ 10^{15}\,\mathrm g $ would be producing photons with energy of order $ 100\,\mathrm{MeV} $ at the present epoch, the observational limit on the $ \gamma $-ray background intensity at $ 100\,\mathrm{MeV} $ immediately implies that their density could not exceed about $ 10^{-8} $ times the critical density \cite{Page:1976wx}.
This suggests that there is little chance of detecting their final explosive phase at the present epoch, at least in the Standard Model of particle physics.
Nevertheless, the $ \gamma $-ray background limit does not preclude PBHs having important cosmological effects \cite{Carr:1976jy}.

\subsubsection{Mass loss and evaporation timescale}

From Eq.~\eqref{eq:BHtemp} the temperature of a black hole with mass $ M \equiv M_{10} \times 10^{10}\,\mathrm g $ is
\begin{equation}
T_\mathrm{BH}
\approx
  1.06\,M_{10}^{-1}\,\mathrm{TeV}\,.
\end{equation}
This assumes that the hole has no charge or angular momentum, since charge and angular momentum are usually assumed to be lost through quantum emission on a shorter timescale than the mass, although this may fail at the Planck scale~\cite{Lehmann:2019zgt}.
The emission is not exactly black-body but depends upon the spin and charge of the emitted particle, the average energy for neutrinos, electrons and photons being $4.22\,T_\mathrm{BH}$\,, $4.18\,T_\mathrm{BH}$ and $5.71\,T_\mathrm{BH}$\,, respectively \cite{Page:1976df}.

The mass loss rate of an evaporating black hole can be expressed as
\begin{equation}
\frac{\mathrm dM_{10}}{\mathrm dt}
= -5.34 \times 10^{-5}\,f(M)\,M_{10}^{-2}\,\mathrm s^{-1}\,.
\end{equation}
Here $ f(M) $ is a measure of the number of emitted particle species, normalised to unity for a black hole with $ M \gg 10^{17}\,\mathrm g $, this emitting only particles which are (effectively) massless: photons, three generations of neutrinos and antineutrinos, and gravitons.
The contribution of each relativistic degree of freedom to $ f(M) $ is \cite{MacGibbon:1991tj}
\begin{equation}
\begin{aligned}
&
f_{s=0} = 0.267\,,
\quad
f_{s=1} = 0.060\,,
\quad
f_{s=3/2} = 0.020\,,
\quad
f_{s=2} = 0.007\,, \\
&
f_{s=1/2} = 0.147~\text{(neutral)}\,,
\quad
f_{s=1/2} = 0.142~\text{(charge $ \pm e $)}\,.
\end{aligned}
\label{eq:spin}
\end{equation}
Holes in the mass range $ 10^{15}\,\mathrm g < M < 10^{17}\,\mathrm g $ emit electrons but not muons, while those in the range $ 10^{14}\,\mathrm g < M < 10^{15}\,\mathrm g $ also emit muons, which subsequently decay into electrons and neutrinos.
The latter range is relevant for the PBHs which are completing their evaporation at the present epoch.

Once $ M $ falls to around $ 10^{14}\,\mathrm g $, a black hole can also begin to emit hadrons.
However, hadrons are composite particles made up of quarks held together by gluons.
For temperatures exceeding the QCD confinement scale, $ \Lambda_\mathrm{QCD} = 250\text{--}300\,\mathrm{MeV} $, one would expect these fundamental particles to be emitted rather than composite particles.
Only pions would be light enough to be emitted below $ \Lambda_\mathrm{QCD} $\,.
Above this temperature, the particles radiated can be regarded as asymptotically free, leading to the emission of quarks and gluons \cite{MacGibbon:1990zk}.
Since there are $ 12 $ quark degrees of freedom per flavour and $ 16 $ gluon degrees of freedom, one would expect the emission rate (i.e., the value of $ f $) to increase suddenly once the QCD temperature is reached.
If one includes just $ u $, $ d $ and $ s $ quarks and gluons, Eq.~\eqref{eq:spin} implies that their contribution to $ f $ is $ 3 \times 12 \times 0.14 + 16 \times 0.06 \approx 6 $, compared to the pre-QCD value of about $ 2 $.
Thus the value of $ f $ roughly quadruples, although there will be a further increase in $ f $ at somewhat higher temperatures due to the emission of the heavier quarks.
After their emission, quarks and gluons fragment into further quarks and gluons until they cluster into the observable hadrons when they have travelled a distance $ \Lambda_\mathrm{QCD}^{-1} \sim 10^{-13}\,\mathrm{cm} $.
This is much larger than the size of the hole, so gravitational effects can be neglected.

If we sum up the contributions from all the particles in the Standard Model up to $ 1\,\mathrm{TeV} $, corresponding to $ M_{10} \sim 1 $, this gives $ f(M) = 15.35 $.
Integrating the mass loss rate over time then gives a lifetime
\begin{equation}
\tau
\approx
  407\,
  \left(\frac{f(M)}{15.35}\right)^{-1}\,M_{10}^3\,\mathrm s\,.
\label{eq:tau}
\end{equation}
The mass of a PBH evaporating at time $ \tau $ after the big bang is then
\begin{equation}
M
\approx
  1.35 \times 10^9\,
  \left(\frac{f(M)}{15.35}\right)^{1/3}\,
  \left(\frac{\tau}{1 \,\mathrm s}\right)^{1/3}\,\mathrm g\,.
\label{eq:lifetime}
\end{equation}
The critical mass for which $ \tau $ equals the age of the Universe is denoted by $ M_* $\,.
For the currently favoured age of $ 13.8\,\mathrm{Gyr} $, one finds
\begin{equation}
M_*
\approx
  1.02 \times 10^{15}\,
  \left(\frac{f_*}{15.35}\right)^{1/3}\,\mathrm g
\approx
  5.1 \times 10^{14}\,\mathrm g\,,
\end{equation}
where the last step assumes $ f_* =1.9 $, the value associated with the temperature $ T_\mathrm{BH}(M_*) = 21\,\mathrm{MeV} $.
At this temperature muons and some pions are emitted, so the value of $ f_* $ accounts for this.
Although QCD effects are initially small for PBHs with $ M = M_* $, only contributing a few percent, they become important once $ M $ falls to
\begin{equation}
M_\mathrm q
\approx
  0.4\,M_*
\approx
  2 \times 10^{14}\,\mathrm g\,,
\label{eq:mq}
\end{equation}
since the peak energy becomes comparable to $ \Lambda_\mathrm{QCD} $ then.
This means that an appreciable fraction of the time-integrated emission from the PBHs evaporating at the present epoch goes into quark and gluon jet products.

It should be stressed that the above analysis is not exact because the value of $ f(M) $ in Eq.~\eqref{eq:lifetime} should really be the weighted average of $ f(M) $ over the lifetime of the black hole.
The more precise calculations of MacGibbon \cite{MacGibbon:1990zk,MacGibbon:1991tj,MacGibbon:2007yq} give the slightly smaller value $ M_* = 5.00 \times 10^{14}\,\mathrm g $.
However, the weighted average is well approximated by $ f(M) $ unless one is close to a particle mass threshold.
For example, since the lifetime of a black hole of mass $ 0.4\,M_* $ is roughly $ 0.25 \times (0.4)^3 = 0.016 $ times that of an $ M_* $ black hole, one expects the value of $ M_* $ to be overestimated by a few percent.
This explains the small difference from MacGibbon's calculation.

\subsubsection{Primary and secondary emission}
\label{sec:spectra}

Particles injected from PBHs have two components: the \emph{primary} component, which is the direct Hawking emission, and the \emph{secondary} component, which comes from the decay of gauge bosons or heavy leptons and the hadrons produced by fragmentation of primary quarks and gluons.
For example, the photon spectrum can be written as
\begin{equation}
\frac{\mathrm d\dot N_\gamma}{\mathrm dE_\gamma}(E_\gamma,M)
= \frac{\mathrm d\dot N^\mathrm{pri}_\gamma}{\mathrm dE_\gamma}(E_\gamma,M)
  + \frac{\mathrm d\dot N^\mathrm{sec}_\gamma}{\mathrm dE_\gamma}(E_\gamma,M)\,,
\end{equation}
where an overdot denotes a time derivative, with similar expressions for other particles.
In order to treat QCD fragmentation, CKSY use the PYTHIA code (version 6), a Monte Carlo event generator constructed to fit hadron fragmentation for centre-of-mass energies $ \sqrt s \lesssim 200\,\mathrm{GeV} $.
The pioneering work of MacGibbon and Webber used the Herwig code but obtained similar results \cite{MacGibbon:1990zk}.

\begin{figure}[ht]
\begin{center}
\includegraphics[width=.5\textwidth]{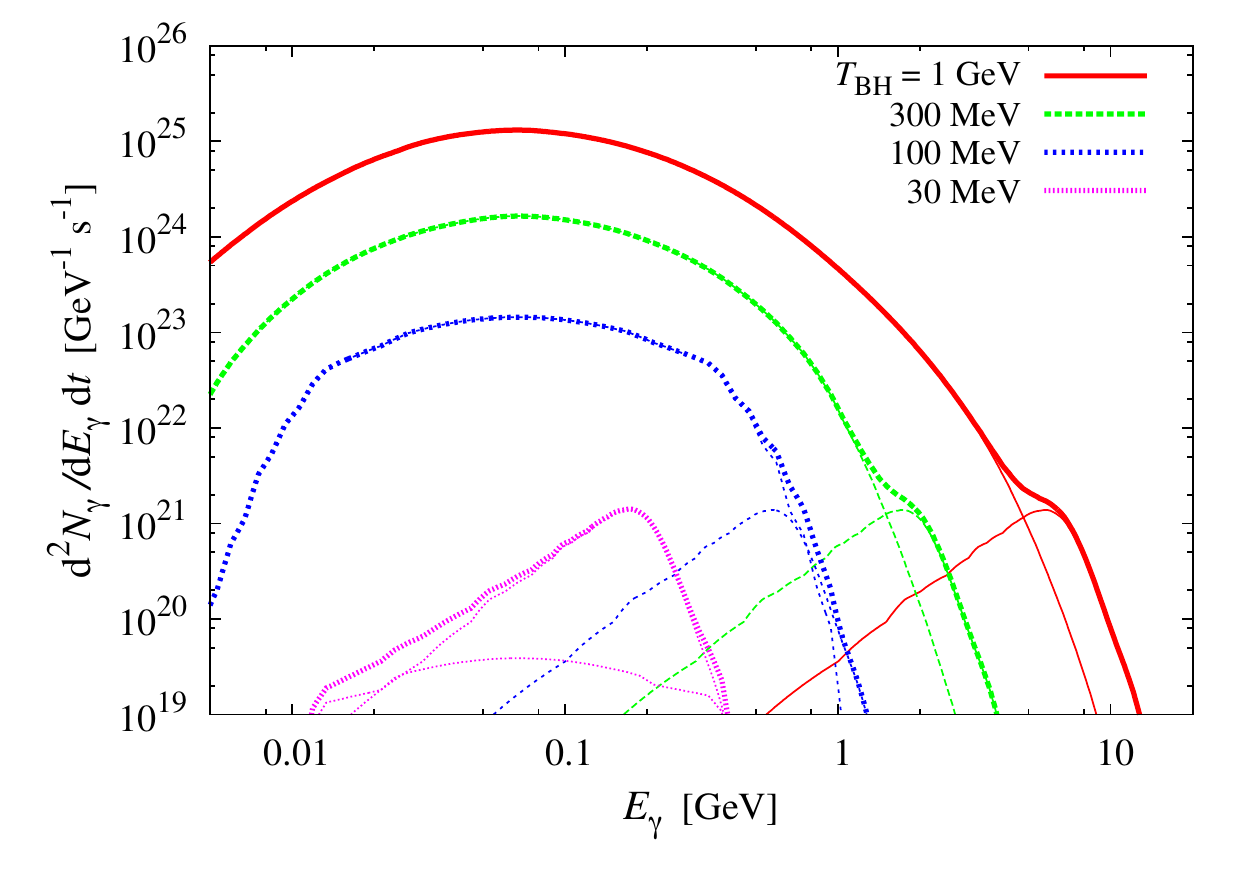}
\end{center}
\caption{Instantaneous emission rate of photons for four typical black hole temperatures, from Ref.~\cite{Carr:2009jm}.
For each temperature, the curve with the right (left) peak represents the primary (secondary) component and the thick curve denotes their sum.}
\label{fig:rate}
\end{figure}

The spectrum of secondary photons is peaked around $ E_\gamma \simeq m_{\pi^0}/2 \approx 68\,\mathrm{MeV} $, because it is dominated by the $ 2\gamma $-decay of soft neutral pions which are practically at rest.
The peak flux can be expressed as
\begin{equation}
\frac{\mathrm d\dot N_\gamma^\mathrm{sec}}{\mathrm dE_\gamma}
(E_\gamma=m_{\pi^0}/2)
\simeq
  2\,\sum_{i=q,g} \mathcal B_{i \to \pi^0}(\overline{E},E_{\pi^0})\,
  \frac{\overline{E}}{m_{\pi^0}}\,
  \frac{\mathrm d\dot N_i^\mathrm{pri}}{\mathrm dE_i}
  (E_i \simeq \overline{E})\,,
\label{eq:peakfluxaprx}
\end{equation}
where $ \mathcal B_{q,g \to\pi^0}(E_\mathrm{jet},E_{\pi^0}) $ is the fraction of the jet energy $ E_\mathrm{jet} $ going into neutral pions of energy $ E_{\pi^0} $\,.
This is of order $ 0.1 $ and fairly independent of jet energy.
If we assume that most of the primary particles have the average energy $ \overline{E} \approx 4.4\,T_\mathrm{BH} $\,, the last factor becomes $\mathrm d\dot N_i^\mathrm{pri}/\mathrm dE_i \approx 1.6 \times 10^{-3}\,\hbar^{-1} $.
Thus the energy dependence of Eq.~\eqref{eq:peakfluxaprx} comes entirely from the factor $ \overline{E}$ and is proportional to the Hawking temperature.
The emission rates of primary and secondary photons for four typical temperatures are shown in Fig.~\ref{fig:rate}.

\begin{figure}[ht]
\begin{center}
\includegraphics[width=.5\textwidth]{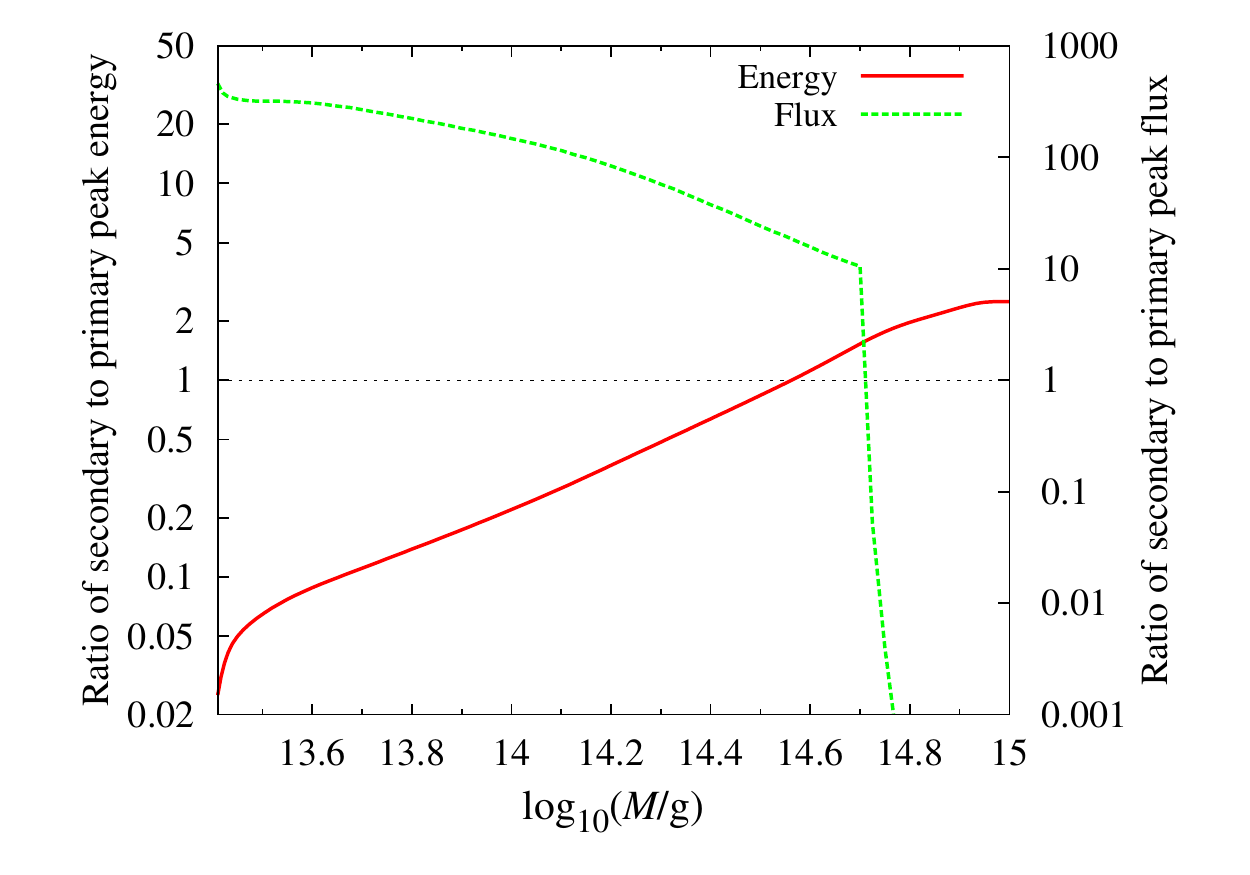}
\end{center}
\caption{Ratios of secondary to primary peak energies (solid) and fluxes (dotted), from Ref.~\cite{Carr:2009jm}.}
\label{fig:ratios}
\end{figure}

It should be noted that the \emph{time-integrated} ratio of the secondary flux to the primary flux increases rapidly once $ M $ goes below $ M_* $\,.
This is because a black hole with $ M = M_* $ will emit quarks efficiently once its mass gets down to the value $ M_\mathrm q $ given by Eq.~\eqref{eq:mq} and this corresponds to an appreciable fraction of its original mass.
On the other hand, a PBH with somewhat larger initial mass, $ M = (1+\mu)\,M_* $\,, will today have a mass \cite{Carr:2009jm}
\begin{equation}
m \equiv M(t_0)
\approx
  (3\,\mu)^{1/3} M_* \quad (\mu \ll 1)\,.
\label{eq:mu}
\end{equation}
Here we have assumed $ f(M) \approx f_* $\,, which should be a good approximation for $ m > M_\mathrm q $ since the value of $ f $ only changes slowly above the QCD threshold.
However, $ m $ falls below $ M_\mathrm q $ for $ \mu < 0.02 $ and if we assume that $f$ jumps discontinuously from $ f_* $ to $ \alpha\,f_* $ at this mass, then Eq.~\eqref{eq:mu} must be reduced by a factor $\alpha^{1/3}$.
The fact that this happens only for $ \mu < 0.02 $ means that the fraction of the black hole mass going into secondaries falls off sharply above $ M_* $\,.
The ratio of the energies at which the spectra peak is
\begin{equation}
\bar E^\mathrm S / \bar E^\mathrm P
\approx
  (68\,\mathrm{MeV})/(600\,m_{14}^{-1}\,\mathrm{MeV})
\approx
  0.6\,(m/M_*)\,,
\end{equation}
while the flux ratio is
\begin{equation}
\biggl.\biggl(\frac{\mathrm d\dot N^\mathrm S}{\mathrm dE}\biggr)_{\bar E^\mathrm S}\biggr/
\biggl(\frac{\mathrm d\dot N^\mathrm P}{\mathrm dE}\biggr)_{\bar E^\mathrm P}
\approx
  1.4\,\left(\frac{m}{M_*}\right)^{-1}\,\mathrm e^{-\chi\,m/M_\mathrm q}
\end{equation}
for some constant $\chi$.
The time-integrated ratios of the peak energies and fluxes are shown in Fig.~\ref{fig:ratios}.
These relations are important when calculating the effects of PBHs evaporating at the present epoch (e.g.\ the $\gamma$-ray background generated by PBHs in the Galactic halo).

\section{Constraints on evaporating PBHs}

\begin{figure}[ht]
\begin{center}
\includegraphics[width=.60\textwidth]{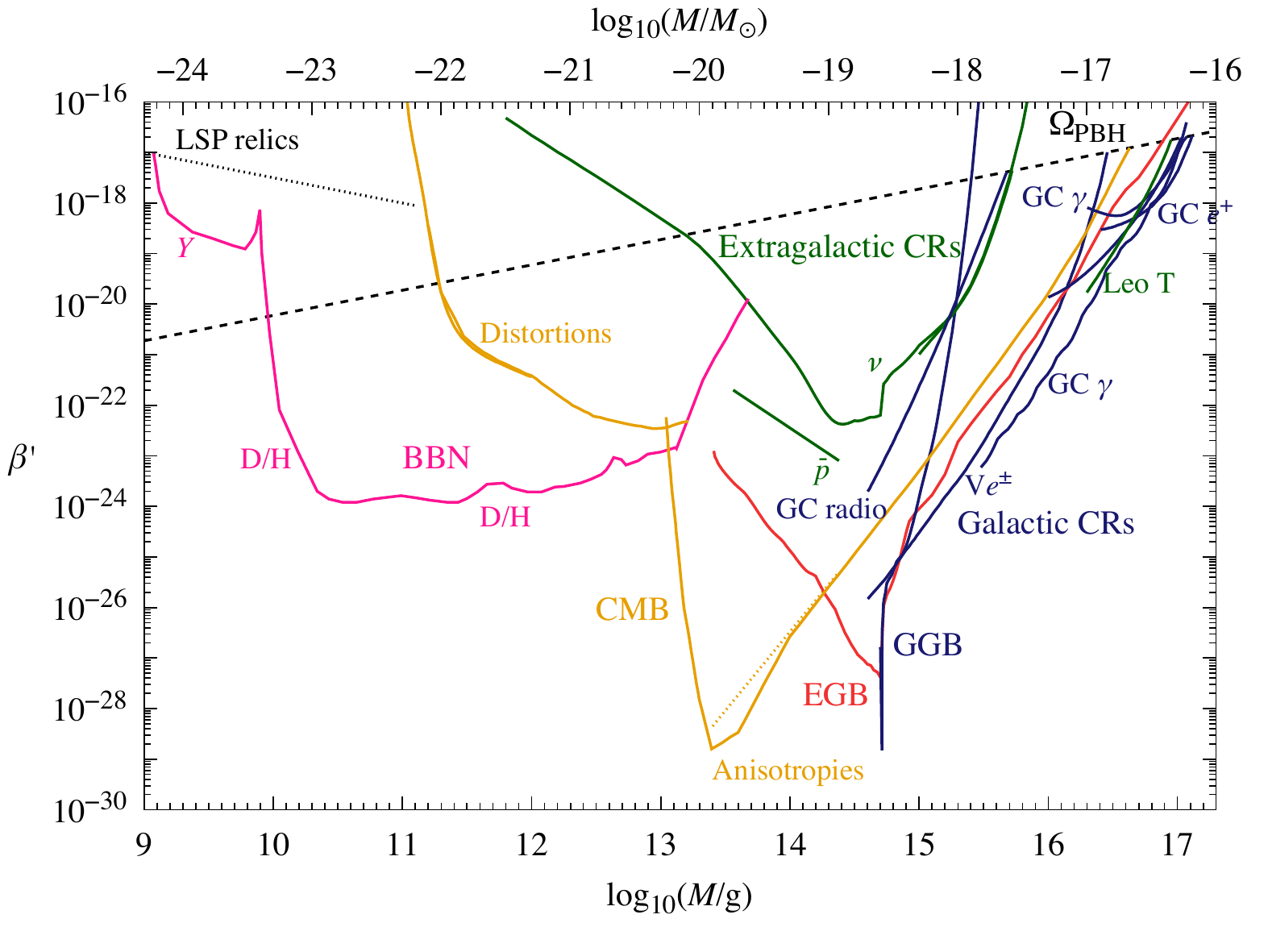}
\end{center}
\caption{Combined BBN (pink) and EGB (red) limits \cite{Carr:2009jm}, compared to other constraints on evaporating PBHs from LSP relics \cite{Lemoine:2000sq} (dotted grey since dependent on LSP mass), the CMB spectral distortions \cite{Acharya:2020jbv,Chluba:2020oip} and anisotropies \cite{Acharya:2020jbv} (yellow), extragalactic antiprotons \cite{Carr:2009jm} and neutrinos \cite{Carr:2009jm,Dasgupta:2019cae} (green), the GGB \cite{Carr:2016hva} and the electrons and positrons observed by Voyager 1 \cite{Boudaud:2018hqb} (blue).
Less secure limits come from observations of primary photons~\cite{Laha:2020ivk,Coogan:2020tuf}, electron-positron annihilation~\cite{DeRocco:2019fjq,Laha:2019ssq} and radio emission from the GC~\cite{Chan:2020zry}(blue since Galactic) and Leo T~\cite{Kim:2020ngi} (green since extragalactic).
The dashed line is the only constraint for no Hawking radiation.}
\label{fig:combined}
\end{figure}

In this section we discuss the constraints on evaporating PBHs.
The results are summarised in Fig.~\ref{fig:combined}, which is an update of Fig.~6 of Ref.~\cite{Carr:2009jm}.
The important point is that the BBN, CMB and extragalactic $ \gamma $-ray background (EGB) limits are the most stringent ones over almost the entire mass range $ 10^9\text{--}10^{17}\,\mathrm g $.
There is just a small range between $10^{14.7}\,\mathrm g$ and $10^{16.1}\,\mathrm g$ where the Galactic $\gamma$-ray background (GGB) and Galactic electron and positron limits dominate.
Obviously none of these constraints would apply if there were no Hawking radiation.
In this case, the only constraint would come from the condition $\Omega_\mathrm{PBH} < 1$, as indicated by the broken line in Fig.~\ref{fig:combined}.
In this context, Raidal \textit{et al.} have suggested that large primordial curvature fluctuations could collapse into horizonless exotic compact objects instead of PBHs \cite{Raidal:2018eoo}.
In this case, they either do not evaporate at all or they do so much more slowly.

\subsection{Big bang nucleosynthesis}

PBHs with $ M \sim 10^{10}\,\mathrm g $ and $ T_\mathrm{BH} \sim 1\,\mathrm{TeV} $ have a lifetime $ \tau \sim 10^3\,\mathrm s $ and therefore evaporate at the epoch of cosmological nucleosynthesis.
The effect of these evaporations on BBN has been a subject of long-standing interest.
Injection of high-energy neutrinos and antineutrinos \cite{1978AZh....55..231V} changes the weak interaction freeze-out time and hence the neutron-to-proton ratio at the onset of BBN, which changes $ {}^4\mathrm{He} $ production.
Since PBHs with $ M = 10^9\text{--}10^{13}\,\mathrm g $ evaporated during or after BBN, the baryon-to-entropy ratio at nucleosynthesis would be increased, resulting in overproduction of $ {}^4\mathrm{He} $ and underproduction of $ \mathrm D $ \cite{Miyama:1978mp}.
Emission of high-energy nucleons and antinucleons \cite{1977PAZh....3..208Z} increases the primordial deuterium abundance due to capture of free neutrons by protons and spallation of $ {}^4\mathrm{He} $.
The emission of photons by PBHs with $ M > 10^{10}\,\mathrm g $ dissociates the deuterons produced in nucleosynthesis \cite{1980MNRAS.193..593L,Ellis:1990nb}.
The limits associated with these effects are shown in Fig.~\ref{fig:bc1}.

Observational data on both the light element abundances and the neutron lifetime have changed since these early papers.
Much more significant, however, have been developments in our understanding of the fragmentation of quark and gluon jets from PBHs into hadrons.
Most of the hadrons created decay almost instantaneously compared to the timescale of nucleosynthesis but long-lived ones (such as pions, kaons and nucleons) remain long enough in the ambient medium to leave an observable signature on BBN.
These effects were first discussed by Kohri and Yokoyama \cite{Kohri:1999ex} for the relatively low mass PBHs evaporating in the early stages of BBN but the analysis was later extended to incorporate the effects of heavier PBHs evaporating after BBN \cite{Carr:2009jm}, the hadrons and high energy photons from these PBHs further dissociating synthesised light elements.

\subsubsection{Modern studies}

High energy particles emitted by PBHs modify the standard BBN scenario in three different ways:
(1) high energy mesons and antinucleons induce extra interconversion between background protons and neutrons even after the weak interaction has frozen out in the background Universe;
(2) high energy hadrons dissociate light elements synthesised in BBN, thereby reducing $ {}^4\mathrm{He} $ and increasing $ \mathrm D $, $ \mathrm T $, $ {}^3\mathrm{He} $, $ {}^6\mathrm{Li} $ and $ {}^7\mathrm{Li} $;
(3) high energy photons generated in the cascade further dissociate $ {}^4\mathrm{He} $ to increase the abundance of lighter elements even more.

The PBH constraints depend on three parameters: the initial baryon-to-photon ratio $ \eta_\mathrm i $\,, the PBH initial mass $ M $ or (equivalently) its lifetime $ \tau $, and the initial PBH number density normalised to the entropy density, $ Y_\mathrm{PBH} \equiv n_\mathrm{PBH}/s $.
From Eq.~\eqref{eq:beta} this is related to the initial mass fraction $ \beta' $ by
\begin{equation}
\beta'
= 5.4 \times 10^{21}\,
  \left(\frac{\tau}{1\,\mathrm s}\right)^{1/2}\,
  Y_\mathrm{PBH}\,.
\end{equation}
The parameters $ \beta' $, $ \tau $ and $ Y_\mathrm{PBH} $ all depend on $ M $ but we suppose a monochromatic mass function in what follows.
The initial baryon-to-photon ratio is set to the present one after allowing for entropy production from PBH evaporations and photon heating due to $ e^+ e^- $ annihilations.

\begin{figure}[ht]
\begin{center}
\includegraphics[scale=0.45]{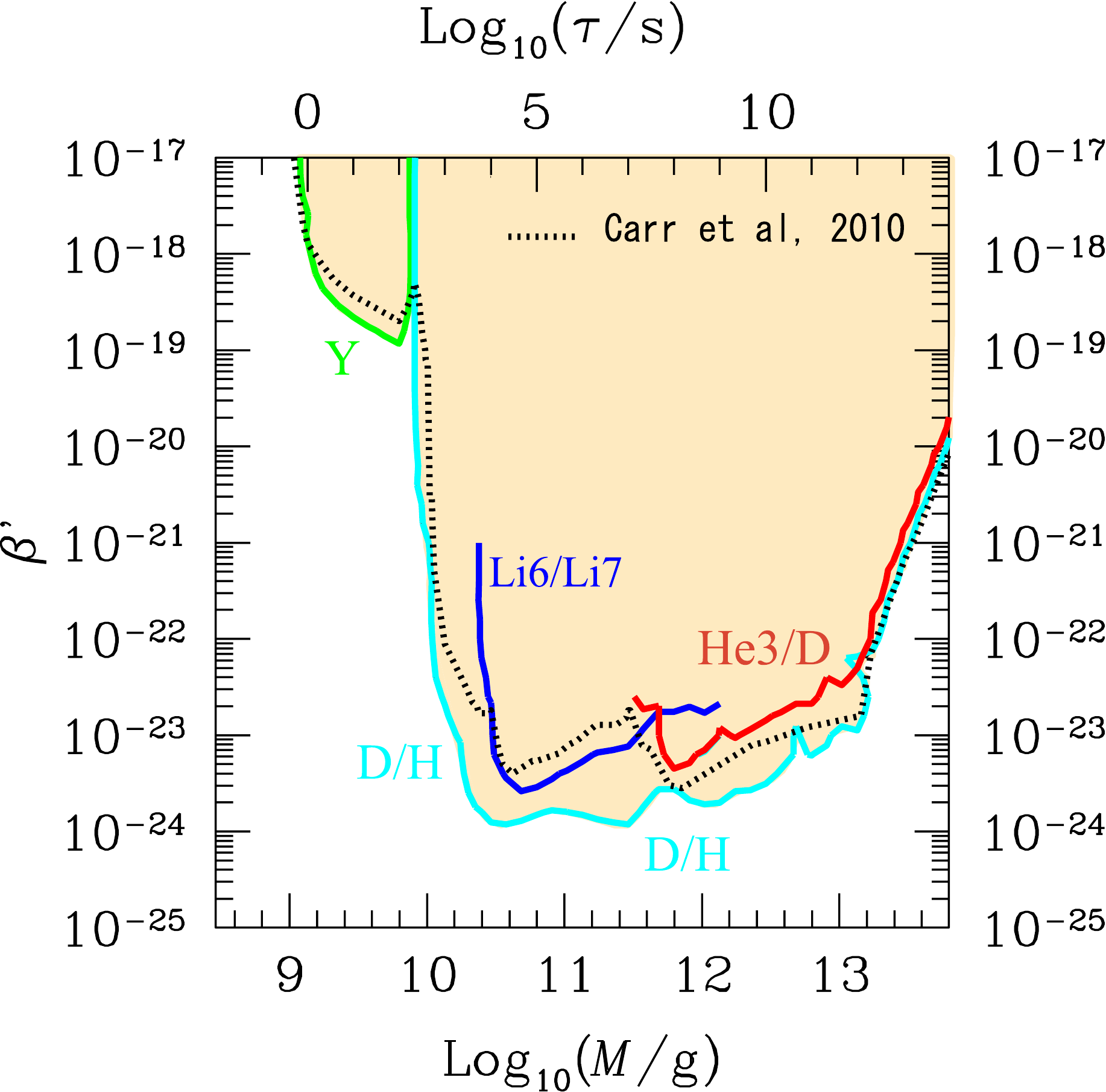}
\end{center}
\caption{Upper bounds on $ \beta'(M) $ at 95\,\% confidence level required for success of BBN model, using data from Refs.~\cite{Kohri:1999ex,Kawasaki:2017bqm,Kawasaki:2020qxm}.
The broken line gives the earlier limit of Ref.~\cite{Carr:2009jm}.}
\label{fig:bbn}
\end{figure}

Figure~\ref{fig:bbn} shows the 95\,\% confidence-level limits which result from these calculations (see CKSY for further details).
PBHs with lifetime shorter than $ 10^{-2}\,\mathrm s $ are free from BBN constraints because they evaporate well before weak freeze-out and leave no trace.
PBHs with $ M \approx 10^9\text{--}10^{10}\,\mathrm g $ and lifetime $ \tau \approx 10^{-2}\text{--}10^2\,\mathrm s $ are constrained by the extra interconversion between protons and neutrons due to emitted mesons and antinucleons, which increases the $ n/p $ freeze-out ratio as well as the final $ {}^4\mathrm{He} $ abundance.
For $ \tau \approx 10^2\text{--}10^7\,\mathrm s $, corresponding to $ M \approx 10^{10}\text{--}10^{12}\,\mathrm g $, hadro-dissociation becomes important and the debris deuterons and non-thermally produced $ {}^6\mathrm{Li} $ put constraints on $ \beta(M) $.
Finally, for $ \tau \approx 10^7\text{--}10^{12}\,\mathrm s $, corresponding to $ M \approx 10^{12}\text{--}10^{13} \,\mathrm g $, energetic neutrons decay before inducing hadro-dissociation.
Instead, photodissociation processes are operative and the most stringent constraint comes from overproduction of $ {}^3\mathrm{He} $.
However, even these effects become insignificant after $ 10^{12}\,\mathrm s $.

These computations involved Monte-Carlo simulations and included experimental and observational uncertainties in the baryon number and reaction and decay rates to obtain error bars in the light element abundances.
The bounds should therefore be much more conservative than the ones obtained without these uncertainties.
We adopted the upper bound on the abundance of ${}^3\mathrm{He}/\mathrm{D}$ rather than ${}^3\mathrm{He}/\mathrm{H}$ since it is more reliable.
Because ${}^3\mathrm{He}$ is both produced and destroyed in low-mass stars, the observed local value of ${}^3\mathrm{He}/\mathrm{H}$ does not imply an upper bound on the primordial component~\cite{Kawasaki:2004qu}.
However, the ${}^3\mathrm{He}/\mathrm{D}$ observed in the solar system should give a reasonable upper bound on the primordial value because this increases with cosmic time~\cite{Kawasaki:2004yh}.
In addition, we included both hadronic and radiative emissions;
these can occur together and cancel each other, which reduces the constraint in overlapping regions.
Reference~\cite{Acharya:2020jbv} claims stronger BBN constraints on evaporating PBHs but does not consider these effects.
For comparison, Fig.~\ref{fig:bbn} shows as a broken line the earlier BBN limit obtained in Ref.~\cite{Carr:2009jm}.
The deuterium limit gives the strongest bound for $M \gtrsim 10^{10}\,\mathrm g$ because hadro-dissociation of helium produces more deuterium, which is severely constrained by more precise recent observational data~\cite{Kawasaki:2017bqm,Kawasaki:2020qxm}.

Keith \textit{et al.}~\cite{Keith:2020jww} have revisited the BBN constraints, making use of more recent deuterium and helium measurements and considering how physics beyond the Standard Model could impact the constraints.
They discuss scenarios which feature large hidden sectors, as well as models of TeV-scale supersymmetry.
Although their constraints seem to be stronger than ours, they depend on limited data taken from figures of Kawasaki \textit{et al.}~\cite{Kawasaki:2017bqm} and this neglects a lot of relevant information, so we do not include them in Fig.~\ref{fig:bbn}.

\subsubsection{Lithium-7}

Overproduction of ${}^7\mathrm{Li}$ by a factor $\sim 3$ is the most serious issue with the standard cosmological model \cite{Coc:2017rpv}.
Astrophysical solutions have so far failed \cite{Korn:2006tv,2015MNRAS.452.3256F}.
Particle decay models reduce lithium via neutron injection but generically overproduce $\mathrm{D}$ \cite{Coc:2014gia}, for which there are increasingly precise constraints.
Therefore more exotic particle decays -- involving neutron-triggered destruction of ${}^7\mathrm{Be}$, followed by $\gamma$-ray destruction of excess $\mathrm{D}$ -- have been suggested, notably those associated with a leptophilic metastable massive particle of mass $\sim 10\,\mathrm{MeV}$ \cite{Goudelis:2015wpa}.
However, PBHs evaporating at $\sim 10^{2}\text{--}10^{4}\,\mathrm s$ might provide a plausible alternative.
For example, PBHs with an extended mass function around $10^{12}\,\mathrm g$ could provide an early injection of neutrons, followed by soft $\gamma$-rays.
This combination could destroy some ${}^7\mathrm{Be}$ at $\sim 10^2\,\mathrm s$, avoiding overproduction of ${}^7\mathrm{Li}$, and subsequently destroy excess $\mathrm{D}$ at $\sim 10^{3}\text{--}10^{4}\,\mathrm s$.

\subsection{Cosmic microwave background}

\subsubsection{Generation of entropy}

The effects of PBH evaporations on the CMB were first analysed by Zel'dovich \textit{et al.} \cite{1977PAZh....3..208Z}.
They pointed out that photons from PBHs smaller than $10^9\,\mathrm g$ are emitted sufficiently early to be completely thermalised and merely contribute to the photon-to-baryon ratio.
The requirement that this does not exceed the observed ratio of around $ 10^9 $ leads to a limit
\begin{equation}
\beta'(M)
< 10^9\,\left(\frac{M}{M_\mathrm{Pl}}\right)^{-1}
\approx
  10^{-5}\,\left(\frac{M}{10^9\,\mathrm g}\right)^{-1}
\quad
(M < 10^9\,\mathrm g)\,,
\label{eq:entropy}
\end{equation}
where $ M_\mathrm{Pl} $ is the Planck mass, so only PBHs below $ 10^4\,\mathrm g $ could generate \emph{all} of the CMB.
This limit is not shown in Fig.~\ref{fig:combined} because it is very weak.

\subsubsection{CMB spectral distortions}

Zel'dovich \text{et al.}~\cite{1977PAZh....3..208Z} also noted that photons from PBHs in the range $ 10^{11}\,\mathrm g < M < 10^{13}\,\mathrm g $, although partially thermalised, will produce noticeable distortions in the CMB spectrum unless
\begin{equation}
\beta'(M)
< \left(\frac{M}{M_\mathrm{Pl}}\right)^{-1}
\approx
  10^{-16}\,\left(\frac{M}{10^{11}\,\mathrm g}\right)^{-1}
\quad
(10^{11}\,\mathrm g < M < 10^{13}\,\mathrm g)\,.
\label{eq:distort}
\end{equation}
This corresponds to the fraction of the density in PBHs being less than unity at evaporation.
In the intermediate mass range, $10^{9}\text{--}10^{11}\,\mathrm g$, there is a transition from limit \eqref{eq:entropy} to the much stronger limit \eqref{eq:distort}.
Subsequently the form of these distortions has been analysed in more detail.
If an appreciable number of photons are emitted after the freeze-out of double-Compton scattering ($ t > 7 \times 10^6\,\mathrm s $), corresponding to $ M > 10^{11}\,\mathrm g $, the distribution of the CMB photons develops a non-zero chemical potential, leading to a $ \mu $-distortion.
If the photons are emitted after the freeze-out of the single-Compton scattering ($ t > 3 \times 10^9\,\mathrm s $), corresponding to $ M > 10^{12}\,\mathrm g $, the distribution is modified by a $ y $-distortion.

These constraints were first calculated in the context of decaying particle models \cite{Sarkar:1984tt,Ellis:1990nb,Hu:1993gc}.
In the PBH context, calculations of Tashiro and Sugiyama \cite{Tashiro:2008sf} showed that the CMB distortion constraints are of order $ \beta'(M) < 10^{-21} $ for some range of $ M $, which is weaker than the BBN constraint but stronger than the constraint given by Eq.~\eqref{eq:distort}.
Later Lucca \textit{et al.}~\cite{Lucca:2019rxf} pointed out some limitations of the Tashiro--Sugiyama analysis and on the basis of some strong assumptions significantly improved the PBH constraints.
Acharya and Khatri~\cite{Acharya:2019xla} and Chluba \textit{et al.}~\cite{Chluba:2020oip} also derived stronger spectral distortion constraints by including secondary emission of electromagnetic particles but they used more conservative assumptions than Lucca \textit{et al.}, whom they criticised for assuming that all the emitted energy goes into non-relativistic $y$ and $\mu$ distortions.
Figure~\ref{fig:combined} uses the $y$ and $\mu$ bounds found in a later paper by Acharya and Khatri~\cite{Acharya:2020jbv} and the $\mu$ bound found by Chluba \textit{et al.}~\cite{Chluba:2020oip}.
PBH evaporations could also be constrained by their effect on the form of the recombination lines in the CMB spectrum \cite{Sunyaev:2009qn}, just as in the annihilating DM scenario \cite{Chluba:2009uv,Slatyer:2009yq}.

\subsubsection{CMB anisotropies}

Another constraint on PBHs evaporating after the time of recombination is associated with the damping of small-scale CMB anisotropies.
This limit was obtained in Ref.~\cite{Carr:2009jm} by modifying an equivalent calculation for decaying particles, first described by Adams \textit{et al.}~\cite{Adams:1998nr}.
According to Zhang \textit{et al.} \cite{Zhang:2007zzh}, the latter constraint can be written in the form
\begin{equation}
\log_{10}\zeta
< -10.8 - 0.50\,x + 0.085\,x^2 + 0.0045\,x^3\,,
\quad
x
\equiv
  \log_{10}\left(\frac{\Gamma}{10^{-13}\,\mathrm s^{-1}}\right)\,,
\label{eq:zeta}
\end{equation}
where $ \Gamma $ is the decay rate, which corresponds to $ \tau(M)^{-1} $ in the PBH case, and $ \zeta $ is the fraction of the CDM in PBHs times the fraction of the emitted energy which goes into heating the matter.
The last factor, which includes the effects of the electrons and positrons as well as the photons, will be denoted by $ f_\mathrm H $ and depends on the redshift \cite{MacGibbon:1991vc,Chen:2003gz}.
Most of the heating will be associated with the electrons and positrons; they are initially degraded by inverse Compton scattering off the CMB photons but after scattering have an energy
\begin{equation}
\gamma^2\,E_\mathrm{CMB}
\approx
  300\,\left(\frac{\gamma}{10^3}\right)^2\,(1+z)\,\mathrm{eV}
\approx
  5 \times 10^5\,\left(\frac{M}{10^{13}\,\mathrm g}\right)^{-2}\,(1+z)\,\mathrm{eV}
\approx
  20\,(1+z)^2\,\mathrm{eV}\,,
\end{equation}
where $\gamma$ is the Lorentz factor and in the last expression we assume that the mass of a PBH evaporating at redshift $z$ in the matter-dominated era is $M \approx M_*\,(1+z)^{-1/2}$.
Since this energy is always above the ionisation threshold for hydrogen ($13.6\,\mathrm{eV}$), we can assume that the heating of the electrons and positrons is efficient before reionisation.
Using Eq.~\eqref{eq:beta} for $ \beta(M) $ and Eq.~\eqref{eq:tau} for $ \tau(M) $, one can now express Eq.~\eqref{eq:zeta} as a limit on $ \beta(M) $.
In the mass range of interest, the rather complicated cubic expression in $ M $ can be fitted by the approximation
\begin{equation}
\beta'(M)
< 3 \times 10^{-30}\,
  \left(\frac{f_\mathrm H}{0.1}\right)^{-1}\,
  \left(\frac{M}{10^{13}\,\mathrm g}\right)^{3.1}
\quad
(2.5 \times 10^{13}\,\mathrm g \lesssim M \lesssim 2.4 \times 10^{14}\,\mathrm g)\,,
\label{eq:cmb}
\end{equation}
where $ f_\mathrm H \approx 0.1 $ is the fraction of emission in electrons and positrons.
Here the lower mass limit corresponds to PBHs evaporating at recombination and the upper one to those evaporating at a redshift $ 6 $~\cite{Fan:2006dp}, after which the ionisation ensures the opacity is too low for the emitted electrons and positrons to heat the matter.

Equation~\eqref{eq:cmb} is stronger than all the other available limits in this mass range but had not been pointed out before CKSY.
Recently it has been studied more carefully by Poulin \textit{et al.} \cite{Poulin:2016anj}, who compute CMB anisotropy constraints on electromagnetic energy injection over a large range of timescales.
They apply their formalism for PBHs in the range $10^{13.5}\,\mathrm g < M < 10^{16.8}\,\mathrm g$, showing that the anisotropy constraints are stronger than the EGB limit below around $10^{14}\,\mathrm g$.
St\"{o}cker \textit{et al.} \cite{Stocker:2018avm} have followed up on this work, using the Boltzmann code CLASS.
Their bounds are several orders of magnitude stronger than the EGB limit in the range $10^{13.5}\,\mathrm g < M < 10^{14.5}\,\mathrm g$ and exclude PBHs with a monochromatic mass distribution in the range $10^{13.5}\,\mathrm g < M < 10^{16.3}\,\mathrm g$ from containing all of the DM.
Poulter \textit{et al.} \cite{Poulter:2019ooo} have extended these constraints to the case in which the PBHs have a non-monochromatic mass function.

Lucca \textit{et al.}~\cite{Lucca:2019rxf} find a stronger anisotropy constraint and their combined anisotropy/spectral distortion constraint appears to dominate other ones over a wide range of masses:
it is stronger than the BBN and EGB limit for $10^{14.4}\,\mathrm g > M > 10^{12.8}\,\mathrm g$, the EGB and GGB limit for $M > 10^{15}\,\mathrm g$ and the Voyager-1 electron and positron limit for $M > 10^{15.6}\,\mathrm g$.
However, Lucca \textit{et al.} use the ``on-the-spot'' limit, which assumes that all of the injected energy goes into thermal bath immediately.
This is a good approximation for spectral distortions generated before recombination but not for CMB anisotropies generated after recombination because non-thermal relativistic particles such as electrons and positrons only gradually transfer their energy into the thermal bath.
This means that the on-the-spot limit gives overly strong bounds.
Figure~\ref{fig:combined} adopts the anisotropy constraint given in the later paper by Acharya and Khatri~\cite{Acharya:2020jbv}, in which they calculate the energy transferred into the thermal plasma without adopting the on-the-spot limit.

\subsection{Extragalactic and Galactic $\gamma$-ray backgrounds}

\subsubsection{Extragalactic $\gamma$-ray background}
\label{sec:EGB}

One of the earliest works that applied the theory of black hole evaporation to astrophysics was carried out by Page and Hawking \cite{Page:1976wx}.
They used the diffuse EGB observations to constrain the mean cosmological number density of PBHs which are completing their evaporation at the present epoch to be less than $ 10^4\,\mathrm{pc}^{-3} $.
This corresponds to an upper limit on $ \Omega_\mathrm{PBH} $ of around $ 10^{-8} $.
The limit was subsequently refined by MacGibbon and Carr \cite{MacGibbon:1991vc} who considered how it is modified by including quark and gluon emission and inferred $ \Omega_\mathrm{PBH} \le (7.6 \pm 2.6) \times 10^{-9}\,h^{-2} $, corresponding to a $2 \sigma$ upper limit of $ 1.3 \times 10^{-8}\,h^{-2} $.
Later they used EGRET observations to derive a slightly stronger limit $ \Omega_\mathrm{PBH} \le (5.1 \pm 1.3) \times 10^{-9}\,h^{-2} $ \cite{Carr:1998fw} or a $2 \sigma$ upper limit of $ 7.7 \times 10^{-9}\,h^{-2} $.
Using the modern value of $ h $ gives $ \Omega_\mathrm{PBH} \le 1.5 \times 10^{-8} $ and this corresponds to $ \beta'(M_*) < 6 \times 10^{-26} $ from Eq.~\eqref{eq:density}.
They also inferred from the form of the $ \gamma $-ray spectrum that PBHs could not provide the \emph{dominant} contribution to the background.

\begin{figure}[ht]
\begin{center}
\includegraphics[width=.5\textwidth]{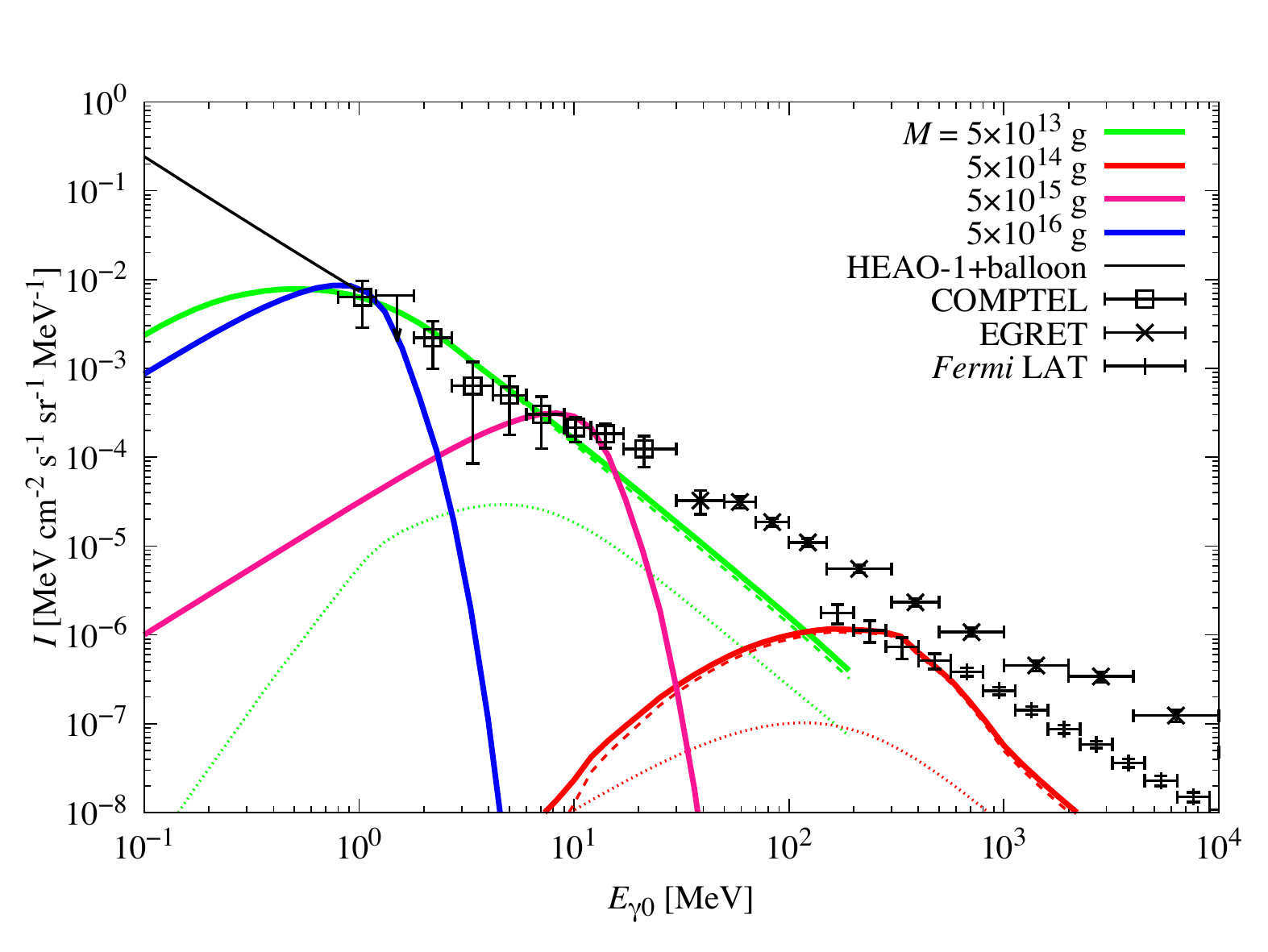}
\end{center}
\caption{Fluxes corresponding to the upper limit on the PBH abundance for various values of $ M $, updated from those in Ref.~\cite{Carr:2009jm} with new observational data from Fermi-LAT~\cite{Ackermann:2014usa}.
All PBHs produce primary photons but those with $ M \lesssim M_* $ also produce secondary photons (dotted lines), giving a stronger constraint on $ \beta $.}
\label{fig:flux}
\end{figure}

We now describe the more precise calculation of CKSY, evaluating the primary and secondary emission according to the prescription of Sec.~\ref{sec:spectra}.
In order to determine the present background spectrum of photons generated by PBH evaporations, we must integrate over the lifetime of the black holes, allowing for the fact that particles generated at earlier cosmological epochs will be redshifted in energy by now.
If the PBHs all have the same initial mass $ M $, and if we approximate the number of emitted photons in the energy bin $ \Delta E_\gamma \simeq E_\gamma $ by $ \dot N_\gamma(E_\gamma) \simeq E_\gamma\,(\mathrm d\dot N_\gamma/\mathrm dE_\gamma $), then the emission rate per volume at cosmological time $ t $ is
\begin{equation}
\frac{\mathrm dn_\gamma}{\mathrm dt}(E_\gamma,t)
\simeq
  n_\mathrm{PBH}(t)\,E_\gamma\,
  \frac{\mathrm d\dot N_\gamma}{\mathrm dE_\gamma}(M(t),E_\gamma)\,,
\end{equation}
where the $ t $-dependence of $ M $ just reflects the evaporation.
Since the photon energy and density are redshifted by factors $ (1+z)^{-1} $ and $ (1+z)^{-3} $, respectively, the present number density of photons with energy $ E_{\gamma 0} $ is
\begin{equation}
n_{\gamma 0}(E_{\gamma 0})
= n_{\mathrm{PBH}0}\,E_{\gamma 0}\,
  \int_{t_\mathrm{min}}^{\min(t_0,\tau)}\!\mathrm dt\,(1+z)\,
  \frac{\mathrm d\dot N_\gamma}{\mathrm dE_\gamma}(M(t),(1+z)\,E_{\gamma 0})\,,
\end{equation}
where $ t_\mathrm{min} $ corresponds to the earliest time at which the photons freely propagate and $ n_{\mathrm{PBH}0} $ is the current PBH number density for $ M > M_* $ or the number density they would have now had they not evaporated for $ M < M_* $\,.
The photon flux is
\begin{equation}
I
\equiv
  \frac{c}{4\pi}\,n_{\gamma 0}\,.
\end{equation}
The calculated present-day fluxes of primary and secondary photons are shown in Fig.~\ref{fig:flux}, where the number density $ n_{\mathrm{PBH}0} $ for each $ M $ has the maximum value consistent with the observations.

Note that the highest energy photons are associated with PBHs of mass $ M_* $\,.
Photons from PBHs with $ M > M_* $ are at lower energies because they are cooler, while photons from PBHs with $ M < M_* $ are at lower energies because (although initially hotter) they are redshifted.
The spectral shape depends on the mass $ M $ and can be easily understood.
PBHs with $ M > M_* $ have a rather sharp peak, well approximated by the instantaneous black-body emission of the primary photons, while PBHs with $ M \le M_* $ have an $ E_{\gamma 0}^{-2} $ fall-off for $ E_{\gamma 0} \gg T_\mathrm{BH}/[1+z(\tau)] $ due to their final phase of evaporation \cite{MacGibbon:1991tj}.

The relevant observations come from HEAO 1 and other balloon observations in the $ 3\text{--}500\,\mathrm{keV} $ range, COMPTEL in the $ 0.8\text{--}30\,\mathrm{MeV} $ range, EGRET in the $ 30 \text{--}100\,\mathrm{MeV} $ range and Fermi-LAT in the $ 100\,\mathrm{MeV}\text{--}820\,\mathrm{GeV} $ range.
All the observations are shown in Fig.~\ref{fig:flux}.
The origin of the diffuse X-ray and $ \gamma $-ray backgrounds is thought to be primarily distant astrophysical sources, such as blazars, and in principle one should remove these contributions before calculating the PBH constraints.
This is the strategy adopted by Barrau \textit{et al.} \cite{Barrau:2003nj}, who thereby obtain a limit $ \Omega_\mathrm{PBH} \le 3.3 \times 10^{-9}$.
CKSY did not attempt such a subtraction, so their constraints on $\beta'(M)$ were more conservative.

\begin{figure}[ht]
\begin{center}
\includegraphics[width=.5\textwidth]{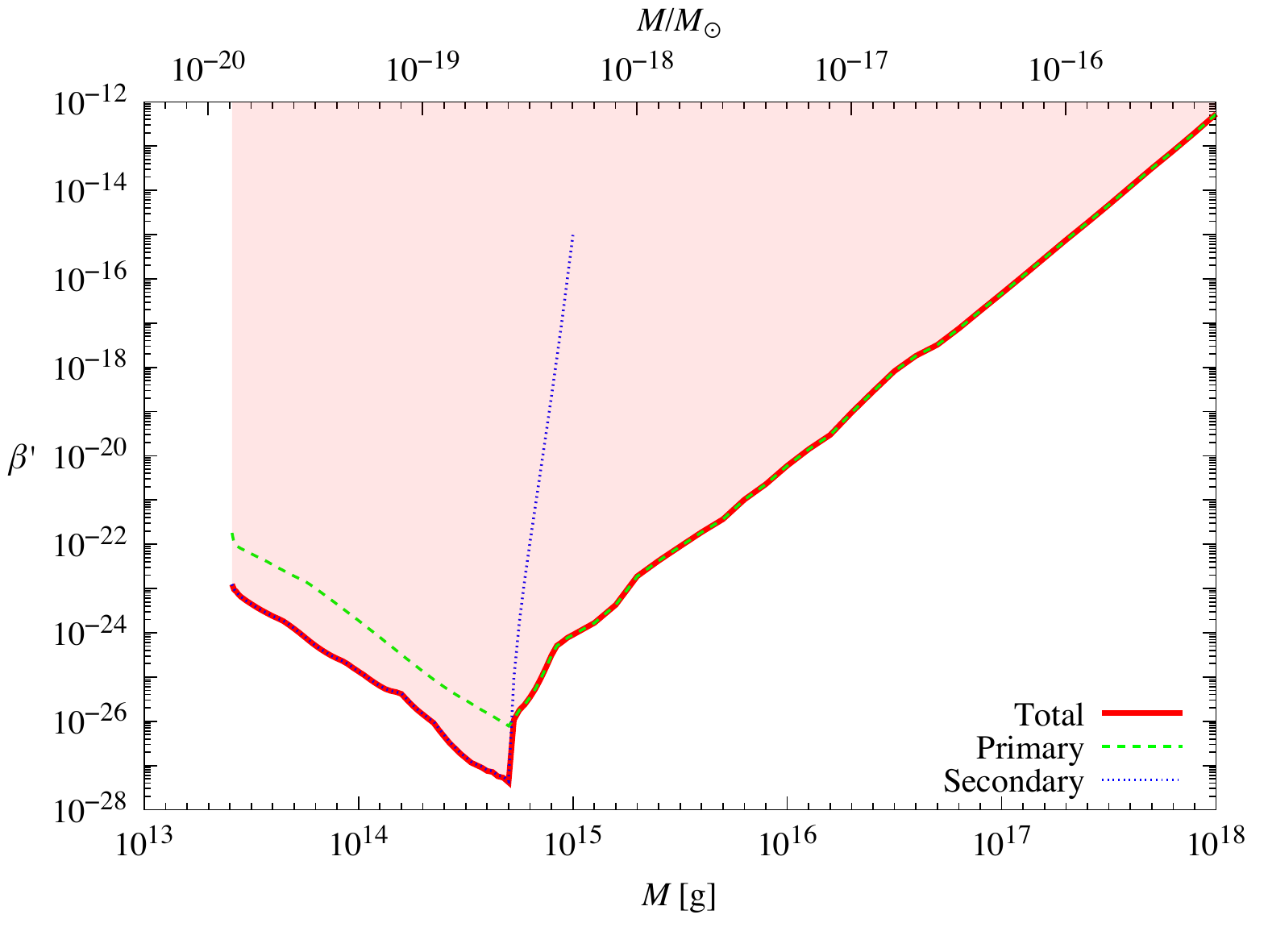}
\end{center}
\caption{Upper bounds on $ \beta'(M) $ from the extragalactic photon background, from Ref.~\cite{Carr:2009jm} but updated, with no other contributors to the background having been subtracted.}
\label{fig:photon}
\end{figure}

In order to analyse the spectra of photons emitted from PBHs, different treatments are needed for PBHs with initial masses below and above $ M_* $\,.
We saw in Sec.~\ref{sec:spectra} that PBHs with $ M > M_* $ can never emit secondary photons at the present epoch, whereas those with $ M \le M_* $ will do so once $ M $ falls below $M_\mathrm q \approx 2 \times 10^{14}\,\mathrm g $.
One can use simple analytical arguments to derive the form of the primary and secondary peak fluxes.
The observed X-ray and $ \gamma $-ray spectra correspond to $ I^\mathrm{obs} \propto E_{\gamma 0}^{-(1+\epsilon)} $ where $ \epsilon $ lies between $ 0.1 $ and $ 0.4 $.
For $ M < M_* $\,, the limit is determined by the secondary flux and one can write the upper bound on $ \beta' $ as
\begin{equation}
\beta'(M)
\lesssim
  5 \times 10^{-28}\,
  \left(\frac{M}{M_*}\right)^{-5/2-2\,\epsilon}
\quad
(M < M_*)\,.
\label{eq:photon1}
\end{equation}
For $ M > M_* $\,, secondary photons are not emitted and one obtains a limit
\begin{equation}
\beta'(M)
\lesssim
  5 \times 10^{-26}\,
  \left(\frac{M}{M_*}\right)^{7/2+\epsilon}
\quad
(M > M_*)\,.
\label{eq:photon2}
\end{equation}
These $ M $-dependences explain qualitatively the slopes in Fig.~\ref{fig:photon}, for which we assume $\epsilon = 0.4$.
The limit drops by a factor of $7$ below the mass $M_*$ because of secondary emission and bottoms out at $ 4 \times 10^{-28} $.
There is also a narrow band $M_{*} < M < 1.005\,M_{*}$ in which PBHs have not yet completed their evaporation even though their current mass is below the mass $M_\mathrm q \approx 0.4\,M_{*}$ at which quark and gluon jets are emitted.
From Eq.~\eqref{eq:density}, the associated limit on the density parameter is $ \Omega_\mathrm{PBH}(M_*) \le 7 \times 10^{-11} $.
This is factor $7$ stronger than the limit in CKSY.

We stress that the discontinuity in the EGB constraint at the mass $M_*$ is entirely a consequence of the (presumably unrealistic) assumption that the PBHs have an exactly monochromatic mass function and even a tiny mass width $\Delta M$ would suffice to smear this out.
Indeed, many derivations of the EGB constraint, including the original Page--Hawking calculation, assume that the PBHs have a power-law mass function.
In this case, the discontinuity is removed and the stronger ($M < M_*$) limit pertains.
Nevertheless, the discontinuity is still interesting in principle and the issue of non-monochromaticity will be even more important when we consider the Galactic background constraint.

Finally, we determine the mass range over which the $ \gamma $-ray background constraint applies.
Since photons emitted at sufficiently early times cannot propagate freely, there is a minimum mass $ M_\mathrm{min} $ below which the above constraint is inapplicable.
The dominant interactions between $ \gamma $-rays and the background Universe in the relevant energy range are pair-production off hydrogen and helium nuclei.
For the opacity appropriate for a 75\,\% hydrogen and 25\,\% helium mix, the redshift below which there is free propagation is given by \cite{MacGibbon:1991vc}
\begin{equation}
1 + z_\mathrm{max}
\approx
  1155\,\left(\frac{h}{0.67}\right)^{-2/3}\,\left(\frac{\Omega_\mathrm b}{0.05}\right)^{-2/3}\,,
\end{equation}
with the baryon density parameter $ \Omega_\mathrm b $ normalised to the modern value.
The condition $ \tau(M_\mathrm{min}) = t(z_\mathrm{max}) $ then gives
\begin{equation}
M_\mathrm{min}
= \left(\frac{t(z_\mathrm{max})}{t_0}\right)^{1/3}\,
  \left(\frac{f(M_\mathrm{min})}{f_*}\right)^{1/3}\,M_*
\approx
  3 \times 10^{13}\,\mathrm g\,.
\end{equation}
The limit is therefore extended down to this mass in Fig.~\ref{fig:photon}.
It goes above the density constraint $ \Omega_\mathrm{PBH}(M) < 0.25 $ for $ M > 7 \times 10^{16}\,\mathrm g $.

Ballesteros \textit{et al.} \cite{Ballesteros:2019exr} have recently improved the EGB bound and extended its mass range by better modeling of the combined AGN and blazar emission in the MeV range.
They also estimate the constraints from any future X-ray experiment which can identify a significantly larger number of astrophysical sources contributing to the diffuse background in this energy range.
However, the conservative upper bound on the PBH abundance given in Fig.~2 of their paper is consistent with that given above.
Coogan \textit{et al.}~\cite{Coogan:2020tuf} have recently studied the generation of soft $\gamma$-rays by asteroid-mass PBHs in nearby astrophysical structures and then used COMPTEL data to limit their DM contribution.
This already provides the strongest constraint up to $10^{17}\,\mathrm g$ and is shown in Fig.~\ref{fig:combined}; the next generation of MeV telescopes will further constrain the PBH parameter space.

In another refinement of the $\gamma$-ray background limit, Arbey \textit{et al.} \cite{Arbey:2019vqx} have extended the EGB constraint on PBHs with masses $10^{13}\text{--}10^{18}\,\mathrm g$ from a monochromatic distribution of Schwarzschild black holes to a non-monochromatic distribution of rotating Kerr black holes, showing that the lower mass window can be closed for near-extremal ones.
Constraints on spinning PBHs associated with superradiance have also been obtained in Ref.~\cite{Fukuda:2019ewf}.
This effect has been studied further by Ferraz \textit{et al.}~\cite{Ferraz:2020zgi}, who show that highly spinning PBHs of mass $\sim 10^{12}\,\mathrm{kg}$, potentially born in a matter-dominated era after inflation, can produce clouds of pions via the superradiant instability with up to nuclear density.
Neutral pion decay and charged pion annihilation then produce an isotropic $\gamma$-ray background comparable to that from Hawking evaporation.

\subsubsection{Galactic $\gamma$-ray background}

If PBHs of mass $ M_* $ are clustered inside our own Galactic halo, as expected, then there should also be a Galactic $ \gamma $-ray background and, since this would be anisotropic, it should be separable from the extragalactic background.
The ratio of the anisotropic to isotropic intensity depends on the Galactic longitude and latitude, the Galactic core radius and the halo flattening.
Some time ago Wright \cite{1996ApJ...459..487W} claimed that such a halo background had been detected in EGRET observations between $ 30\,\mathrm{MeV} $ and $ 120\,\mathrm{MeV} $ \cite{Sreekumar:1997un} and attributed this to PBHs.
His detailed fit to the data, subtracting various other known components, required the PBH clustering factor to be $ (2\text{--}12) \times 10^5\,h^{-1} $, comparable to that expected, and the local PBH explosion rate to be $ \mathcal R = 0.07\text{--}0.42 \,\mathrm{pc}^{-3}\,\mathrm{yr}^{-1} $.
A later analysis of EGRET data between $ 70\,\mathrm{MeV} $ and $ 150\,\mathrm{MeV} $, assuming a variety of distributions for the PBHs, was given by Lehoucq \textit{et al.} \cite{Lehoucq:2009ge}.
In the isothermal model, which gives the most conservative limit, the Galactic $ \gamma $-ray background requires $ \mathcal R \le 0.06 \,\mathrm{pc}^{-3}\,\mathrm{yr}^{-1} $.
They claimed that this corresponds to $ \Omega_\mathrm{PBH}(M_*) \le 2.6 \times 10^{-9} $, which from Eq.~\eqref{eq:density} implies $ \beta'(M_*) < 1.4 \times 10^{-26} $, a factor of $30$ above the extragalactic background constraint \eqref{eq:photon1}.
Lehoucq \textit{et al.} themselves claimed that it corresponds to $ \beta'(M_*) < 1.9 \times 10^{-27} $ because they used an inaccurate formula relating $ \Omega_\mathrm{PBH} $ and $ \beta $.

It should be stressed that the Lehoucq \textit{et al.} analysis does not constrain PBHs of \emph{initial} mass $ M_* $ because these no longer exist.
Rather it constrains PBHs of \emph{current} mass $ M_* $ and, from Eq.~\eqref{eq:mu} with $ m = M_* $\,, this corresponds to an initial mass of $ 1.26\,M_* $\,.
This contrasts to the situation with the extragalactic background, where the strongest constraint on $ \beta(M) $ comes from the time-integrated contribution of the $ M_* $ black holes.
So the black holes generating the Galactic background correspond to the partially-evaporated ``tail'' population and the Galactic limit is sensitive to the non-monochromaticity in the PBH mass function.
In the present context, we assume this is narrow, so that $ \beta(M) $ can be defined as the integral over the entire mass width.

To examine which value of $ M $ corresponds to the strongest limit on $ \beta(M)$, CKSY noted that Eq.~\eqref{eq:mu} implies that the emission from PBHs with initial mass $ (1+\mu)\,M_* $ currently peaks at an energy $ E \approx 100\,(3\,\mu)^{-1/3}\,\mathrm{MeV} $ for $ \mu < 1 $, which is in the range $ 70\text{--}150\,\mathrm{MeV} $ for $ 0.7 > \mu > 0.08 $.
The peak energy is above $ 150\,\mathrm{MeV} $ for $ \mu < 0.08 $, so the $ \gamma $-ray band is in the Rayleigh--Jeans region, where the flux of an individual hole scales as $ m^3 \propto \mu $, so the limit on $ \beta(M) $ scales as $ \mu^{-1} $.
For $ 0.7 > \mu > 0.08 $, the current number flux of photons from each PBH scales as $ m^{-1} \propto \mu^{-1/3} $, so the effective limit on $ \beta(M) $ scales as $ \mu^{1/3} $.
The observed $ \gamma $-ray band enters the Wien part of the spectrum for $ \mu > 0.7 $, so the limit on $ \beta(M) $ weakens exponentially for $ M > 1.7\,M_* $\,.
Hence the largest contribution to the Galactic background and the strongest constraint on $ \beta(M) $ comes from PBHs with $ M \approx 1.08\,M_* $ and has the form indicated in Fig.~\ref{fig:combined}.

Subsequently, the problem was studied in much greater detail by CKSY2 \cite{Carr:2016hva}.
To go beyond the Lehoucq \textit{et al.} analysis, we included several important effects.
First, we distinguished between the initial mass $M = (1+\mu)\,M_*$ and the current mass $m$, this being given by
\begin{equation}
m
=
\begin{cases}
\left[(\mu+1)^3 -1 + (1-\alpha^{-1})\,q^3\right]^{1/3}\,M_*
& (\mu \geq \mu_\mathrm c) \\
(3\,\alpha\,\mu)^{1/3}\,(1 + \mu + \mu^2/3)^{1/3}\,M_*
& (0 \leq \mu \leq \mu_\mathrm c)\,,
\end{cases}
\end{equation}
where
\begin{equation}
\mu_\mathrm c
\approx
  q^3/(3\,\alpha)
= 0.005\,(\alpha/4)^{-1}\,(q/0.4)^3
\end{equation}
is the value of $ \mu $ corresponding to the mass $ M_\mathrm c $ at which secondary emission eventually becomes important.
Second, we distinguished between primary and secondary emission (see Fig.~\ref{fig:ratios}).
Third, we distinguished between initial mass function $\mathrm dn/\mathrm dM$ and the current one,
\begin{equation}
\frac{\mathrm dn}{\mathrm dm}
= \left[
    \frac{1}{\alpha}\,
    \left(\frac{m}{M_*}\right)^2\,
    \left(\frac{\mathrm dn}{\mathrm dM}\right)_*\,,
    \left(\frac{m}{M_*}\right)^2\,
    \left(\frac{\mathrm dn}{\mathrm dM}\right)_*\,,
    \left(\frac{\mathrm dn}{\mathrm dM}\right)
  \right]
\quad\textnormal{for}\quad
[m < M_\mathrm q\,, M_\mathrm q < m < M_*\,, m > M_*]\,,
\end{equation}
where $M_\mathrm q$ is given by Eq.~\eqref{eq:mq}, so the main GGB contribution comes from the $\mathrm dn/\mathrm dm \propto m^2$ low mass tail.

\begin{figure}[ht]
\begin{center}
\includegraphics[width=.45\textwidth]{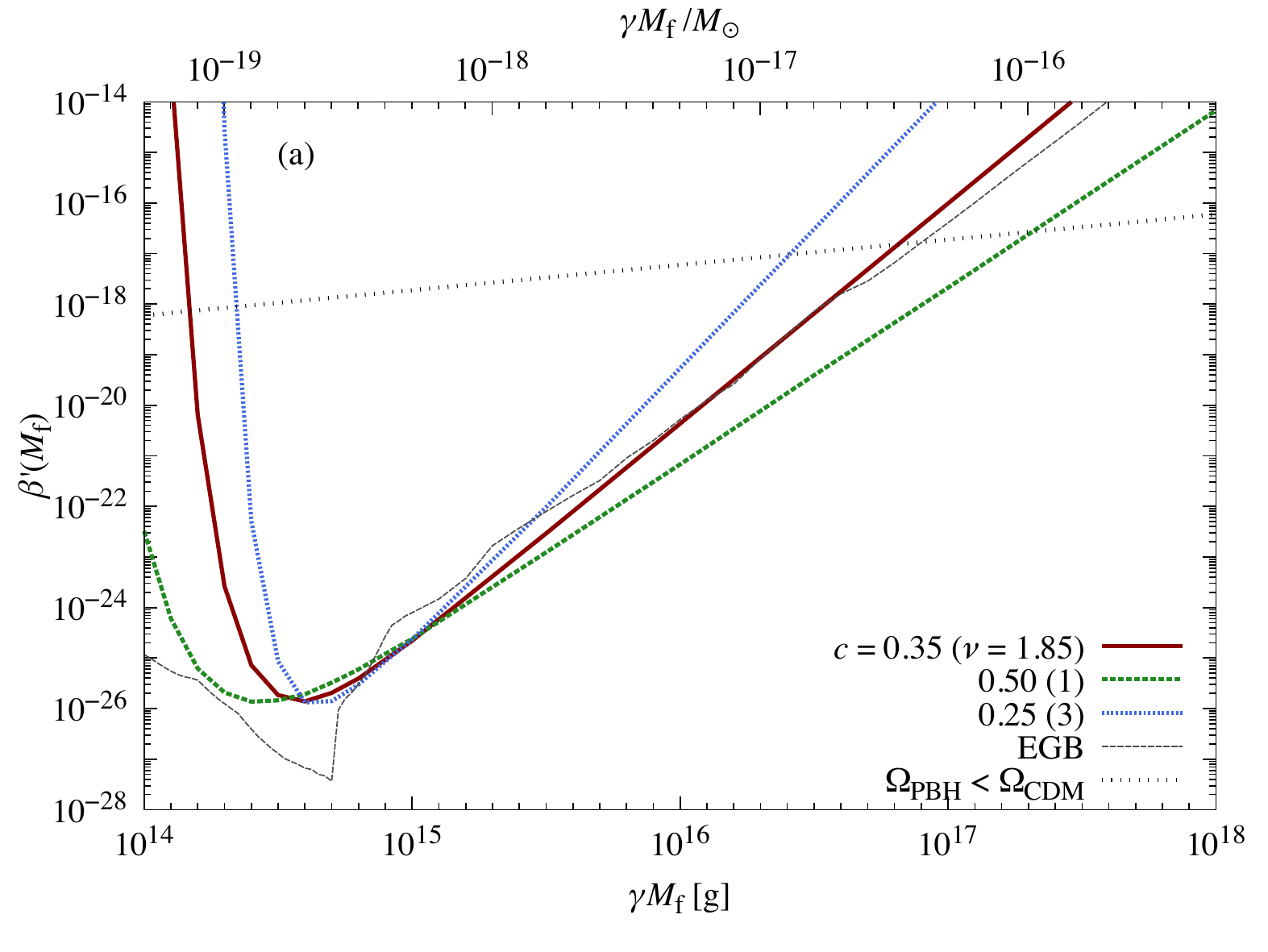}
\includegraphics[width=.45\textwidth]{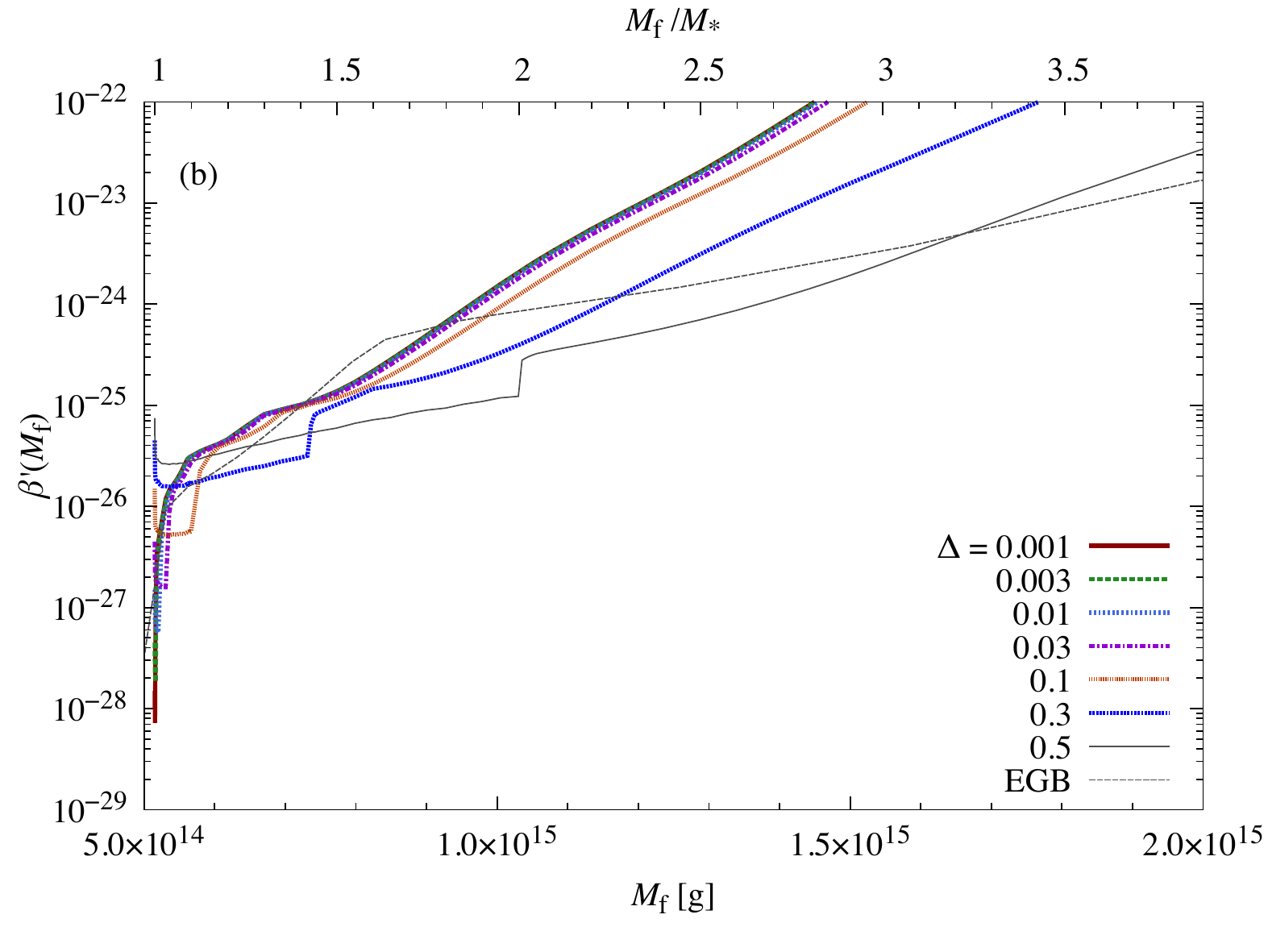}
\end{center}
\caption{Constraint imposed by Galactic $\gamma$-ray background for an extended mass function (a), as defined by Eq.~\eqref{eq:extended}, or a nearly monochromatic mass function (b), as defined by Eq.~\eqref{eq:tophat}, from Ref.~\cite{Carr:2016hva}.
The extragalactic $\gamma$-ray background (EGRB) limit comes from CKSY.}
\label{fig:ul-mono}
\end{figure}

Following Lehoucq \textit{et al.}, we took the halo density profile to have the form
\begin{equation}
\rho_\mathrm{PBH}(R)
= \frac{f\,\rho_\mathrm s}
       {(R/R_\mathrm s)^\gamma\,[1+(R/R_\mathrm s)^\alpha]^{(\beta-\alpha)/\alpha}}\,,
\end{equation}
with a set of best-fit parameters $ \gamma = 1.24, \alpha = 1, \beta = 4-\gamma = 2.86, R_\mathrm s = 28.1\,\mathrm{kpc}\,, \rho_\mathrm s = 3.50 \times 10^{-3}\,M_\odot\,\mathrm{pc}^{-3} $.
We described the directional dependence with the function
\begin{equation}
g(\boldsymbol n)
= \frac{1}{r_\mathrm{gal}}\,
  \int_0^{r_\mathrm{gal}}\!\mathrm dr\,
  \frac{\rho_\mathrm{PBH}(R(\boldsymbol n,r))}{\bar\rho_\mathrm{PBH}}\,,
\end{equation}
with $ r_\mathrm{gal} = 100\,\mathrm{kpc} $ being our distance from the edge of the halo.
We then compared the predicted intensity with Fermi-LAT observations and inferred constraints on $\beta(M)$ and $n_\mathrm{PBH}(M)$.
In Fig.~\ref{fig:ul-mono}(a), we show the constraint for power-law extended mass function:
\begin{equation}
\frac{\mathrm dn}{\mathrm dM}
= \frac{\mathrm dn}{\mathrm dM}_*\,
  \left(\frac{M}{M_*}\right)^{\nu}
\quad
(M < M_\mathrm f)\,,
\label{eq:extended}
\end{equation}
this including the critical-collapse case with $\nu = 1.85$.
In Fig.~\ref{fig:ul-mono}(b), we show the constraint for a nearly monochromatic ``top hat'' initial mass function:
\begin{equation}
\frac{\mathrm dn}{\mathrm dM}
=
\begin{cases}
\frac{n}{\Delta M_\mathrm f}
& ((1-\Delta)M_\mathrm f < M < M_\mathrm f) \\
0
& (M < (1-\Delta)\,M_\mathrm f\,, M_\mathrm f < M)
\end{cases}
\label{eq:tophat}
\end{equation}
with fractional width $ \Delta < 1 $ and total comoving number density $ n $.
Here one requires $M_\mathrm f > M_*$ and each limit has a kink at the threshold of secondary emission.
The depth of the limit in Fig.~\ref{fig:ul-mono} is sensitive to $\Delta$ and becomes arbitrarily small as $\Delta \rightarrow 0$.
However, a very small value of $\Delta$ is clearly unphysical and the extragalactic and Galactic limits are comparable for $\Delta \sim 1$.

\subsection{Extragalactic and Galactic cosmic rays}

\subsubsection{Extragalactic antiprotons}

Galactic antiprotons can constrain PBHs only in a very narrow range around $ M \approx M_* $ since they diffusively propagate to the Earth on a time much shorter than the cosmic age.
However, there will also be a spectrum of cosmic ray antiprotons from PBHs sufficiently light that they evaporated well before the epoch of galaxy formation ($ \sim 1\,\mathrm{Gyr} $).
Although such pregalactic PBHs would not be clustered, the antiprotons they emitted could still be around today and may occasionally enter the Galaxy and be detectable at Earth, so we require that this does not happen frequently enough to exceed the observational limits.
We assume that the maximum mass relevant for this constraint corresponds to the PBHs evaporating at galaxy formation:
\begin{equation}
M
\lesssim
  \left(\frac{1\,\mathrm{Gyr}}{t_0}\right)^{1/3}\,
  \left(\frac{f(M)}{f_*}\right)^{1/3}\,M_*
\approx
  2 \times 10^{14}\,\mathrm g\,.
\end{equation}
This mass-scale is not necessarily relevant because even the antiprotons generated \emph{inside} galaxies may escape on a cosmological timescale and become part of the extragalactic background.
However, the leakage time depends on the size of the cosmic ray region, which is rather uncertain, and could well be comparable to $ t_0 $\,.
We therefore adopt the above upper mass limit.
The minimum mass relevant for this constraint is determined as follows.
The mean rate of $ \bar p p $ annihilations in the cosmological background, where the target $ p $'s are in hydrogen and helium nuclei, is
\begin{equation}
\Gamma_{\bar pp}(t)
= \langle\sigma_{\bar pp}\,v_{\bar p}\rangle\,n_p(t)
\approx
  2 \times 10^{-22}\,\left(\frac{t}{t_0}\right)^{-2}\,\mathrm s^{-1}\,,
\end{equation}
where we have used $ \langle\sigma_{\bar pp}\,v_{\bar p}\rangle/c \approx 4 \times 10^{-26}\,\mathrm{cm}^2 $ and $ n_p(t_0) \approx 2 \times10^{-7}\,\mathrm{cm}^{-3} $.
The condition that an antiproton emitted by a PBH with lifetime $ \tau $ survives until now is then
\begin{equation}
\int_\tau^{t_0}\!\Gamma_{\bar pp}(t')\,\mathrm dt'
\approx
  \left(\frac{t_0}{\tau}\right)\,
  \left(\frac{t_0}{5 \times 10^{21}\,\mathrm s}\right)
\approx
  \frac{1.3\,\mathrm{Myr}}{\tau}
< 1\,.
\end{equation}
Using Eq.~\eqref{eq:lifetime}, we infer that the extragalactic $ \bar p $ limit applies for
\begin{equation}
M
\gtrsim
  \left(\frac{1.3\,\mathrm{Myr}}{t_0}\right)^{1/3}\,
  \left(\frac{f(M)}{f_*}\right)^{1/3}\,
  M_*
\approx
  4 \times 10^{13}\,\mathrm g\,.
\end{equation}
We have seen there is no primary emission of antiprotons from PBHs and the analog of Eq.~\eqref{eq:peakfluxaprx} implies that the current number density from secondary emission of PBHs with mass $ M $ is
\begin{equation}
n_{\bar p}(M, t_0)
= \frac{n(\tau)\,\tau}{(1+z(\tau))^3}\,
  \frac{\mathrm d\dot N_{\bar p}}{\mathrm d\ln E_{\bar p}}
  (E_{\bar p}=100\,\mathrm{MeV})
\approx
  6 \times 10^{11}\,
  \beta'(M)\,
  \left(\frac{f(M)}{f_*}\right)^{-1}\,
  \left(\frac{M}{M_*}\right)^{1/2}\,
  \mathrm{cm}^{-3}\,.
\end{equation}
Since the antiproton-to-proton ratio $ r $ is required to be less than $ 10^{-5} $, one finds the upper bound
\begin{equation}
\beta'(M)
\lesssim
  3 \times 10^{-24}\,
  \left(\frac{r}{10^{-5}}\right)\,
  \left(\frac{f(M)}{f_*}\right)\,
  \left(\frac{M}{M_*}\right)^{-1/2}
\quad
(4 \times 10^{13}\,\mathrm g \lesssim M \lesssim 2 \times 10^{14}\,\mathrm g)\,.
\end{equation}
This constraint is shown by the green line in Fig.~\ref{fig:combined} and turns out to be much weaker than the Galactic antiproton constraint (discussed below).

\subsubsection{Emission of other particles}

\begin{figure}[ht]
\begin{center}
\includegraphics[width=.5\textwidth]{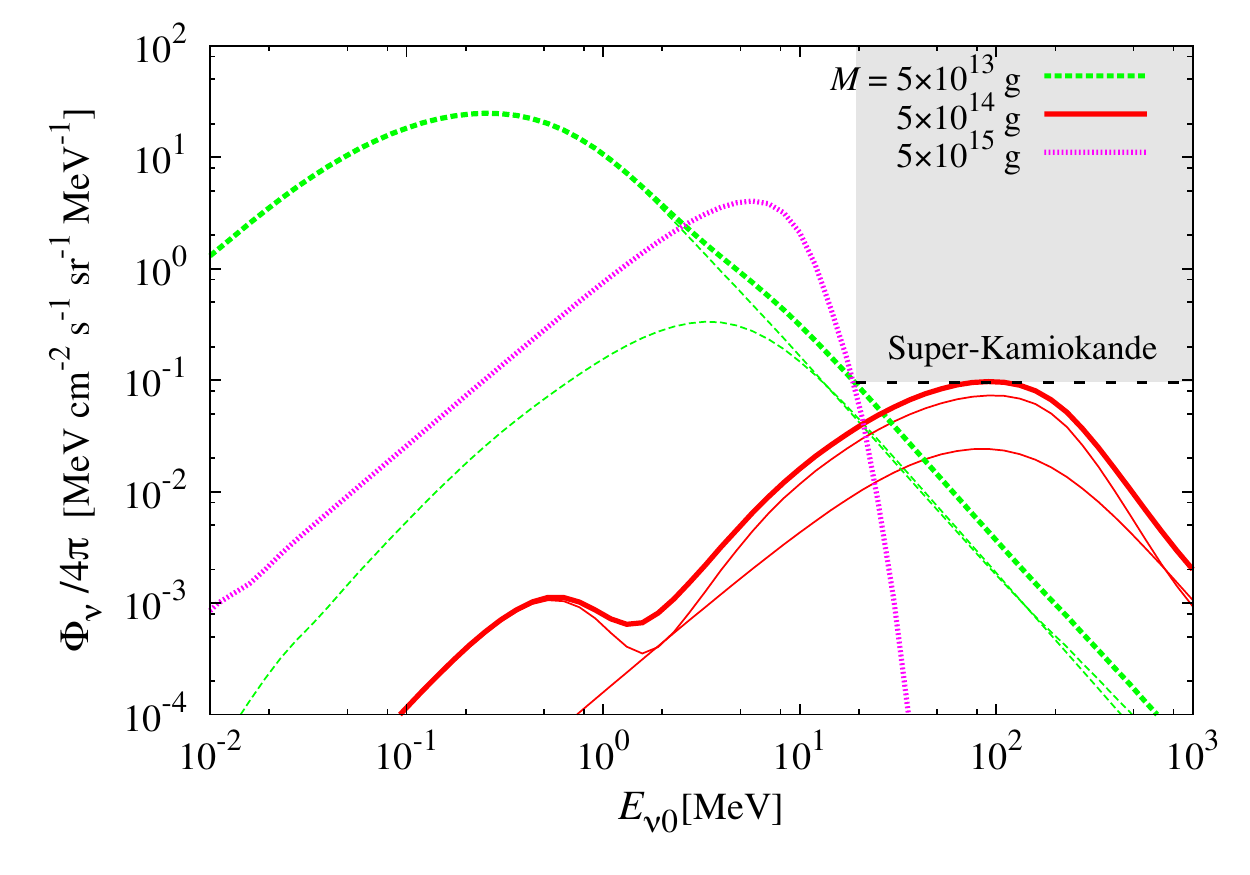}
\end{center}
\caption{An illustration of the maximum $ \bar\nu_e $ flux allowed by the SK limit for three PBH masses, the shaded region being excluded, from Ref.~\cite{Carr:2009jm}.
For $ M \lesssim M_* $\,, the lower (upper) curves represent primary (secondary) components and the thick curves denote their sum.
$ M \gtrsim M_* $ holes emit primary neutrinos only.}
\label{fig:nu}
\end{figure}

Neutrinos may either be emitted directly as black-body radiation (primaries) or they may result from the decay of emitted pions, leptons, neutrons and anti-neutrons (secondaries).
As a result, their emission spectra are similar to those for photons up to a normalisation factor.
The neutrino background can in principle constrain PBHs whose lifetime exceeds the time of neutrino decoupling ($ \tau \gtrsim 1\,\mathrm s $).
This corresponds to a minimum mass of $ M_{\mathrm{min},\nu} \approx 10^9\,\mathrm g $.
However, the low-energy neutrinos which we have to use are poorly limited by observations.
In Super-Kamiokande (SK), the null detection of relic $ \bar\nu_e $'s implies $ \Phi_{\bar\nu_e} \le 1.2\,\mathrm{cm}^{-2}\,\mathrm s^{-1} $ above the threshold $ E_{\bar\nu_e 0} = 19.3\,\mathrm{MeV} $ \cite{Malek:2002ns}.
As seen in Fig.~\ref{fig:nu}, this energy corresponds to the high-energy tail for light PBHs.
The constraint associated with the relic neutrinos is shown by the green line in Fig.~\ref{fig:combined} and is much weaker than the BBN and photon background limits.
Similar calculations have been made by Bugaev \textit{et al.} \cite{Bugaev:2002yt,Bugaev:2008gw} and Dasgupta \textit{et al.} \cite{Dasgupta:2019cae}.
However, they assume that the PBHs have a continuous mass function and they link their model with a particular inflationary scenario.

Lunardini and Perez-Gonzalez \cite{Lunardini:2019zob} show that the neutrino emission depends on whether they are Dirac or Majorana.
In the Dirac case, PBHs radiate right-handed and left-handed neutrinos in equal amounts, thus increasing the effective number of neutrino species $N_\mathrm{eff}$\,.
They derive a new bound on $\beta'(M)$ for $M$ in the range $4 \times 10^{7}\text{--}10^9\,\mathrm g$.
Future measurements of $N_\mathrm{eff}$ may constrain the initial fraction for $M$ as low as $1\,\mathrm g$.
If an excess in $N_\mathrm{eff}$ is found, PBHs with Dirac neutrinos could provide an explanation of it.

Evaporating PBHs should produce any other particles predicted in theories beyond the Standard Model.
The number of PBHs is therefore limited by both the abundance of stable massive particles \cite{Green:1999yh} or the decay of long-lived ones \cite{Khlopov:2004tn}.
In supersymmetry or supergravity, the lightest supersymmetric particle (LSP) is stable and becomes a candidate for the DM.
If LSPs are produced by the evaporation of PBHs, in order not to exceed the observed CDM density at present, one obtains the upper bound \cite{Lemoine:2000sq}:
\begin{equation}
\beta'(M)
\lesssim
  10^{-18}\,\left(\frac{M}{10^{11}\,\mathrm g}\right)^{-1/2}\,
  \left(\frac{m_\mathrm{LSP}}{100\,\mathrm{GeV}}\right)^{-1}
\quad
(M < 10^{11}\,(m_\mathrm{LSP}/100\,\mathrm{GeV})^{-1}\,\mathrm g)\,.
\label{eq:stable}
\end{equation}
This constraint is shown in Fig.~\ref{fig:combined} but depends on the mass of the LSP and is therefore subject to considerable uncertainty \cite{Kohri:2005wn,Kawasaki:2008qe}.
In addition, unstable particles such as the gravitino or neutralino might be produced by evaporating PBHs.
The decay of these particles into lighter ones also affects BBN \cite{Kawasaki:2004yh,Kawasaki:2004qu} and this gives another constraint \cite{Khlopov:2004tn}:
\begin{equation}
\beta'(M)
\lesssim
  5 \times 10^{-19}\,
  \left(\frac{M}{10^{9}\,\mathrm g}\right)^{-1/2}\,
  \left(\frac{Y_\mathrm{PBH}}{10^{-14}}\right)\,
  \left(\frac{x_\phi}{0.006}\right)^{-1}
\quad
(M < 10^{9}\,\mathrm g)\,,
\end{equation}
where $ Y_\mathrm{PBH} $ is the PBH number density to entropy density ratio and $ x_\phi $ is the fraction of the luminosity going into quasi-stable massive particles, both being normalised to reasonable values.
This limit is not shown explicitly in Fig.~\ref{fig:combined} but it has a similar form to Eq.~\eqref{eq:stable}.

PBHs could dominate the density of the Universe before they evaporate and particles emitted by them might then contribute to the DM or baryon asymmetry~\cite{Fujita:2014hha,Lennon:2017tqq,Morrison:2018xla}.
Recently Auffinger \textit{et al.}~\cite{Auffinger:2020afu} have considered the emission of warm DM candidates by PBHs in the range $ 10^{-5}\text{--}10^9\,\mathrm g $.
If such particles provide the DM, this could inhibit structure formation, which gives an upper limit on $\beta$ in this mass range.
The limit is weakened for increasing particle spin but PBH domination is ruled out for spin values up to 2.
Hooper and Krnjaic~\cite{Hooper:2020otu} have suggested that the cosmic baryon asymmetry may have been generated by PBH emission of GUT Higgs or gauge bosons.
The PBHs evaporate after the EW phase transition and thereby evade the sphaleron wash-out encountered in the usual GUT baryogenesis model.
They identify a range of scenarios in which this could work for $M < 10^8\,\mathrm g$.

\subsubsection{Reionisation and 21\,cm signature}

The Planck 2018 results give the optical depth as $ \tau \sim 0.05 $ \cite{Aghanim:2019ame} for CMB photons emitted from the last scattering surface and observations of Gunn--Peterson absorption troughs imply that reionisation of the Universe occurred at $ z \sim 6 $~\cite{Fan:2006dp}.
Thus PBHs cannot be so numerous that they reionise the Universe earlier than $ z \sim 6 $.
In principle, this leads to a constraint on PBHs with
\begin{equation}
M
\ge
  M_*\,\left(\frac{t_\mathrm{dec}}{t_0}\right)^{1/3}\,\left(\frac{f(M)}{f_*}\right)^{1/3}
\approx
  2 \times 10^{13}\,\mathrm g\,.
\end{equation}
On the other hand, Belotsky \textit{et al.} have attributed the reionisation of the pregalactic medium to evaporating PBHs.
They show that PBHs with a monochromatic mass distribution around $5 \times 10^{16}\,\mathrm g$ could ensure this \cite{Belotsky:2014twa}.
They also argue that their reionisation effect and DM contribution can be simultaneously enhanced with an extended mass distribution around this mass \cite{Belotsky:2017vsr}.
More recently, Cang \textit{et al.}~\cite{Cang:2020aoo} have studied how Hawking radiation from PBHs in the $10^{15}\text{--}10^{17}\,\mathrm g$ range with various mass spectra can change the cosmic recombination history by ionizing and heating the intergalactic medium, this modifying the CMB anisotropies.
They find that future experiments can improve current Planck bounds by about two orders of magnitudes.
All monochromatic distributions above $4 \times10^{16}\,\mathrm g$ that are allowed by current CMB data could be excluded.

An increase in the ionisation of the intergalactic medium would also produce a 21\,cm signature.
Mack and Wesley \cite{Mack:2008nv} have argued that future observations of 21\,cm radiation from high redshift neutral hydrogen could place important constraints on PBHs in the mass range $ 5 \times 10^{13}\,\mathrm g < M < 10^{17}\,\mathrm g $, essentially because photons emitted from PBHs during $ 30 < z < 300 $ peak in the energy range in which the intergalactic medium has low optical depth.
Any process which heats the intergalactic medium in this period will produce a signal but the ionising flux of photons, electrons and positrons from PBHs would generate a distinctive feature in the 21\,cm brightness temperature.
PBHs with $ 5 \times 10^{13}\,\mathrm g < M < 10^{14}\,\mathrm g $ evaporate in $ 30 < z < 90 $ and would raise the 21\,cm brightness temperature, thereby reducing the absorption seen against the CMB.
PBHs with $ M \sim 10^{14}\,\mathrm g $ would raise the spin temperature above the CMB, so that the 21\,cm line appears in emission rather than absorption.
PBHs with $ 10^{14}\,\mathrm g < M < 10^{17}\,\mathrm g $ would have a less pronounced effect.
The Mack--Wesley limit therefore bottoms out at around $ 10^{14}\,\mathrm g $ and has the form
\begin{equation}
\beta'(M)
< 3 \times 10^{-29}\,
  \left(\frac{M}{10^{14}\,\mathrm g}\right)^{7/2}
\quad
(M > 10^{14}\,\mathrm g)\,,
\end{equation}
which is well below the $\gamma$-ray background limit.
Recently EDGES (Experiment to Detect the Global Epoch of Reionization Signature) reported the detection of a 21\,cm absorption signal stronger than astrophysical expectations.
Clark \textit{et al.} \cite{Clark:2018ghm} have inferred a constraint on any energy injection into the intergalactic medium by evaporating PBHs because the heating of the neutral hydrogen gas would weaken the 21\,cm absorption signal.
For $e \gamma \gamma$ final states they find strong 21\,cm bounds that can be more stringent than the current extragalactic diffuse photon bounds.
However, none of these reionisation or 21\,cm limits is shown in Fig.~\ref{fig:combined} because of various astrophysical uncertainties.

\subsubsection{Galactic electrons and positrons}

The evaporation of PBHs with $M>10^{17}\,\mathrm g$ is expected to inject sub-GeV electrons and positrons into the Galaxy.
These particles are shielded by the solar magnetic field for Earth-bound detectors, but not for Voyager 1, which is now beyond the heliopause.
Boudaud and Cirelli \cite{Boudaud:2018hqb} use its data to constrain the number of PBHs in the Galaxy and, as indicated in Fig.~\ref{fig:combined}, find $ \beta' < 10^{-20} $ for $M < 10^{16}\,\mathrm g$ for one particular model.
Their limits are based on local Galactic measurements and thus complement those derived from cosmological observations.
Dutta \textit{et al.}~\cite{Dutta:2020lqc} point out that mildly relativistic e$^{\pm}$ from PBHs in the mass range $10^{15}\text{--}10^{17}\,\mathrm g$ generate radio emission through inverse-scattering on CMB photons and suggest that SKA will be able to detect such emission from the nearby ultra-faint dwarf galaxies, Segue I and Ursa Major II, as well as the globular cluster $\omega$-Cen and the Coma cluster.

Kim~\cite{Kim:2020ngi} obtains a constraint by noting that fast electrons from evaporating PBHs heat up the interstellar medium through Coulomb interactions.
Using observations of the dwarf galaxy Leo T, he obtains an upper bound for $M < 10^{17}\,\mathrm g$ which is stronger than either the extragalactic or Galactic $\gamma$-ray background limit.
This limit is shown by the ``Leo T'' line in Fig.~\ref{fig:combined}.
Laha \textit{et al.}~\cite{Laha:2020vhg} have re-analysed this effect and find that the limit from Leo T is weaker than claimed by Kim but that it is strengthened if the PBHs are spinning.
High-energy particles from PBH evaporations would emit radio synchrotron radiation in the presence of a strong magnetic field, so Chan and Lee~\cite{Chan:2020zry} use radio data from the Galactic centre to constrain the number of PBHs there.
Using three different mass distributions (monochromatic, lognormal and power-law), they show that PBHs can provide only a very small fraction of DM in the mass range below $10^{17}\,\mathrm g$.
This limit is shown by the ``GC radio'' line in Fig.~\ref{fig:combined}.

\subsubsection{Galactic antiprotons}

Since the ratio of antiprotons to protons in cosmic rays is less than $ 10^{-4} $ over the energy range $ 100\,\mathrm{MeV}\text{--}10\,\mathrm{GeV} $, whereas PBHs should produce them in equal numbers, PBHs could only contribute appreciably to the antiprotons \cite{Turner:1981ez,Kiraly:1981ci}.
It is usually assumed that the observed antiprotons are secondary particles, produced by spallation of the interstellar medium by primary cosmic rays.
However, the spectrum of secondary antiprotons should show a steep cut-off at kinetic energies below $ 2\,\mathrm{GeV} $, whereas the spectrum of PBH antiprotons should continue down to $ 0.2\,\mathrm{GeV} $.
Also any primary antiproton fraction should tend to $ 0.5 $ at low energies.
Both these features provide a distinctive signature of any PBH contribution.

The black hole temperature must be much larger than $1\,\mathrm{GeV}$ to generate antiprotons, so the local cosmic ray flux from PBHs should be dominated by the ones just entering their explosive phase at the present epoch.
Such PBHs should be clustered inside our halo, so any charged particles emitted will have their flux enhanced relative to the extragalactic spectra by a factor $ \zeta $ which depends upon the halo concentration factor and the time for which particles are trapped inside the halo by the Galactic magnetic field.
This time is rather uncertain and also energy-dependent.
At $ 100\,\mathrm{MeV} $ one expects roughly $ \zeta \sim 10^3 $ for electrons or positrons and $ \zeta \sim 10^4 $ for protons or antiprotons \cite{Carr:1998fw}.

MacGibbon and Carr \cite{MacGibbon:1991vc} originally calculated the PBH density required to explain the interstellar antiproton flux at $ 1\,\mathrm{GeV} $ and found a value somewhat larger than the density associated with the $ \gamma $-ray background limit.
After the BESS balloon experiment measured the antiproton flux below $ 0.5\,\mathrm{GeV} $ \cite{Yoshimura:1995sa}, Maki \textit{et al.} \cite{Maki:1995pa} tried to fit this in the PBH scenario by using Monte Carlo simulations of cosmic ray propagation.
They found that the local antiproton flux would be mainly due to PBHs exploding within a few kpc and used the observational data to infer a limit on the local PBH explosion rate of $ \mathcal R < 0.017\,\mathrm{pc}^{-3}\,\mathrm{yr}^{-1} $.
A later attempt to fit the antiproton data came from Barrau \textit{et al.} \cite{Barrau:2002ru}, who concluded that PBHs with $ \beta'(M_*) \approx 5 \times 10^{-28} $ would be numerous enough to explain the observations by BESS95 \cite{Yoshimura:1995sa}, BESS98 \cite{Orito:1999re}, CAPRICE \cite{Boezio:2001ac} and AMS \cite{Jacholkowska:2007fb}.
However, this was based on the assumption that the PBHs have a spherically symmetric isothermal profile with a core radius of $ 3.5\,\mathrm{kpc} $.
A different clustering assumption would lead to a different constraint on $ \beta'(M_*) $.

PBHs might also be detected by their antideuteron flux.
Barrau \textit{et al.} \cite{Barrau:2002mc} argue that AMS (Alpha Magnetic Spectrometer) and GAPS (General AntiParticle Spectrometer) would be able to detect the antideuterons from PBH explosions if their local density were sufficiently large and inferred potential limits $ \beta'(M_*) < 1.5 \times 10^{-26}\,(\zeta/10^4)^{-1} $ and $ 8.2 \times 10^{-28}\,(\zeta/10^4)^{-1} $.
More recent data come from the BESS Polar-I experiment, which flew over Antarctica in December 2004 \cite{Abe:2008sh}.
The reported antiproton flux lies between that of BESS(95+97) and BESS(2000) and the $ \bar p/p $ ratio ($ r \approx 10^{-5} $) is similar to that reported by BESS(95+97).
Although the latter indicated an excess flux below $ 400\,\mathrm{MeV} $, this was not found in the BESS Polar-I data.
However, given the magnitude of the error bars, we might still expect a constraint on the local PBH number density similar to that discussed above, in which case the constraint on $ \beta(M_*) $ becomes
\begin{equation}
\beta'(M_*)
< 3.9 \times 10^{-25}\,
  \left(\frac{\zeta}{10^4}\right)^{-1}\,,
\end{equation}
where we have used Eq.~\eqref{eq:density}.
For reasonable values of $ \zeta $, this is much weaker than the $ \gamma $-ray background limit, but the value of $ \zeta $ is anyway very uncertain and so this limit is not shown explicitly in Fig.~\ref{fig:combined}.

\subsubsection{Galactic centre radiation}

Several authors have suggested that the $ 511\,\mathrm{keV} $ annihilation line radiation from the Galactic centre could result from the positrons emitted by the PBHs there \cite{Bugaev:2002yt,Bugaev:2008gw,Bambi:2008kx}, as first suggested by Okele and Rees \cite{1980A&A....81..263O}.
Most recently, DeRocco and Graham \cite{DeRocco:2019fjq} and Laha \cite{Laha:2019ssq} have constrained $10^{16}\text{--}10^{17}\,\mathrm g$ PBHs from measurements of this line by SPI/INTEGRAL, excluding models in which they constitute all the DM.
Laha \textit{et al.}~\cite{Laha:2020ivk} have also used this data to constrain PBHs radiating primary photons and shown that -- depending on their mass function and other astrophysical uncertainties -- PBHs smaller than $10^{18}\,\mathrm g$ could contribute less than 1\,\% of the DM.
In Fig.~\ref{fig:combined}, the positron annihilation-line and continuum $\gamma$-ray constraints are indicated by the ``GC $e^+$'' and ``GC $\gamma$'' lines, respectively.
The former is particularly interesting if the PBHs are spinning \cite{Dasgupta:2019cae}.
In this case, the limit is stronger in the higher mass range and stronger than the diffuse $\gamma$-ray limit for $M$ exceeding a few times $10^{16}\,\mathrm g$.
However, it should be stressed that Galactic constraints depend on the assumed radial profile, as well as the local DM density and propagation uncertainties, whereas the extragalactic bounds are free from these uncertainties.

\subsection{PBH explosions}

The extragalactic $\gamma$-ray background limit implies that the PBH explosion rate $ \mathcal R $ could be at most $ 10^{-6}\,\mathrm{pc}^{-3}\,\mathrm{yr}^{-1} $ if the PBHs are uniformly distributed or $ 10\,\mathrm{pc}^{-3}\,\mathrm{yr}^{-1} $ if they are clustered inside galactic halos \cite{Page:1976wx}.
The latter figure might be compared to the Lehoucq \textit{et al.} Galactic $ \gamma $-ray limit of $ 0.06\,\mathrm{pc}^{-3}\,\mathrm{yr}^{-1} $ \cite{Lehoucq:2009ge} and the Maki \textit{et al.} antiproton limit of $ 0.02\,\mathrm{pc}^{-3}\,\mathrm{yr}^{-1} $ \cite{Maki:1995pa}.
We now compare these limits to the direct observational constraints on the explosion rate.
These come both from $\gamma$-ray bursts and high-energy cosmic ray showers.

\subsubsection{Gamma-ray bursts and photosphere effects}

In the Standard Model of particle physics, where the number of elementary particle species never exceeds around $ 100 $, it has been appreciated for a long time that the likelihood of detecting the final explosive phase of PBH evaporations is very low \cite{Semikoz:1994uz}.
However, the physics of the QCD phase transition is still uncertain and the prospects of detecting explosions would be improved in less conventional particle physics models~\cite{1979Natur.277..199P}.
For example, in a Hagedorn-type picture \cite{Carter:1976di}, where the number of particle species exponentiates at the quark-hadron temperature, the upper limit on $ \mathcal R $ is reduced to $ 0.05\,\mathrm{pc}^{-3}\,\mathrm{yr}^{-1} $ \cite{Fichtel:1994sf}, which is comparable to the antiproton limit.

Even without the Hagedorn effect, something dramatic may occur at the QCD temperature since the number of species emitted increases substantially~\cite{Halzen:1991uw}.
For this reason, Cline and colleagues have long argued that the formation of a fireball at the QCD temperature could explain some of the short period $ \gamma $-ray bursts (i.e.\ those with duration less than $ 100\,\mathrm{ms} $) \cite{Cline:1992ps,Cline:1995sz,Cline:1996zg}.
In Ref.~\cite{Cline:2001tq} they claim to find $ 42 $ BATSE candidates of this kind and the fact that their distribution matches the spiral arms suggests that they are Galactic.
In Ref.~\cite{Cline:2005xb} they report a class of short-period KONUS bursts which has a much harder spectrum than usual and identify these with exploding PBHs.
In Ref.~\cite{Cline:2006nv} they find a further $ 8 $ candidates in the Swift data.
Overall they claim that the BATSE, KONUS and Swift data correspond to a $ 4.5 \sigma $ effect and that several events exhibit the time structure expected of PBH evaporations \cite{Cline:2009ni}.
One distinctive feature of $ \gamma $-ray bursts generated by PBH explosions would be a temporal delay between the high and low energy pulses, an effect which might be detectable by Fermi LAT \cite{Ukwatta:2009xk}.
Multimessenger techniques of searching for signals from evaporating PBHs are discussed in Ref.~\cite{Petkov:2019edm}.

It is clearly important to understand how likely PBHs are to resemble $ \gamma $-ray bursts from a theoretical perspective.
The usual assumption that there is no interaction between emitted particles \cite{1984PhLB..143...92O} was refuted by Heckler \cite{Heckler:1995qq}, who claimed that QED (Quantum Electrodynamics) interactions could produce an optically thick photosphere once the black hole temperature exceeds $ T_\mathrm{BH} = 45\,\mathrm{GeV} $.
He proposed that a similar effect may operate at the lower temperature $ T_\mathrm{BH} \approx 200\,\mathrm{MeV} $ due to QCD effects \cite{Heckler:1997jv}.
Variants of these models and their astrophysical implications have been studied by various authors \cite{Cline:1998xk,Kapusta:2000xt,Daghigh:2001gy,Daghigh:2002fn,Daghigh:2006dt}.

MacGibbon \textit{et al.} \cite{MacGibbon:2007yq} have identified a number of physical and geometrical effects which may invalidate these claims.
First, the particles must be causally connected in order to interact and this means that the standard cross-sections are reduced (viz.\ the particles are created at a finite time and do not go back to the infinite past).
Second, because of the Landau--Pomeranchuk--Migdal effect, a scattered particle requires a minimum distance to complete each bremsstrahlung interaction, with the consequence that there is unlikely to be more than one complete interaction per particle near the black hole.
They conclude that the emitted particles do not interact sufficiently to form a QED photosphere and that the conditions for QCD photosphere formation could only be temporarily satisfied (if at all) when the black hole temperature is of order $ \Lambda_\mathrm{QCD} $\,.
Even in this case, the strong damping of the Hawking production of QCD particles around this threshold may suffice to suppress it.
In any case, they claim that no QCD photosphere persists once the black hole temperature climbs above $ \Lambda_\mathrm{QCD} $\,.

A rather different way of producing a $ \gamma $-ray burst is to assume that the outgoing charged particles form a plasma due to turbulent magnetic field effects at sufficiently high temperatures \cite{1996MNRAS.283..626B}.
However, MacGibbon \textit{et al.} argue that this too is implausible.
They conclude that the observational signatures of a cosmic or Galactic halo background of PBHs or an individual high-temperature black hole remain essentially those of the standard Hawking model, with little change to the detection probability.
Nevertheless, perhaps the best strategy is to accept that our understanding of such effects is incomplete and focus on the empirical aspects of the $ \gamma $-ray burst observations.

\subsubsection{High energy cosmic rays showers}

At much higher energies, several groups have looked for $ 1\text{--}100\,\mathrm{TeV} $ photons from PBH explosions using cosmic ray detectors.
However, in this case, the constraints are also strongly dependent on the theoretical model \cite{Bugaev:2007py}.
In the Standard Model the upper limits on the explosion rate are $ 5 \times 10^8\,\mathrm{pc}^{-3}\,\mathrm{yr}^{-1} $ from the CYGNUS array \cite{Alexandreas:1993zx}, $ 8 \times 10^6\,\mathrm{pc}^{-3}\,\mathrm{yr}^{-1} $ from the Tibet array \cite{1996A&A...311..919T}, $ 1 \times 10^6\,\mathrm{pc}^{-3}\,\mathrm{yr}^{-1} $ from the Whipple Cherenkov telescope \cite{Linton:2006yu}, and $ 8 \times 10^8\,\mathrm{pc}^{-3}\,\mathrm{yr}^{-1} $ from the Andyrchy array \cite{Petkov:2008rz}.
These limits correct for the effects of burst duration and array ``dead time.''
Such limits are far weaker than the ones associated with observations at $ 100\,\mathrm{MeV} $.
They would be even weaker in the QED photosphere model advocated by Heckler, since there are then far fewer TeV particles \cite{Petkov:2008rv}.
For example, Bugaev \textit{et al.} \cite{Bugaev:2009ad} have used Andyrchy data to obtain an upper limit of $ 1 \times 10^9\,\mathrm{pc}^{-3}\,\mathrm{yr}^{-1} $ in the Daghigh--Kapusta model \cite{Kapusta:2000xt,Daghigh:2001gy,Daghigh:2002fn,Daghigh:2006dt} and $ 5 \times 10^9\,\mathrm{pc}^{-3}\,\mathrm{yr}^{-1} $ in the Heckler model \cite{Heckler:1995qq}.
Because of the uncertainties, we do not show any of these limits in Fig.~\ref{fig:combined}.

The most recent constraints on PBH explosions come from observations of the High Altitude Water Cherenkov Observatory, which is sensitive to $\gamma$-rays with energies of $300\,\mathrm{GeV}$ to $100\,\mathrm{TeV}$.
With its large instantaneous field-of-view of $2$ steradians and a duty cycle over 95\,\%, the observatory is well suited to perform an all-sky search for PBH bursts and Albert \textit{et al.} \cite{Albert:2019qxd} have obtained an upper limit on the local PBH burst rate density of $3400\,\mathrm{pc}^{-3}\,\mathrm{yr}^{-1}$ at the 99\,\% confidence level, far stronger than any existing electromagnetic limit.

\subsection{More speculative effects}

\subsubsection{Higgs instability}

Evaporating PBHs can act as seeds for a cosmological phase transition.
The analysis of this problem was pioneered by Hiscock \cite{Hiscock:1987hn}, who assumed that the black hole mass remains constant throughout the transition.
The more general case, without this assumption, was studied by Gregory \textit{et al.}~\cite{Gregory:2013hja}, who concluded that small PBHs can catalyze a phase transition.
This argument was applied to the stability of the Higgs vacuum in Refs.~\cite{Burda:2015isa,Burda:2015yfa,Burda:2016mou}, where it was shown that a PBH with mass up to $\sim 10^{9}\,M_\mathrm{Pl}$ may destabilize the vacuum before evaporating.
The thermal effects of the Hawking radiation of such PBHs has been studied by a number of authors \cite{Mukaida:2017bgd,Kohri:2017ybt,Hayashi:2020ocn} but with conflicting conclusions due to different interpretations and different choices of the quantum state around the black hole.
In principle, one can constrain the abundance of such small PBHs by percolation analysis of vacuum decay~\cite{Dai:2019eei}.
For example, Kohri and Matsui \cite{Kohri:2017ybt} have obtained an upper bound,
\begin{equation}
\beta'(M)
< 10^{-21}\,
  \left(\frac{M}{10^{9}\,\mathrm g}\right)^{3/2} \quad (M < M_*)\,.
\end{equation}
However, we cannot draw a definite conclusion without a proper formulation and understanding of quantum tunneling in the presence of gravity, so this limit is not included in Fig.~\ref{fig:combined}.

\subsubsection{Planck mass relic constraints}

If PBH evaporations leave stable Planck-mass relics, these might also contribute to the DM.
This was first pointed out by MacGibbon \cite{MacGibbon:1987my} and has subsequently been explored in the context of inflationary scenarios by numerous authors \cite{Barrow:1992hq,Carr:1994ar,Green:1997sz,Alexeyev:2002tg,Chen:2002tu,Barrau:2003xp,Chen:2004ft,Nozari:2005ah}.
If the relics have a mass $ \kappa\,M_\mathrm{Pl}$ and reheating occurs at a temperature $ T_\mathrm R $\,, this also being the maximum temperature after inflation if reheating is rapid, then the requirement that they have less than the critical density today implies \cite{Carr:1994ar}
\begin{equation}
\beta'(M)
< 8 \times 10^{-28}\,\kappa^{-1}\,\left(\frac{M}{M_\mathrm{Pl}}\right)^{3/2}
\label{relics}
\end{equation}
for the mass range
\begin{equation}
\left(\frac{T_\mathrm {Pl}}{T_\mathrm{R}}\right)^{2}
< \frac{M}{M_\mathrm{Pl}}
< 10^{11}\,\kappa^{2/5}\,.
\end{equation}
We would now require the density to be less than $ \Omega_\mathrm{CDM} \approx 0.25 $, which strengthens the original limit by a factor of $ 4 $.
The lower mass limit arises because PBHs generated before reheating are diluted exponentially.
The upper mass limit arises because PBHs larger than this dominate the total density before they evaporate, in which case the final cosmological baryon-to-photon ratio is determined by the baryon-asymmetry associated with their emission.
This might be compared to lower limit imposed by the trans-Planckian censorship conjecture \cite{Cai:2019igo}.
Alexander and M\'esz\'aros \cite{Alexander:2007gj} have advocated an extended inflationary scenario in which evaporating PBHs naturally generate the DM, the entropy and the baryon asymmetry of the Universe.
This triple coincidence applies providing inflation ends at $ t \sim 10^{-23}\,\mathrm s $, so the PBHs would need to have an initial mass $ M \sim 10^6\,\mathrm g $.

It is usually assumed that Planck-mass relics would be undetectable apart from their gravitational effects.
However, Lehmann \textit{et al.} \cite{Lehmann:2019zgt} point out that they may carry electric charge, making them visible to terrestrial detectors.
They evaluate constraints and detection prospects in detail and show that this scenario, if not already ruled out by monopole searches, can be explored within the next decade using planned experimental equipment.

Although Planck-mass relics might be a natural consequence of quantum gravity effects, stable PBH relics could in principle be much larger than this.
For example, given a suitable dark sector, Bai and Orlofsky \cite{Bai:2019zcd} argue that PBHs lighter than $10^9\,\mathrm g$ but above the Planck mass could become stable against evaporation by being charged under a new dark gauged U(1) and becoming near-extremal.
If such relics exist, their mergers would have interesting observational signature in the form of high-energy neutrinos and $\gamma$-rays.
PBH evaporation might also cease as a result of near-extremal rotation, and de Freitas Pacheco and Silk~\cite{deFreitasPacheco:2020wdg} argue that this might apply for PBHs formed in an early post-inflation matter-dominated epoch during preheating.
If one invokes an early accretion phase, this could naturally lead to asteroid-scale DM.

\section{Constraints on non-evaporated PBHs}
\label{sec:Constraints}

\begin{figure}[ht]
\begin{center}
\includegraphics[width=.85\textwidth]{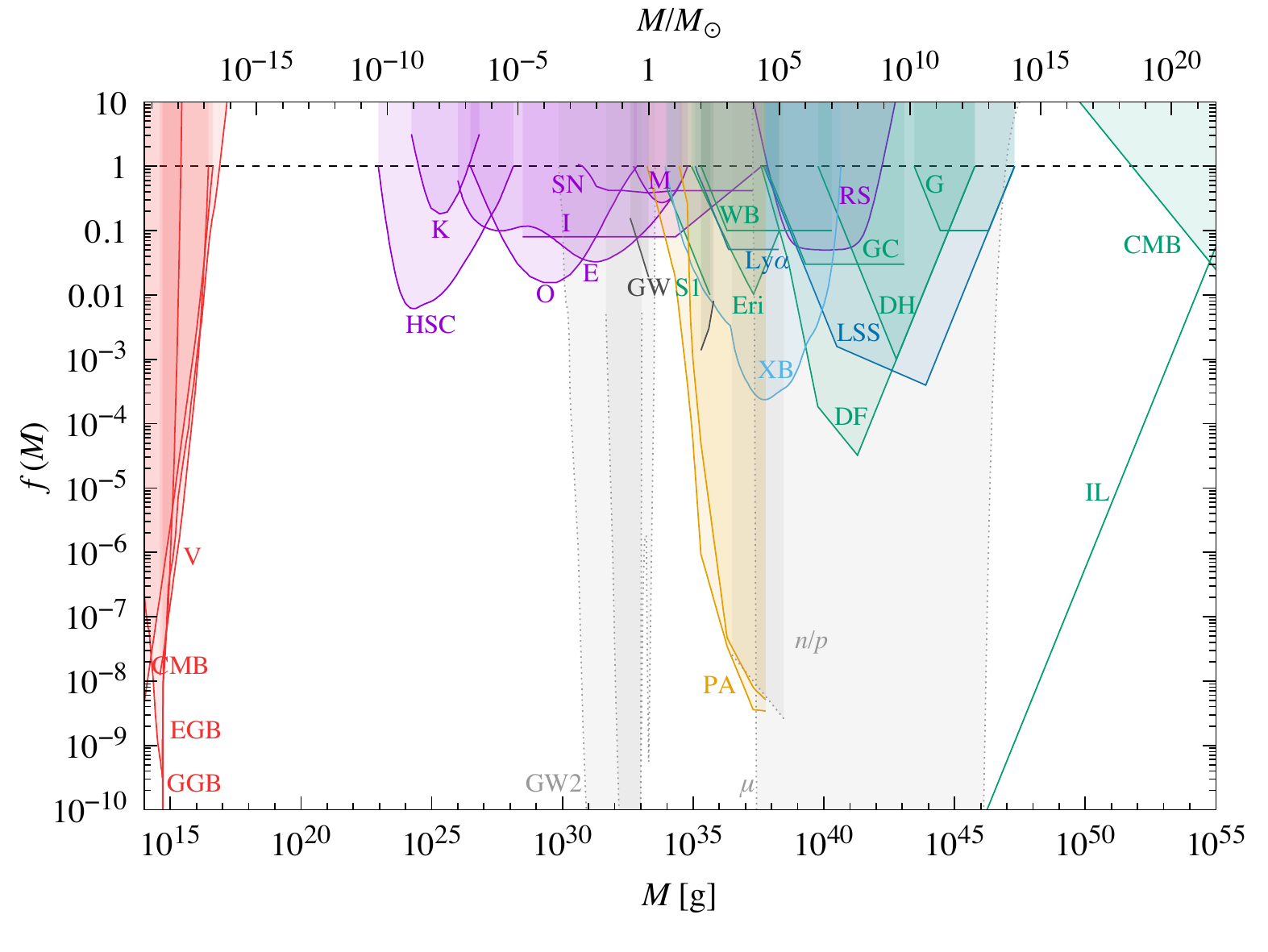}
\end{center}
\caption{Constraints on $f(M)$ from evaporation (red), lensing (magenta), dynamical effects (green), gravitational waves (black), accretion (light blue), CMB distortions (orange), large-scale structure (dark blue) and background effects (grey).
Evaporation limits come from the extragalactic $\gamma$-ray background (EGB), CMB anisotropies (CMB), the Galactic $\gamma$-ray background (GGB) and Voyager-1 $e^\pm$ limits (V).
Lensing effects come from microlensing of stars in M31 by Subaru (HSC), in the Magellanic Clouds by MACHO (M) and EROS (E), in the local neighbourhood by Kepler (K), in the Galactic bulge by OGLE (O) and the Icarus event in a cluster of galaxies (I), microlensing of supernovae (SN) and quasars (Q), and millilensing of compact radio sources (RS).
Dynamical limits come from disruption of wide binaries (WB) and globular clusters (GC), heating of stars in the Galactic disk (DH), survival of star clusters in Eridanus II (Eri) and Segue 1 (S1), infalling of halo objects due to dynamical friction (DF), tidal disruption of galaxies (G), and the CMB dipole (CMB).
Accretion limits come from X-ray binaries (XB), CMB anisotropies measured by Planck (PA) and gravitational waves from binary coalescences (GW).
Large-scale structure constraints come from the Lyman-$\alpha$ forest (Ly$\alpha$) and various other cosmic structures (LSS).
Background constraints come from CMB spectral distortion ($\mu$), 2nd order gravitational waves (GW2) and the neutron-to-proton ratio ($n/p$).
The incredulity limit (IL) corresponds to one hole per Hubble volume.
These constraints are broken down into different categories in subsequent figures, these including some less certain limits which are omitted here.}
\label{fig:latest}
\end{figure}

We next review the various constraints associated with PBHs which are too large to have evaporated completely by now, updating the equivalent discussion in Refs.~\cite{Carr:2009jm,Carr:2016drx}.
All the limits assume that PBHs cluster in galactic halos in the same way as other forms of CDM unless they are too large to reside in halos.
In this section we also assume the PBHs have a monochromatic mass function, in the sense that they span a mass range of at most $\Delta M \sim M$.
In this case, the fraction $f(M)$ of the halo in PBHs is related to $\beta'(M)$ by
\begin{equation}
f(M)
\equiv
  \frac{\Omega_\mathrm{PBH}(M)}{\Omega_\mathrm{CDM}}
\approx
  3.79\,\Omega_\mathrm{PBH}(M)
= 3.81 \times 10^8\,\beta'(M)\,
  \left(\frac{M}{M_\odot}\right)^{-1/2}\,,
\label{eq:f}
\end{equation}
where we have taken $ \Omega_\mathrm{CDM} = 0.264$ from Ref.~\cite{Aghanim:2018eyx}.
For an extended or multi-modal mass function, which we discuss later, one can define the total PBH contribution to the DM, $f_\mathrm{PBH}$, by integrating over $M$ and there is then a constraint $f_\mathrm{PBH} < 1$.
Our limits on $f(M)$ are summarised in Fig.~\ref{fig:latest}, which is an updated version of Fig.~8 of Ref.~\cite{Carr:2009jm}.
The main constraints derive from (partial) PBH evaporations, gravitational lensing, numerous dynamical effects, PBH accretion and gravitational wave observations.
Since Fig.~\ref{fig:latest} is complicated, the constraints for each of these categories are shown separately in subsequent figures.
We note that there are several limits in most mass ranges.

It must be stressed that the constraints have varying degrees of certainty and most come with caveats.
For some, the observations are well understood (e.g.\ the CMB and gravitational lensing data) but there are uncertainties in the black hole physics.
For others, the observations themselves are not fully understood or depend upon additional astrophysical assumptions;
we will give explicit examples of this below.
Many of the constraints depend on physical parameters which are not indicated explicitly, and some of them depend on various cosmological and astrophysical assumptions.
Sometimes the limit can be assigned a confidence level (CL) but this is often hard to assess and not indicated in the original paper.
Note that some of the limits are extended into the $f >1$ domain, although this is obviously unphysical.
This is because some limits will shift downward as the observational data improve, so it is useful to know their form.

Similar figures can be found in numerous other papers;
see, for example, Table~1 of Josan \textit{et al.} \cite{Josan:2009qn}, Fig.~4 of Mack \textit{et al.} \cite{Mack:2006gz}, Fig.~9 of Ricotti \textit{et al.} \cite{Ricotti:2007au}, Fig.~1 of Capela \textit{et al.} \cite{Capela:2013yf}, Fig.~1 of Clesse \& Garc\'ia-Bellido \cite{Clesse:2016vqa} and Fig.~10 of Sasaki \textit{et al.} \cite{Sasaki:2018dmp}.
However, we believe this is the most comprehensive treatment available to date (i.e.\ prior to 2021).
We group the limits by type and discuss those within each type in order of increasing mass.
Sometimes we express the limit in terms of an analytic function $f_{\mathrm{max}}(M)$ over some mass range but this is not always possible.

\subsection{Evaporation constraints}

For PBHs somewhat larger than $M_*$\,, one can use the $\gamma$-ray background to constrain the value of $f(M)$.
This can be derived from the constraint on $\beta (M)$ derived in Sec.~\ref{sec:EGB} by using Eq.~\eqref{eq:f}.
For $M > 2 M_{*}$\,, one can neglect the change of mass and the time-integrated spectrum $\mathrm dN^{\gamma}/\mathrm dE$ of photons from each PBH is just obtained by multiplying the instantaneous spectrum by the age of the Universe $t_{0}$\,.
For PBHs of mass $M$, the discussion in the appendix of Ref.~\cite{Carr:2009jm} gives
\begin{equation}
\frac{\mathrm dN^\gamma}{\mathrm dE}
\propto
\begin{cases}
E^{3}\,M^{3}
& ( E < M^{-1} ) \\
E^{2}\,M^{2}\,\mathrm e^{-E M}
& ( E > M^{-1} )\,,
\end{cases}
\end{equation}
where we put $\hbar = c = G = 1$.
This peaks at $E \sim M^{-1}$ with a value independent of $M$.
The number of background photons per unit energy per unit volume from all PBHs is then obtained by integrating over the mass function:
\begin{equation}
\mathcal E(E)
= \int_{M_{\mathrm{min}}}^{M_{\mathrm{max}}} \!\mathrm dM\,
  \frac{\mathrm dn}{\mathrm dM}\,
  \frac{\mathrm dN^\gamma}{\mathrm dE}(M,E)\,,
\end{equation}
where $M_{\mathrm{min}}$ and $M_{\mathrm{max}}$ specify the mass limits.
For a monochromatic mass function, this gives
\begin{equation}
\mathcal E(E)
\propto
  f(M) \times
\begin{cases}
E^{3}\,M^{2}
& ( E < M^{-1} ) \\
E^{2}\,M\,\mathrm e^{-E M}
& ( E > M^{-1} )
\end{cases}
\end{equation}
and the associated intensity is
\begin{equation}
I(E)
\equiv
  \frac{E\,\mathcal E(E)}{4 \pi}
\propto
  f( M ) \times
\begin{cases}
E^{4}\,M^{2}
& ( E < M^{-1} ) \\
E^{3}\,M\,\mathrm e^{- E M}
& ( E > M^{-1} )
\end{cases}
\end{equation}
with units $ \mathrm s^{-1}\,\mathrm{sr}^{-1}\,\mathrm{cm}^{-2} $.
This peaks at $E \sim M^{-1} $ with a value $I^{\mathrm{max}}(M) \propto f(M)\,M^{-2}$.
The observed extragalactic intensity is $I^\mathrm{obs} \propto E^{-(1+\epsilon)} \propto M^{1+\epsilon} $ where $\epsilon$ lies between $0.1$ (the value favoured in Ref.~\cite{Sreekumar:1997un}) and $ 0.4 $ (the value favoured in Ref.~\cite{Strong:2004ry}).
Hence requiring $I^{\mathrm{max}}(M) \le I^\mathrm{obs}(M) $ gives \cite{Carr:2009jm}
\begin{equation}
f(M)
\lesssim
  2 \times 10^{-8}\,
  \left(\frac{M}{M_{*}}\right)^{3 + \epsilon}
\quad
( M > M_{*} = 5 \times 10^{14} \mathrm g )\,.
\end{equation}
As expected, this is equivalent to condition \eqref{eq:photon2}, which is represented in Fig.~\ref{fig:photon}.
We have seen that the Galactic $\gamma$-ray background may give a stronger limit but this depends sensitively on the form of the mass function.

\begin{figure}[ht]
\begin{center}
\includegraphics[width=.60\textwidth]{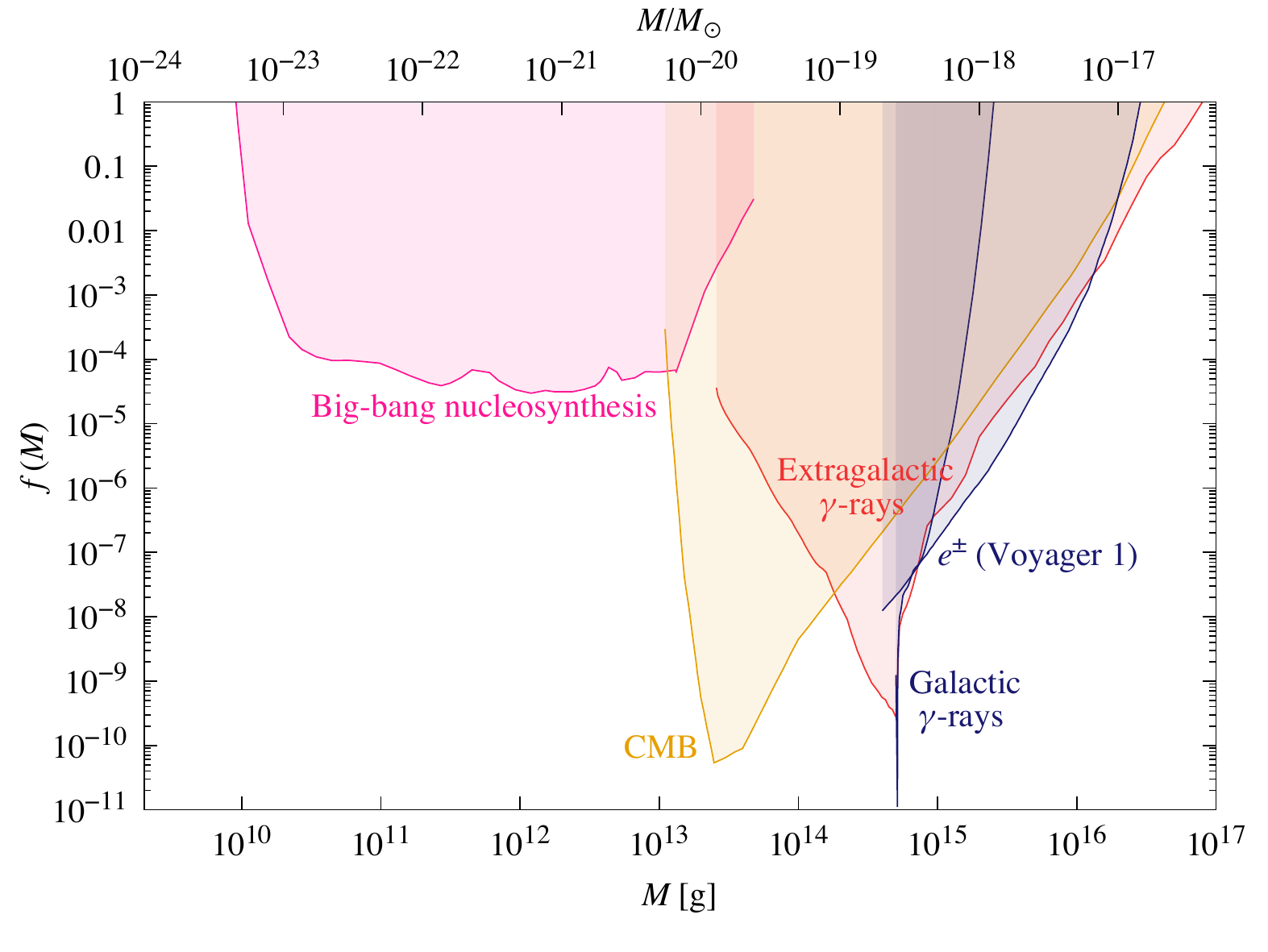}
\end{center}
\caption{Evaporation constraints from BBN~\cite{Carr:2009jm}, CMB spectral distortions and anisotropies~\cite{Acharya:2020jbv,Chluba:2020oip}, extragalactic $\gamma$-rays~\cite{Carr:2009jm}, Galactic $\gamma$-rays~\cite{Carr:2016hva} and Voyager-1 e$^\pm$~\cite{Boudaud:2018hqb}, based on Fig.~\ref{fig:combined} and Eq.~\eqref{eq:f}.}
\label{fig:evap}
\end{figure}

PBHs smaller than $10^{15}\,\mathrm g$ have evaporated completely and therefore cannot contribute to the DM.
The function $f(M)$ is not defined in this range, so the abundance is usually described in terms of the collapse fraction $\beta(M)$.
Nevertheless, one can still formally relate $\beta(M)$ to $f(M)$ using Eq.~\eqref{eq:f}.
The dominant constraints in Fig.~\ref{fig:combined} are therefore also represented in Fig.~\ref{fig:evap}, including the Voyager-1 limits \cite{Boudaud:2018hqb}.
However, we do not show constraints which depend on the DM profile in the Galactic centre or dwarf galaxies because these are uncertain.

\subsection{Lensing constraints}

\begin{figure}[ht]
\begin{center}
\includegraphics[width=.60\textwidth]{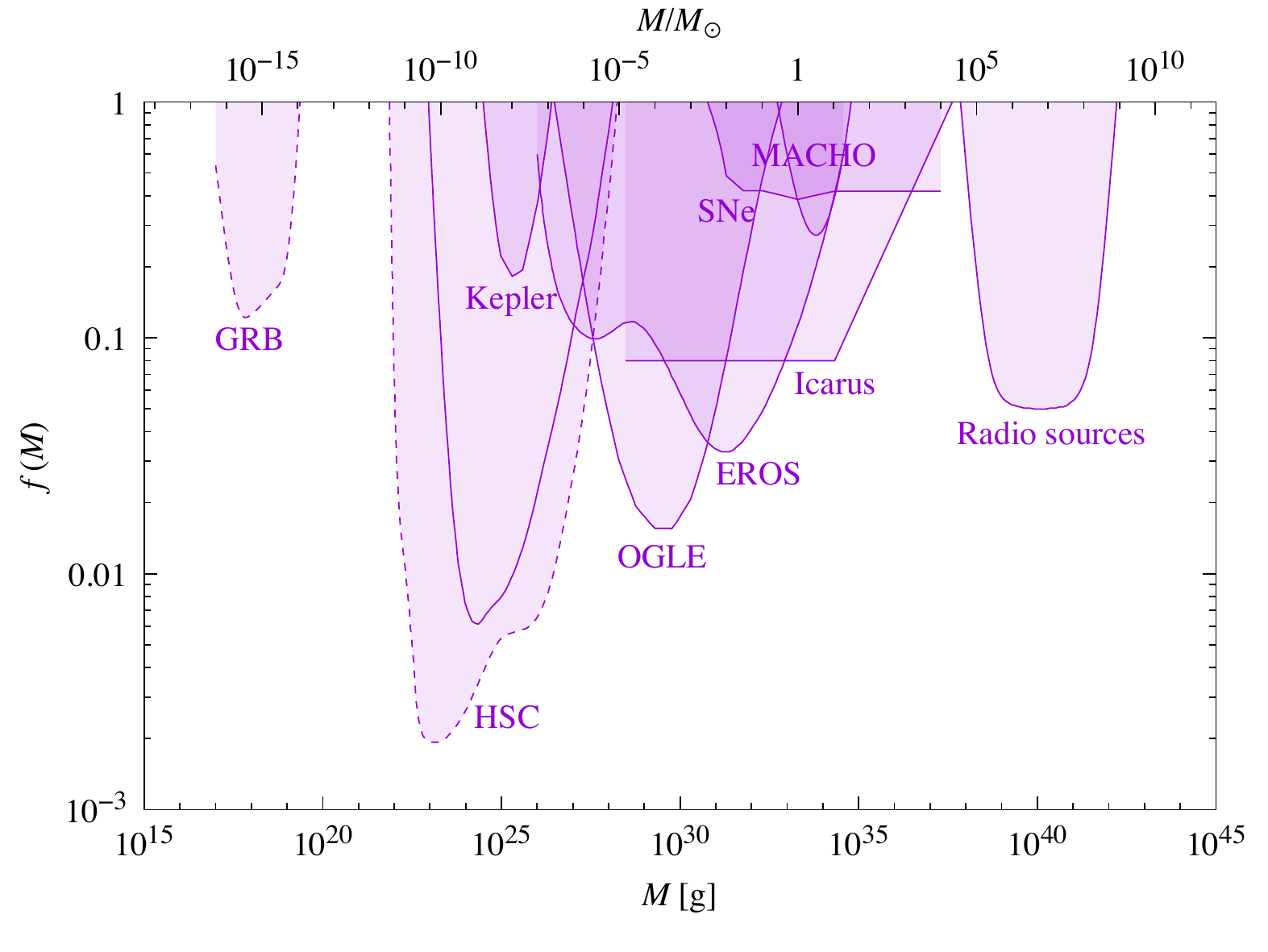}
\end{center}
\caption{Lensing constraints from HSC~\cite{Smyth:2019whb}, Kepler~\cite{Griest:2013esa,Griest:2013aaa}, OGLE~\cite{Niikura:2019kqi}, EROS~\cite{Tisserand:2006zx}, MACHO~\cite{Allsman:2000kg}, Icarus~\cite{Oguri:2017ock}, SNe~\cite{Zumalacarregui:2017qqd} and radio sources~\cite{Wilkinson:2001vv}, with the disputed GRB constraint~\cite{Barnacka:2012bm} and the original HSC constraint \cite{Niikura:2017zjd} shown by broken lines.}
\label{fig:lensing}
\end{figure}

The lensing constraints on $f(M)$ are summarised in Fig.~\ref{fig:lensing}.
Where possible, we use 95\,\% CL constraints but one must distinguish between limits based on positive detections and null detections.
Claimed positive detections come from OGLE in the low mass range \cite{Niikura:2019kqi}, from MACHO and quasar microlensing \cite{1993Natur.366..242H} in the solar mass range and from millilensing of AGN jets \cite{Vedantham:2017kyb} in the high mass range.

\subsubsection{Femtolensing and picolensing}

It has been argued that constraints on PBHs with very low mass come from the femtolensing of $\gamma$-ray bursts (GRBs).
Assuming the bursts are at a redshift $z \sim 1$, early studies \cite{Marani:1998sh,Nemiroff:2001bp} claimed to exclude $f = 1$ in the mass range $10^{-16}\text{--}10^{-13}\,M_{\odot}$ and later work \cite{Barnacka:2012bm} gave a limit which can be approximated as
\begin{equation}
f(M)
<
0.1 \quad ( 5 \times 10^{16}\,\mathrm g < M < 10^{19}\,\mathrm g )\,.
\label{femto}
\end{equation}
The femtolensers could either be PBHs or ultra-compact DM minihalos (e.g.\ made up of QCD axions).
However, Katz \textit{et al.} \cite{Katz:2018zrn} have reviewed this argument, taking into account the extended nature of the source as well as wave-optic effects, and argued that most GRBs are inappropriate for femtolensing searches due to their large sizes.
A small fraction, characterised by fast variability, might be small enough to be useful but many such bursts would be needed to achieve meaningful constraints.
The femtolensing limit \eqref{femto} is therefore not shown in Fig.~\ref{fig:latest} and indicated with a broken line in Fig.~\ref{fig:lensing}.
However, Jung and Kim \cite{Jung:2019fcs} have recently discussed another technique which probes the asteroid mass range, using the lensing parallax of GRBs observed simultaneously by spatially separated detectors.

\subsubsection{Microlensing of stars}
\label{sec:ml}

Microlensing (ML) observations of stars in the Magellanic Clouds (LMC and SMC) probe the fraction of the Galactic halo in compact objects in the mass range $10^{-7}\,M_{\odot} < M < 10\,M_{\odot}$ \cite{Paczynski:1985jf}.
The optical depth of the halo towards the LMC and SMC, defined as the probability that any given star is amplified by at least $1.34$ at a given time, is 
\begin{equation}
\tau^{(\mathrm{SMC})}_{\mathrm L}
= 1.4\,\tau^{(\mathrm{LMC})}_{\mathrm L}
= 6.6 \times 10^{-7}\,f(M)
\end{equation}
for the S halo model \cite{Alcock:2000kd}.
Although the initial motivation for ML surveys was to search for brown dwarfs with $M < 0.08\,M_{\odot}$\,, the possibility that the halo is dominated by these objects was soon ruled out by the MACHO experiment.
Indeed, this excluded all the DM being in PBHs over the mass range $0.1\text{--}0.9\,M_{\odot}$ and the upper limit was later extended to $30 M_{\odot}$ \cite{Allsman:2000kg}.
The associated constraint is shown in Fig.~\ref{fig:lensing}.
However, MACHO did observe ML and claimed their 13--17 events were consistent with compact objects of $M \sim 0.5\,M_{\odot}$ contributing 20\,\% of the halo mass \cite{Alcock:2000ph}.
This raised the possibility that some of the DM could be PBHs formed at the QCD phase transition \cite{Jedamzik:1996mr,Widerin:1998my,Jedamzik:1999am}.
Similar claims were later made by the POINT-AGAPE collaboration, who detected six ML events in a survey of M31 \cite{CalchiNovati:2005cd}.
They argued that this was more than could be expected from self-lensing alone and concluded that 20\,\% of the halo mass in the direction of M31 could be PBHs in the range $0.5\text{--}1\,M_{\odot}$\,.

The EROS collaboration monitored brighter stars in a wider solid angle and thus obtained more stringent constraints on $ f(M) $, arguing that some of the MACHO events were due to self-lensing.
Specifically, they excluded $ 6 \times 10^{-8}\,M_\odot < M < 15\,M_\odot $ objects from having $f \sim 1$ \cite{Tisserand:2006zx}.
This gives the 95\,\% CL limit shown in Fig.~\ref{fig:lensing}, which may be approximated as 
\begin{equation}
f(M)
<
\begin{cases}
1
& ( 6 \times 10^{-8}\,M_{\odot}< M < 15\,M_{\odot} ) \\
0.1
& ( 10^{-6}\,M_{\odot}< M < 1\,M_{\odot} ) \\
0.04
& ( 10^{-3}\,M_{\odot}< M < 0.1\,M_{\odot} )\,.
\end{cases}
\end{equation}
There are also constraints from the OGLE experiment, which monitored both the Galactic Bulge and Magellanic Clouds.
In particular, OGLE-III data towards the LMC \cite{Wyrzykowski:2010mh} and SMC \cite{Wyrzykowski:2011tr} gave stronger results than indicated above in the high mass range.
Combining SMC constraints from EROS and OGLE-III~\cite{Novati:2013fxa} excludes $f \sim 1$ over the range $10^{-5}\text{--}10^2 M_{\odot}$\,, with the limit bottoming out with $f \sim 0.1$ at $10^{-2} \, M_{\odot}$\,.
We do not include this limit in Fig.~\ref{fig:lensing} because it depends on some unidentified detections being attributed to self-lensing.

Recently Niikura \textit{et al.} \cite{Niikura:2019kqi} have used five years of observations of stars in the Galactic bulge by OGLE to constrain the PBH abundance.
The ML light-curves span a distribution of timescales over the range $t_E \approx [1, 300]\,\mathrm{days}$ and can be well modeled as being due to ordinary stars or stellar remnants.
However, there are also six ultra-short events with $t_E \approx [0.1, 0.3]\,\mathrm{day}$.
These might be attributed to free-floating planets but they could also be generated by planetary-mass PBHs with $f \sim 0.01$.
On the null hypothesis that the OGLE data contain no PBH events, the 95\,\% CL upper bound may be approximated as
\begin{equation}
f(M)
<
\begin{cases}
1
& (10^{-7}\,M_{\odot}< M < 1\,M_{\odot} ) \\
0.1
& ( 10^{-6}\,M_{\odot}< M < 10^{-2}\,M_{\odot} ) \\
0.01
& (M \sim 10^{-4}\,M_{\odot} )\, .
\end{cases}
\end{equation}
and this limit is shown in Fig.~\ref{fig:lensing}.
Wyrzykowski and Mandel \cite{Wyrzykowski:2019jyg} have used Gaia data on the distances and proper motions of non-blended sources to improve the mass estimates in the parallax ML events found in 8 years of OGLE-III observations towards the Galactic Bulge.
They identify $18$ new events and the derived distribution of masses for the lenses is consistent with a continuous distribution of stellar remnant masses.
However, they also find $8$ candidates in the theoretically expected mass gap of $2\text{--}5\,M_{\odot}$ between neutron stars and black holes.
These might conceivably be PBHs \cite{Carr:2019kxo}, although they could also be stellar black holes if they were born with large recoil velocities.

The interpretation of the MACHO, EROS and OGLE results is very sensitive to the properties of the Milky Way halo~\cite{Green:2017qoa}.
All the above constraints assume a semi-isothermal density sphere but Calcino \textit{et al.} \cite{Calcino:2018mwh} argue that this is inconsistent with the Milky Way rotation curve and does not incorporate uncertainties in the shape of the halo.
With their new model, which is non-spherical and accounts for these features, the ML constraints for LMC weaken for masses around $10\,M_{\odot}$ but tighten at lower masses.
They claim that a wide PBH mass distribution could also shift the constraint towards smaller masses and that spatial clustering of the PBHs may remove the constraint altogether at $M \sim 1\text{--}10\,M_{\odot}$ since there would be some probability that a given source is not lensed at all.
Hawkins has come to a similar conclusion, arguing that the recent low-mass Galactic halo models would relax the constraints and allow the halo to consist entirely of solar-mass PBHs \cite{Hawkins:2015uja}.

Kepler data have considerably improved the limits in the low mass range \cite{Griest:2013esa,Griest:2013aaa}, implying
\begin{equation}
f(M)
< 0.3 \qquad ( 2 \times 10^{-9}\,M_{\odot} < M < 10^{-7}\,M_{\odot} )\,.
\end{equation}
This limit is indicated on Fig.~\ref{fig:lensing}.
Niikura \textit{et al.} \cite{Niikura:2017zjd} have carried out a dense-cadence 7-hour observation of M31 with the Subaru Hyper Suprime-Cam (HSC) to search for ML by PBHs of similar mass in the halos of the Milky Way or M31.
If such PBHs make up a significant fraction of DM, they expect to find many ML events.
However, they identify only a single candidate event, which translates into the most stringent current upper bound on the PBH abundance in the mass range $10^{-11}\,M_{\odot} < M < 10^{-6}\,M_{\odot}$\,.
The black hole size becomes less than photon wavelength and the Einstein radius becomes smaller than the source size at the lower bound.
However, Smyth \textit{et al.} \cite{Smyth:2019whb} point out that Niikura \textit{et al.} underestimate this effect because they assume a source size of a solar radius, whereas the sources are likely to be larger than this.
They find that the HSC constraints are weaker by up to almost three orders of magnitude in some cases, increasing the mass below which PBHs can provide the DM from $10^{-11}\,M_{\odot}$ to $10^{-10}\,M_{\odot}$\,.
The Niikura \textit{et al.} limit is therefore shown by a broken line in Fig.~\ref{fig:lensing}.
Sugiyama \textit{et al.} \cite{Sugiyama:2019dgt} have also studied this effect and point out that, if a denser-cadence monitoring of a sample of white dwarfs over a year were available, then one could probe the wave-optics effect on a ML light-curve and push the PBH constraint down to $10^{-11}\,M_{\odot}$\,.

Going beyond the local neighbourhood, recent observations of fast transient events in massive galaxy clusters, interpreted as individual stars in giant arcs highly magnified due to caustic crossing, can also be used to constrain PBHs.
For example, Oguri \textit{et al.} \cite{Oguri:2017ock} find that the lens mass associated with the first event for MACS J1149 must be in the range $0.1\text{--}4 \times 10^3\,M_{\odot}$\,.
The derived lens properties are consistent with a ML event produced by such a star contributing to the intra-cluster light but DM models with $f > 0.1$ in compact objects in the mass range $10^{-5} \text{--}10^2\,M_{\odot}$ would predict too low magnification.
This corresponds to the ``Icarus'' line in Fig.~\ref{fig:lensing}.

\subsubsection{Microlensing of supernovae}

PBHs cause most lines of sight to be demagnified relative to the mean, with a long tail of higher magnifications.
This effect was first applied to type Ia supernovae (SNe) by Metcalf and Silk~\cite{Metcalf:2006ms}, who inferred constraints on compact objects in the mass range $10^{-2} \text{--}10^{10}\,M_{\odot}$\,.
More recently, Zumalac\'arregui and Seljak \cite{Zumalacarregui:2017qqd} have used the lack of lensing of SNe to constrain PBHs, modelling the effects of large-scale structure, allowing for a non-Gaussian model for the intrinsic SNe luminosity distribution and addressing potential systematic errors.
Using current JLA SNe data, they derive a 95\,\% CL bound $f<0.35$, which rules out PBHs comprising all the DM with $5 \sigma$ significance.
The finite size of SNe limits the validity of the results to $M > 10^{-2}\,M_{\odot}$ but this closes a previously open PBH mass range.
This limit is shown in Fig.~\ref{fig:lensing}.

Garc\'ia-Bellido \textit{et al.} \cite{Garcia-Bellido:2017imq} argue that this limit can be weakened if the PBHs have an extended mass function or are clustered, concluding that the bound on the fraction of DM in PBHs is $f < 1.09\,(1.38)$ for the JLA (Union 2.1) catalogues.
This would be compatible with PBHs providing all the DM in the LIGO band.
However, Zumalac\'arregui and Seljak dispute this claim, pointing out that part of the discrepancy arises through a (later corrected) error in the radius of the SNe photosphere and arguing that the dependence on mass only enters if the mass function has considerable support in the $M<10^{-2}M_{\odot}$ region where the constraints become weaker.

\subsubsection{Quasar microlensing and millilensing}

Early studies of the ML of quasars \cite{1994ApJ...424..550D} seemed to exclude all the DM being in objects with $10^{-3}\,M_{\odot} < M < 60\,M_{\odot}$\,.
However, this limit does not apply in the $\Lambda$CDM picture and so is not shown in Fig.~\ref{fig:latest}.
At around the same time, Hawkins \cite{1993Natur.366..242H} claimed evidence for a critical density of Jupiter-mass objects from observations of quasar ML and associated these with PBHs formed at the quark-hadron transition.
However, his later analysis yielded a lower density (cf.\ $\Omega_\mathrm{DM} \approx 0.3$) and a larger mass of around $1 \,M_{\odot}$ \cite{Hawkins:2006xj}.
The status of this claim remains unclear \cite{Zackrisson:2003wu}.
Studies of quasar ML by Mediavilla \textit{et al.} initially gave a limit \cite{2009ApJ...706.1451M}
\begin{equation}
f(M)
< 1
\quad ( 10^{-3}\,M_{\odot}< M < 60\,M_{\odot} )
\end{equation}
but they later found positive evidence for ML, this indicating that 20\,\% of the total mass is in compact objects in the mass range $0.05\text{--}0.45\,M_{\odot}$ \cite{Mediavilla:2017bok}.
Although they attributed this to a normal stellar component, Hawkins has argued that one requires PBHs to explain the observations \cite{Hawkins:2020zie}.
Since it is difficult to express the result of Ref.~\cite{Mediavilla:2017bok} as a quantitative upper bound on $f(M)$, we do not include it in Figs.~\ref{fig:lensing}.

In a higher mass range, Vedantham \textit{et al.} \cite{Vedantham:2017kyb} claim to have detected long-term radio variability -- what they term Symmetric Achromatic Variability -- in the light-curves of active galactic nuclei (AGN).
They propose that this arises from gravitational millilensing of features in AGN jets moving relativistically through gravitational lensing caustics created by $10^3\text{--}10^6\,M_{\odot}$ condensates or black holes within intervening galaxies.
The lower end of this mass range has been inaccessible with previous gravitational lensing observations.
This offers a powerful probe of the cosmological matter distribution on these mass scales, as well as micro-arcsecond resolution of AGN, a factor of 30--100 greater than is possible with Very Long Baseline Interferometry (VLBI).
Banik \textit{et al.}~\cite{Banik:2018tyb} consider the kink-like distortions of razor-thin lensing arcs caused by free-floating black holes along the line-of-sight to AGN observed in the radio.
The null detection of such distortions by VLBI and the Atacama Large Millimeter/submillimeter Array (ALMA) potentially constrains the density of PBHs with $M >10^3\,M_{\odot}$ to be $f< 10^{-5}$.

Millilensing of compact radio sources \cite{Wilkinson:2001vv} gives a limit which can be approximated as
\begin{equation}
f(M)
<
\begin{cases}
( M / 2 \times 10^{4}\,M_{\odot} )^{-2}
& ( M < 10^{5}\,M_{\odot} ) \\
0.06
& ( 10^{5}\,M_{\odot} < M < 10^{8}\,M_{\odot} ) \\
(M / 4 \times 10^{8}\,M_{\odot} )^{2}
& ( M > 10^{8}\,M_{\odot} )\,.
\end{cases}
\end{equation}
Though weaker than other constraints in this mass range, we include this limit in Fig.~\ref{fig:lensing}.
Hezaveh \textit{et al.} \cite{Hezaveh:2016ltk} have studied substructure in the matter density near galaxies, using ALMA observations of the strong lensing system SDP.81.
They find evidence for a $10^9\,M_{\odot}$ subhalo near one of the images and derive constraints on the abundance of DM subhalos down to around $10^7\,M_{\odot}$ (the mass of the smallest detected satellites in the Local Group).
There are hints of additional substructure but the results are consistent with the $\Lambda$CDM prediction.
Although they do not explicitly constrain PBHs, so we do not include their results in Fig.~\ref{fig:lensing}, this suggests $f < 0.04$ for $10^7 < M/M_{\odot} < 10^9$.

\subsubsection{Microlensing of Mira variables, pulsars and fast radio bursts}

Finally, we mention a number of other lensing effects which have been discussed in the literature and may eventually constrain the DM fraction in PBHs.
The first two effects have been primarily considered in the context of searching for the black hole remnants of ordinary stars but the mass range could also be relevant for PBHs.
The other effects are specifically relevant to PBHs.
The associated constraints are not shown in Fig.~\ref{fig:lensing} since they are only potential rather than actual limits.

Abrams and Takada~\cite{Abrams:2020jvs} suggest monitoring $10^9$ long-duration ($> 100\,\mathrm{days}$) ML events in the Galactic bulge using the Large Synoptic Survey Telescope.
This would correspond to black holes larger than $30\,M_{\odot}$\,, the dominant source of ML for a Salpeter mass function.
Karami \textit{et al.} \cite{Karami:2016rjp} discuss the detection of ML of Mira variables with VLBI.
These stars are sufficiently bright and compact to permit direct imaging using the Very Long Baseline Array (VLBA).
ML by Galactic black holes would be relatively common and features in the associated images would be sufficiently well resolved to fully reconstruct the lens properties, enabling the measurement of mass, distance, and tangential velocity of the lensing object to a precision of better than 15\,\%.
Future radio ML surveys conducted with upcoming radio telescopes combined with modest improvements in the VLBA could increase the rate of Galactic black hole events to roughly $10\,\mathrm{yr}^{-1}$, sufficient to double the number of known stellar mass black holes in a couple years and determining the distribution functions of stellar mass black holes.

Other interesting constraints come from pulsar timing.
Schutz and Liu \cite{Schutz:2016khr} claim the non-detection of the 3rd order Shapiro time delay as PBHs move around the Galactic halo will soon constrain those with mass in the range $1\text{--}10^3\,M_{\odot}$\,.
Indeed, observations with the Square Kilometre Array (SKA) potentially preclude $f$ exceeding $0.01 \text{--}0.1$ in this mass range.
Munoz \textit{et al.} \cite{Munoz:2016tmg} have pointed out that similar limits could come form the lensing of fast radio bursts (FRBs).
The potential constraint in this case bottoms out at a mass of around $10\,M_{\odot}$ with a value $f \sim 0.01$.
Other advances in this context could include studies of interference pattern in the frequency spectrum \cite{Katz:2019qug} and searches for the lensing of mini-FRBs \cite{Laha:2018zav}.
Pulsars are sensitive to acceleration and gravitational redshifts induced by any transiting PBHs and Dror \textit{et al.} \cite{Dror:2019twh} have studied the sensitivity of pulsar timing arrays (PTAs) to this effect.
They find that the SKA could exclude $f \sim 1$ over the mass range $10^{-12}\text{--}10^{2}\,M_{\odot}$ and the constraint would be $f < 0.01$ in the range $10^{-4}\text{--}10^{-2}\,M_{\odot}$\,.
PTAs can potentially probe a substantially lower density of such objects because of the large effective radius over which they can be observed.

Jow \textit{et al.} \cite{Jow:2020rcy} have studied the importance of wave effects for the ML of point-like radio bursts.
The frequency dependence of wave effects breaks degeneracies in the usual geometric optics limit and constructive interference leads to a generic increase in the ML cross-section in the wave-optics regime compared to the geometric-optics regime.
For example, fast radio bursts (FRBs) and pulsars will be lensed in the full wave-optics regime by isolated masses in the range $10^{-7}\text{--}10^{-4}\,M_\odot$\,, with the interference pattern itself determining the lens mass.

Bai and Orlofsky \cite{Bai:2018bej} point out that X-ray pulsars with higher photon energies and smaller sizes are good sources for ML in the $10^{-16}\text{--}10^{-11}\,M_{\odot}$ window.
Among existing X-ray pulsars, the most promising is SMC X-1 because of its apparent brightness and large distance.
Their analysis of existing data from the RXTE telescope suggests that $f \sim 1$ in PBHs is close to being excluded in the asteroid mass range.
Future observations of this source by X-ray telescopes with larger effective areas could close this mass window entirely.

\subsection{Dynamical constraints}

A variety of dynamical constraints come into play at higher mass scales and many of them involve the destruction of various astronomical objects by the passage of nearby PBHs.
As shown by Carr and Sakellariadou \cite{Carr:1997cn}, if the PBHs have density $\rho$ and velocity dispersion $V$, while the objects have mass $M_c$\,, radius $R_c$\,, velocity dispersion $V_c$ and survival time $t_L$\,, then the constraint has the general form:
\begin{equation}
f(M)
<
\begin{cases}
M_c\,V/(G\,M\,\rho\,t_L\,R_c)
& ( M < M_c\,(V/V_c) ) \\
M_c/(\rho\,V_c\,t_L\,R_c^2)
& ( M_c\,(V/V_c) < M < M_c\,(V/V_c)^3) \\
M\,V_c^2/( \rho\,R_c^2\,V^3\,t_L)\,\exp[(M/M_c)\,(V_c/V)^3]
& ( M > M_c\,(V/V_c)^3)\,.
\end{cases}
\label{eq:carsaklim}
\end{equation}
The three limits correspond to disruption by multiple encounters, one-off encounters and non-impulsive encounters, respectively, and we assume $V > V_c$\,.
The fraction is thus constrained over the mass range
\begin{equation}
\frac{M_c\,V}{G\,\rho_\mathrm{DM}\,t_L\,R_c}
< M
< M_c\,\left(\frac{V}{V_c}\right)^3\,,
\end{equation}
where $\rho_\mathrm{DM}$ is the DM
density.
The mass limits correspond to the values of $M$ for which one would have $f = 1$, so PBHs could provide \emph{all} the DM only if $M$ is below the lower limit or above the higher limit.
Note that various numerical factors are omitted in these expressions.

Constraint \eqref{eq:carsaklim} applies providing there is at least one PBH within the relevant environment, be that a galactic halo, a cluster of galaxies or the Universe itself.
This is termed the ``incredulity limit'' and for an environment of mass $M_\mathrm E$\, it corresponds to the condition
\begin{equation}
f(M)
> f_\mathrm{IL}
\equiv
  ( M / M_{\mathrm{E}} )\,,
\end{equation}
where $M_\mathrm E$ is around $3 \times 10^{12}\,M_{\odot}$ for halos, $10^{14}\,M_{\odot}$ for clusters and $10^{22}\,M_{\odot}$ for the Universe.
However, this is only relevant if the exponential upper cut-off in Eq.~\eqref{eq:carsaklim} lies to the right of the incredulity limit and this applies for $M_\mathrm E < \rho_\mathrm{DM}\,t_L\,t_c^2\,V^3$.
Otherwise Eq.~\eqref{eq:carsaklim} is unaffected.

\begin{figure}[ht]
\begin{center}
\includegraphics[width=.60\textwidth]{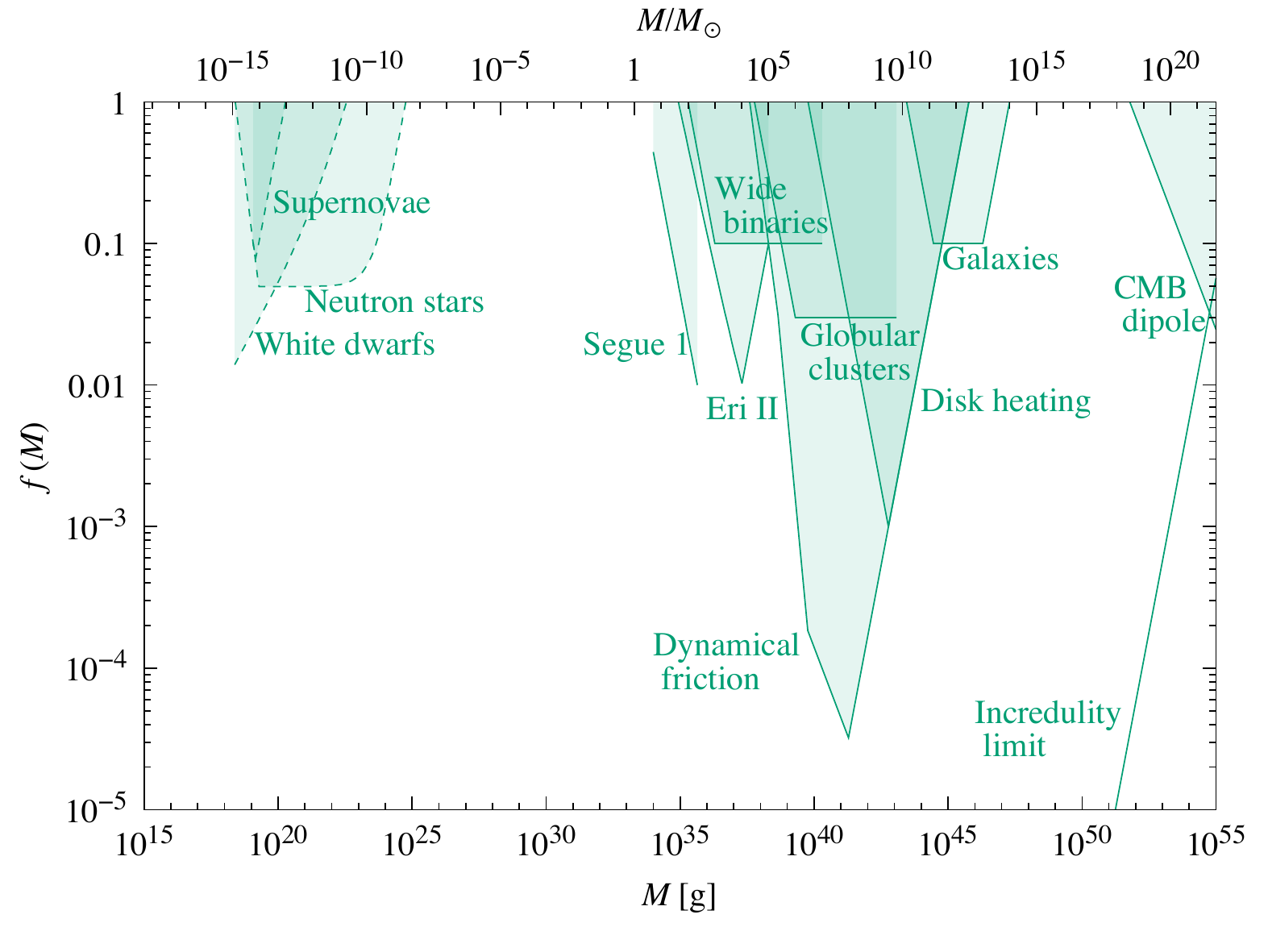}
\end{center}
\caption{Dynamical constraints from Segue I~\cite{Koushiappas:2017chw}, Eridanus II~\cite{Zoutendijk:2020rti}, wide binaries~\cite{Monroy-Rodriguez:2014ula} and, based on Ref.~\cite{Carr:2018rid}, from globular clusters, disk heating, dynamical friction, galaxy disruption in clusters, the CMB dipole and the incredulity limit.
The disputed white dwarf \cite{Capela:2012jz} and neutron star ~\cite{Capela:2013yf} and supernovae~\cite{Graham:2015apa} limits are shown by broken lines.}
\label{fig:dynamics}
\end{figure}

The dynamical limits are summarised in Fig.~\ref{fig:dynamics} and most of them apply to objects in galactic halos.
In this case, the upper limit on the mass of the objects which provide the DM lies in the range $10\text{--}10^{6}\,M_{\odot}$\,, so this is particularly relevant for intermediate mass objects.
The limits would also apply to forms of DM other than PBHs.
For example, Raidal \textit{et al.} have suggested that large primordial curvature fluctuations could collapse into horizonless exotic compact objects \cite{Raidal:2018eoo}.
The DM might also be \emph{clusters} of PBHs \cite{1987ApJ...316...23C,Belotsky:2015psa} and this is suggested by the claim by Dokuchaev \textit{et al.} \cite{Dokuchaev:2004kr} and Chisholm \cite{Chisholm:2005vm} that PBHs could form in tight clusters, giving a local overdensity well in excess of that provided by the halo concentration alone.
Overdense regions might also form ultra-compact minihalos (UCMHs), with the UCMH constraints on the density perturbations being stronger than the PBH constraints in some mass ranges \cite{Bringmann:2011ut}.

\subsubsection{Terrestrial and solar system encounters}

The effects of PBH collisions on the Earth have been a subject of long-standing interest, ever since Ryan suggested this could explain the Tunguska event~\cite{1973Natur.245...88J}.
For example, Khriplovich \textit{et al.} \cite{Khriplovich:2008er} have examined whether terrestrial collisions could be detected acoustically.
The seismic energy deposited by a $10^{15}\,\mathrm g$ PBH is comparable to a 4th magnitude earthquake but the time between collisions is very long ($10^{7}\,\mathrm{yr}$) even if such PBHs provide all the DM.
Encounters with larger (Earth-threatening) PBHs are even rarer but the rate at which non-collisional close encounters could be detected by seismic activity is two orders of magnitude larger.
Luo \textit{et al.} \cite{Luo:2012pp} have proposed searching for PBH transits through or near the Earth by studying the seismic waves generated in Earth's interior.
These would have two signatures: the almost simultaneous arrival of the waves everywhere on the Earth's surface and the excitation of unusual spheroidal modes with a characteristic frequency-spacing.
We do not show any of these constraints in Fig.~\ref{fig:dynamics}, because they are only important for masses so low that the PBHs would have evaporated.
However, the equivalent limits for other forms of DM (e.g.\ snowballs) are shown in Ref.~\cite{Carr:1997cn}.

Scholtz and Unwin \cite{Scholtz:2019csj} propose that a planetary-mass PBH could be captured by the Solar System and thereby explain Planet 9.
This possibility is discussed further by Siraj and Loeb~\cite{Siraj:2020upy}, while Arbey and Auffinger~\cite{Arbey:2020urq} suggest that a fly-by spacecraft might even measure its Hawking radiation in the radio band.
Looking further to the future, gravitational-wave observatories in space might detect the dynamical effects of PBHs.
For example, LISA could detect PBHs in the mass range $10^{14}\text{--}10^{20}\,\mathrm g$ by measuring the gravitational impulse induced by any nearby ones~\cite{Adams:2004pk,Seto:2004zu}.

\subsubsection{Stars, neutron stars and white dwarfs}

Zhilyaev \cite{Zhilyaev:2007rx} has suggested that PBHs could be captured by stars and eventually collide with them.
For those in the mass range $10^{-3}\text{--}10^{-2}\,M_{\odot}$\,, such collisions could produce the small class of $\gamma$-ray bursts emitting thermal bremsstrahlung in the range $20\text{--}300\,\mathrm{keV}$.
Roncadelli \textit{et al.}~\cite{Roncadelli:2009qj} have considered halo PBHs being captured and swallowed by stars in the Galactic disc.
The stars would eventually be accreted by the holes, producing radiation and a population of subsolar black holes which could only be of primordial origin.
Every disc star would contain such a black hole if the DM were in PBHs smaller than $3 \times 10^{26}\,\mathrm g$.
Since the time-scale on which a star captures a PBH scales as $\tau_{\mathrm{cap}} \propto n_{\mathrm{PBH}}^{-1} \propto M\,f(M)^{-1}$, requiring this to exceed the age of the Galactic disc implies~\cite{Carr:2009jm}
\begin{equation}
f(M)
< M / 3 \times 10^{26}\,\mathrm g \,,
\label{eq:neutron}
\end{equation}
where we have used the normalisation of Roncadelli \textit{et al.} This corresponds to a \emph{lower} limit on the mass of objects providing the DM.
A similar analysis of the collisions of PBHs with main-sequence stars, red-giant cores, white dwarfs and neutron stars by Abramowicz \textit{et al.} \cite{Abramowicz:2008df} gives a lower limit of $10^{20}\,\mathrm g$.
Capela \textit{et al.} have also considered the capture of PBHs by white dwarfs \cite{Capela:2012jz} and neutron stars \cite{Capela:2013yf} and find that the survival of these objects implies a limit which Ref.~\cite{Carr:2016drx} approximates as 
\begin{equation}
f(M)
< \frac{M}{4.7\times 10^{24}\,\mathrm g}\,
  \left(1-\exp\left[-\frac{ M }{ 2.9 \times 10^{23}\,\mathrm g }\right]\right)^{-1}
\quad \left(2.5 \times 10^{18} \mathrm g < M < 10^{25}\,\mathrm g\right)\,.
\end{equation}
This is similar to Eq.~\eqref{eq:neutron} at the high-mass end but flattens off at intermediate masses.
The lower cut-off at $2 \times 10^{18}\,\mathrm g$ arises because PBHs lighter than this will not have time to consume the neutron stars during the age of the Universe.
This explains the form of the `neutron star' limit in Fig.~\ref{fig:dynamics}.
Pani and Loeb \cite{Pani:2014rca} have argued that this excludes PBHs from providing the dark matter throughout the sublunar window.
However, this argument is sensitive to the dark matter density at the centers of globular clusters.
This was taken to be $10^{4}\,\mathrm{GeV}\,\mathrm{cm}^{-3}$ but it is now known to be much lower than this for particular globular clusters \cite{Ibata:2012eq,Bradford:2011aq}.
Since these limits are no longer believed \cite{Capela:2014qea,Defillon:2014wla}, they are not included in Fig.~\ref{fig:latest} but they are shown by broken lines in Fig.~\ref{fig:dynamics}.

Graham \textit{et al.} \cite{Graham:2015apa} argue that PBHs can trigger white dwarf (WD) explosions.
Their transit through a WD causes localised heating through dynamical friction and this can initiate runaway thermonuclear fusion, causing the WD to explode as a supernova.
The shape of the observed WD distribution rules out PBHs with masses $10^{19}\text{--}10^{20}\,\mathrm g$ providing the local DM, while those in the range $10^{20}\text{--}10^{22}\,\mathrm g$ are constrained by the observed supernova rate.
This raises the intriguing possibility that a class of supernova may be triggered through rare events induced by DM rather than the conventional mechanism of accreting white dwarfs exploding upon reaching the Chandrasekhar mass.
However, Montero-Camacho \textit{et al.} argue that this argument is inapplicable \cite{Montero-Camacho:2019jte}, so the supernovae limit is shown by a broken line in Fig.~\ref{fig:dynamics}.

Several authors have suggested that PBH interactions with a neutron star (NS) could explain various astronomical phenomena.
There are too many astrophysical uncertainties in these models to infer clear constraints on $f(M)$ but they all depend on the fact that a PBH captured by a NS sinks to the center and consumes it from the inside.
For example, Takhistov \textit{et al.} \cite{Takhistov:2017nmt,Takhistov:2017bpt} argue that collisions between NSs and PBHs could generate a population of solar-mass PBHs, leading to a new class of $\gamma$-ray bursts and micro-quasars.
Abramowicz and Bejger \cite{Abramowicz:2017zbp} argue that collisions with PBHs of $10^{23}\,\mathrm g$ may explain the phenomenology of FRBs, in particular their millisecond durations and large luminosities.
Contrary to the usual explanation for FRBs, no large magnetic fields are needed in their model.

Fuller \textit{et al.} \cite{Fuller:2017uyd} show that some or all of the inventory of r-process nucleosynthesis can be produced in such interactions if PBHs with masses $10^{-14}\text{--}10^{-8}\,M_{\odot}$ make up just a few percent of the DM.
When such a PBH is swallowed by a rotating millisecond NS, they argue that the resulting spin-up ejects $\sim 0.1\text{--}0.5\,M_{\odot}$ of relatively cold neutron-rich material.
This can also produce a kilonova-type afterglow and an FRB but not significant gravitational radiation or neutrinos, allowing such events to be differentiated from the compact-object mergers detectable by gravitational-wave observatories.
Ejected matter is also heated by beta decay, which leads to positron emission consistent with the observed $511\,\mathrm{keV}$ line from the Galactic Center.
They claim that this scenario is compatible with pulsar and NS statistics, as well as with the r-process evolution history within the Galaxy.

\subsubsection{Wide binaries}

Binary stars with wide separation are vulnerable to disruption from encounters with PBHs or any other type of compact object \cite{1985ApJ...290...15B,1987ApJ...312..367W}.
Observations of wide binaries in the Galaxy therefore constrain the abundance of halo PBHs.
By comparing the result of simulations with observations, Yoo \textit{et al.} \cite{Yoo:2003fr} originally ruled out compact objects with $ M > 43\,M_\odot $ from providing most of the halo mass.
However, Quinn \textit{et al.} \cite{Quinn:2009zg} later performed a more careful analysis of the radial velocities of these binaries and found that one of the widest-separation ones was spurious.
The resulting 95\,\% CL constraint became
\begin{equation}
f(M)
<
\begin{cases}
(M/500\,M_\odot)^{-1}
& (500\,M_\odot < M \lesssim 10^3\,M_\odot ) \\
0.5
& (10^3\,M_\odot \lesssim M < 10^8\,M_\odot)\,.
\end{cases}
\end{equation}
It flattens off above $10^{3}\,M_{\odot}$ because one-off disruption then dominates, as implied by Eq.~\eqref{eq:carsaklim}.
More recent studies by Monroy-Rodr\'iguez and Allen~\cite{Monroy-Rodriguez:2014ula} have reduced the lower mass limit from $500\,M_{\odot}$ to $21\text{--}78\,M_{\odot}$\,, so the limit is taken to be $100\,M_{\odot}$ in Fig.~\ref{fig:dynamics} and $f(M)$ then flattens off at around $0.1$.
The narrow window between the ML lower bound and the wide-binary upper bound is therefore shrinking and may in principle be eliminated altogether.
More positively, Tian \textit{et al.} \cite{2020ApJS..246....4T} have recently studied more than 4,000 halo wide binaries in the Gaia survey and detected a break in their separation distribution that may be indicative of PBHs with $M > 10 M_{\odot}$\,.

\subsubsection{Disruption of globular clusters and dwarf galaxies}

An argument similar to the binary disruption one shows that the survival of globular clusters against tidal disruption by passing PBHs gives a limit
\begin{equation}
f(M)
<
\begin{cases}
( M / 3 \times 10^{4}\,M_{\odot} )^{-1}
& ( 3 \times 10^{4}\,M_{\odot} < M < 10^{6}\,M_{\odot} ) \\
0.03
& ( 10^{6}\,M_{\odot} < M < 10^{11}\,M_{\odot} )\,,
\end{cases}
\end{equation}
although this depends sensitively on the mass and the radius of the cluster.
The limit flattens off above $ 10^{6}\,M_{\odot}$ because one-off encounters then dominate.
The upper limit of $ 3 \times 10^{4}\,M_{\odot}$ on the mass of objects dominating the halo is consistent with the numerical calculations of Moore \cite{1993ApJ...413L..93M}.

In a related limit, Brandt \cite{Brandt:2016aco} claims that a mass above $5\,M_{\odot}$ is excluded by the fact that a star cluster near the centre of the dwarf galaxy Eridanus II has not been disrupted by halo objects.
His constraint can be written as \cite{Carr:2016drx}
\begin{equation}
f(M)
\lesssim
\begin{cases}
( M / 3.7\,M_{\odot} )^{-1} / [ 1.1 - 0.1\ln( M / M_{\odot} ) ]
& ( M < 10^{4}\,M_{\odot} ) \\
( M / 10^{6}\,M_{\odot} )
& ( M > 10^{4}\,M_{\odot} )\,,
\end{cases}
\label{eq:eribound}
\end{equation}
where the density and velocity dispersion of the DM at the centre of the galaxy are taken to be $0.1\,M_{\odot}\,\mathrm{pc}^{-3}$ and $5\,\mathrm{km}\,\mathrm s^{-1}$, respectively, and the age of the star cluster is taken to be $3\,\mathrm{Gyr}$.
This limit is shown in Fig.~\ref{fig:dynamics}.
The second expression in Eq.~\eqref{eq:eribound} was not included in Ref.~\cite{Brandt:2016aco} but corresponds to having at least one black hole in the dwarf galaxy (the incredulity limit).
Zoutendijk \textit{et al.} \cite{Zoutendijk:2020rti} update Brandt's constraint with new data on Eridanus II and present more secure evidence that the stellar overdensity seen towards it is a star cluster belonging to that galaxy.

Koushiappas and Loeb \cite{Koushiappas:2017chw} have studied another type of effect of PBHs on the dynamical evolution of stars in dwarf galaxies.
They find that mass segregation leads to a depletion of stars in the center of such galaxies and the appearance of a ring in the projected stellar surface density profile.
Using Segue 1 as an example, they show that current observations of the projected surface stellar density imply $f < 0.04$ at 99.9\,\% CL for masses above around $10\,M_{\odot}$\,.
Their limit is shown in Fig.~\ref{fig:dynamics} and has the expected $M^{-1}$ form.

Several groups have obtained PBH constraints from observations of ultra-faint dwarf galaxies (UFDGs), although they do not express their results as limits on $f(M)$.
Zhu \textit{et al.} \cite{Zhu:2017plg} follow the joint evolution of stars and PBHs in UFDGs with a Fokker--Planck code for different DM parameters and then use a Markov chain approach to constrain the PBH properties.
They find that two-body relaxation between the stars and PBHs drives up the stellar core size and increases the central stellar velocity dispersion.
Using the observed half-light radius and stellar velocity dispersion as joint constraints, they infer that the PBHs must have a mass range of $2\text{--}14\,M_{\odot}$ if they are to provide a substantial fraction of the DM.
Boldrini \textit{et al.} \cite{Boldrini:2019isx} use an N-body code to show that the PBHs can induce a cusp-to-core transition in UFDGs via the dynamical friction of the DM particles on PBHs in the mass range $25\text{--}100\,M_{\odot}$ and this requires an upper limit $f < 0.01$.
Stegmann \textit{et al.} \cite{Stegmann:2019wyz} have improved these limits by considering a much larger sample of UFDGs and allowing the PBHs to have a lognormal mass distribution.
The analysis of the half-light radii and velocity dispersions resulting from their simulations appears to exclude PBHs with $\mathcal{O}(1\text{--}100)\,M_{\odot}$ from providing all the DM.

\subsubsection{Disc heating and tidal streams}

Other dynamical limits come into play at higher mass scales.
Halo objects will overheat the stars in the Galactic disc unless one has \cite{Carr:1997cn}
\begin{equation}
f( M )
<
\begin{cases}
( M / 3 \times 10^{6}\,M_{\odot} )^{-1}
& ( M < 3 \times 10^{9}\,M_{\odot} ) \\
( M / M_{\mathrm{halo}} )
& ( M > 3 \times 10^{9}\,M_{\odot} )\,,
\end{cases}
\label{eq:disc}
\end{equation}
where the lower expression is the incredulity limit.
The upper limit of $3 \times 10^{6}\,M_{\odot}$ agrees with the more precise calculations by Lacey and Ostriker \cite{1985ApJ...299..633L}, although they argued that black holes with $2 \times 10^{6}\,M_{\odot}$ could \emph{explain} some features of disc heating.
Constraint \eqref{eq:disc} bottoms out at $M \sim 3 \times 10^{9}\,M_{\odot}$ with a value $f \sim 10^{-3}$ and is shown in Fig.~\ref{fig:dynamics}.
Evidence for a similar effect may come from the claim of Totani \cite{Totani:2009af} that elliptical galaxies are puffed up by dark halo objects of $10^{5}\,M_{\odot}$\,, although this limit is not shown in Fig.~\ref{fig:dynamics}.

Tidal streams in the Milky Way are sensitive probes of the population of low-mass DM subhalos predicted by CDM simulations.
This was originally considered for subhalos made of thermal relics \cite{Banik:2018pjp} but the argument can also be applied to constrain PBHs.
Bovy \textit{et al.} \cite{Bovy:2016irg} have developed a method for calculating the perturbed distribution function of a stream segment, computing the perturbed density for subhalo distributions down to $10^5\,M_{\odot}$\,.
They apply this formalism to the Pal 5 stream and $10$ DM subhalos within $20\,\mathrm{kpc}$ of the Galactic center and with masses in the range $10^{6.5}\text{--}10^9\,M_{\odot}$\,.
If one applies this argument to PBHs and assumes that the Pal 5 stream is $5\,\mathrm{Gyr}$ old, their limit corresponds to $f(M) < 0.002$ in this mass range.
However, Bovy \textit{et al.} do not infer this constraint themselves, so it is not shown in our figures.

\subsubsection{Dynamical friction effect on halo objects}

Another limit in this mass range arises because halo objects will be dragged into the nucleus of our own Galaxy by the dynamical friction of the spheroid stars and halo objects themselves (if they have an extended mass function), this leading to excessive nuclear mass unless \cite{Carr:1997cn}
\begin{equation}
f(M)
<
\begin{cases}
( M / 2 \times 10^{4}\,M_{\odot} )^{-10/7}\,( r_{\mathrm c} / 2\,\mathrm{kpc} )^{2}
& ( M < 5 \times 10^{5}\,M_{\odot} ) \\
( M / 4 \times 10^{4}\,M_{\odot} )^{-2}\,( r_{\mathrm c} / 2\,\mathrm{kpc} )^{2}
& ( 5 \times 10^{5}\,M_{\odot} < M < 2 \times 10^{6}\,( r_{\mathrm c} / 2\,\mathrm{kpc} )\,M_{\odot} ) \\
( M / 0.1\,M_{\odot} )^{-1/2}
& ( 2 \times 10^{6}\,(r_{\mathrm c} / 2\,\mathrm{kpc} )\,M_{\odot} < M < 10^{8}\,M_{\odot} ) \\
( M / M_{\mathrm{halo}} )
& ( M > 10^{8}\,M_{\odot} )\,.
\end{cases}
\end{equation}
The last expression is the incredulity limit and first three correspond to the drag being dominated by spheroid stars (low $M$), halo objects (high $M$) and some combination of the two (intermediate $M$).
The limit is sensitive to the halo core radius $ r_\mathrm c $ but with the normalisation of $ 2\,\mathrm{kpc}$ one requires $f < 1$ for $M < 2 \times 10^{4}\,M_{\odot}$\,.
The limit bottoms out at $M \sim 10^{8}\,M_{\odot}$ with a value $f \sim 3 \times 10^{-5}$ and is shown in Figs.~\ref{fig:dynamics}.
However, there is a caveat here in that holes drifting into the nucleus might be ejected by the slingshot mechanism if there is already a binary black hole there \cite{Hut:1992iy}.
This possibility was explored by Xu and Ostriker \cite{Xu:1994vb}, who obtained a larger upper limit of $3 \times 10^{6}\,M_{\odot}$\,.

\subsubsection{Disruption and tidal distortion of galaxies in clusters}

A similar argument to that used for globular clusters shows that the survival of galaxies in clusters against tidal disruption by giant cluster PBHs gives a limit~\cite{Carr:1997cn}
\begin{equation}
f(M)
<
\begin{cases}
( M / 
10^{10}\,M_{\odot} )^{-1}
& ( 
10^{10}\,M_{\odot} < M < 10^{11}\,M_{\odot} ) \\
0.1
& ( 10^{11}\,M_{\odot} < M < 10^{13}\,M_{\odot} ) \\
( M / 2 \times 10^{14}\,M_{\odot}  )
& ( M > 10^{13}\,M_{\odot})\,,
\end{cases}
\end{equation}
where $2 \times 10^{14}\,M_{\odot}$ in the last term corresponds to the mass of a typical cluster.
This limit is shown in Fig.~\ref{fig:dynamics}.
There is also a constraint from the lack of unexplained tidal distortions of galaxies in clusters.
If the fraction of distorted galaxies at any time is $\Delta_{\mathrm g}$, this limit takes the form~\cite{Carr:1997cn}
\begin{equation}
f( M )
< \lambda^{-3}\,
  \Delta_{\mathrm g}\,
  \left(\frac{ \rho_{\mathrm g} }{ \rho_{\mathrm c} }\right)
\approx
  20\,
  \left(\frac{ \lambda }{ 3 }\right)^{-3}\,
  \Delta_{\mathrm g}\,,
\end{equation}
where $\rho_{\mathrm g}$ and $\rho_{\mathrm c}$ are the densities of the galaxy and cluster, respectively, and $\lambda \approx 3$ represents the difference between distortion and disruption.
So one has an interesting limit providing the fraction of unexplained distorted galaxies is less than 5\,\%.
This limit only applies for $M > \lambda^{-3} M_{\mathrm g} \approx 3 \times 10^{9}\,M_{\odot}$ since otherwise the tidal distortion radius is smaller than $R_{\mathrm g}$\,.
In particular, van den Bergh~\cite{1969Natur.224..891V} concluded from observations of the Virgo cluster, which has a mass of $10^{15}\,M_{\odot}$ and only 4 of 73 cluster members showing unexplained tidal distortions, that the DM could not be in the form of compact object in the mass range $10^{9}\text{--}10^{13}\,M_{\odot}$\,.
This limit is not shown in Fig.~\ref{fig:dynamics} since it is sensitive to poorly determined astrophysical parameters.

\subsubsection{Intergalactic PBHs}

If there were a population of huge intergalactic PBHs with density parameter $\Omega_D(M)$\,, each galaxy would have a peculiar velocity due to its gravitational interaction with the nearest one \cite{etde_6856669}.
If the objects were smoothly distributed, the typical distance between them would be
\begin{equation}
d
\approx
  40\,\Omega_D(M)^{-1/3}\,(M/10^{16}\,M_{\odot})^{1/3}\,(h/0.7)^{-2/3}\,\mathrm{Mpc}\,.
\end{equation}
This would also be the expected distance of the nearest black hole to the Milky Way.
Over the age of the Universe, this should induce a peculiar velocity
\begin{equation}
V_\mathrm{pec}
\approx
  G\,M\,f(\Omega_0)\,t_0/d^2\,,
\end{equation}
where $\Omega_0 \approx 0.3$ is the total density parameter and $f(\Omega_0) \approx \Omega_0^{0.6}$.
Since the CMB dipole anisotropy shows that the peculiar velocity of our Galaxy is only $400\,\mathrm{km}\,\mathrm s^{-1}$, one infers
\begin{equation}
\Omega_D
< (M/2 \times 10^{17}\,M_{\odot})^{-1/2}\,(t_0/14\,\mathrm{Gyr})^{-3/2}\,(\Omega_0/0.3)^{-0.9}\,(h/0.7)^{-2}\,.
\label{eq:dipole}
\end{equation}
This scenario is interesting only if there is at least one such object within the observable universe and this corresponds to the cosmological incredulity limit
\begin{equation}
\Omega_D(M)
> 3 \times 10^{-8}\,(M/10^{16}\,M_{\odot})\,(t_0/14\,\mathrm{Gyr})^{-3}\,(h/0.7)^{-2}\,,
\label{eq:cil}
\end{equation}
where we have taken the horizon scale to be $d \approx 3\,c\,t_0 \approx 10\,h^{-1}\,\mathrm{Gpc}$.
This intersects Eq.~\eqref{eq:dipole} at $M \approx 10^{21}\,M_{\odot}$\,, so this corresponds to the largest possible intergalactic dark object within the visible Universe.
Constraints~\eqref{eq:dipole} and \eqref{eq:cil} are shown in Fig.~\ref{fig:dynamics} and close off Fig.~\ref{fig:latest} at the right.

\subsection{Cosmic structure constraints}

The dynamical constraints show that PBHs larger than $10^2\,M_{\odot}$ cannot provide the DM but Carr and Silk \cite{Carr:2018rid} point out that such PBHs could generate cosmic structures through the `seed' or `Poisson' effect even if $f$ is small.
The seed effect was first pointed out by Hoyle and Narlikar \cite{1966RSPSA.290..177H} and subsequently studied in Refs.~\cite{1972ApJ...177L..79R,1984MNRAS.206..801C}.
The Poisson effect was first pointed out by M\'esz\'aros \cite{Meszaros:1975ef} and subsequently studied in Refs.~\cite{1977A&A....56..377C,1983ApJ...275..405F,1983ApJ...268....1C}.
This raises the issue of the maximum mass of a PBH and whether the $10^6\text{--}10^{10}\,M_{\odot}$ black holes in galactic nuclei could be of primordial origin.
It is sometimes argued that the success of the BBN scenario implies that PBHs can only form before $1\,\mathrm s$, corresponding to $M < 10^5\,M_{\odot}$\,.
However, the fraction of Universe collapsing into PBHs at that time is only $10^{-6}\,(t/1\,\mathrm s)^{1/2}$, so the effect on BBN should be tiny.
As discussed later, the most important constraint in this mass range comes from the limit on the CMB $\mu$ distortion.

If the Universe contains a population of PBHs of mass $M$, then a region of mass $\mathcal M$ may contain just one of them (giving a seed effect) or many of them (giving a Poisson effect).
The associated initial fluctuation is
\begin{equation}
\delta_\mathrm i
\approx
\begin{cases}
M/\mathcal M
& (\text{seed}) \\
(f\,M/\mathcal M)^{1/2}
& (\text{Poisson})\,.
\end{cases}
\end{equation}
If $f=1$, the Poisson effect dominates for all $\mathcal M$.
If $f \ll 1$, the seed effect dominates for $\mathcal M < M/f$, the limit corresponding to one PBH per region of mass $\mathcal M$ (i.e.\ the incredulity limit).
In either case, the fluctuation grows as $z^{-1}$ from the epoch of matter domination ($z_\mathrm{eq} \approx 4000$), so the mass binding at redshift $z_\mathrm B$ is
\begin{equation}
\mathcal M
\approx
\begin{cases}
4000\,M\,z_\mathrm B^{-1}
& (\text{seed}) \\
10^7\,f\,M\,z_\mathrm B^{-2}
& (\text{Poisson})\,.
\end{cases}
\label{eq:bind}
\end{equation}
If PBHs provide the DM ($f = 1$), the dynamical limits discussed in the previous section require $M < 10^3\,M_{\odot}$ and so the Poisson effect could only bind a scale $\mathcal M < 10^{10}\,z_\mathrm B^{-2}\,M_{\odot}$\,, which is necessarily subgalactic.
This assumes the PBHs have a monochromatic mass function.
If it is extended, the situation is more complicated \cite{Carr:2018rid} because the mass of the effective seed for a region depends on the mass of that region ($\mathcal M$).

\begin{figure}[ht]
\begin{center}
\includegraphics[width=.60\textwidth]{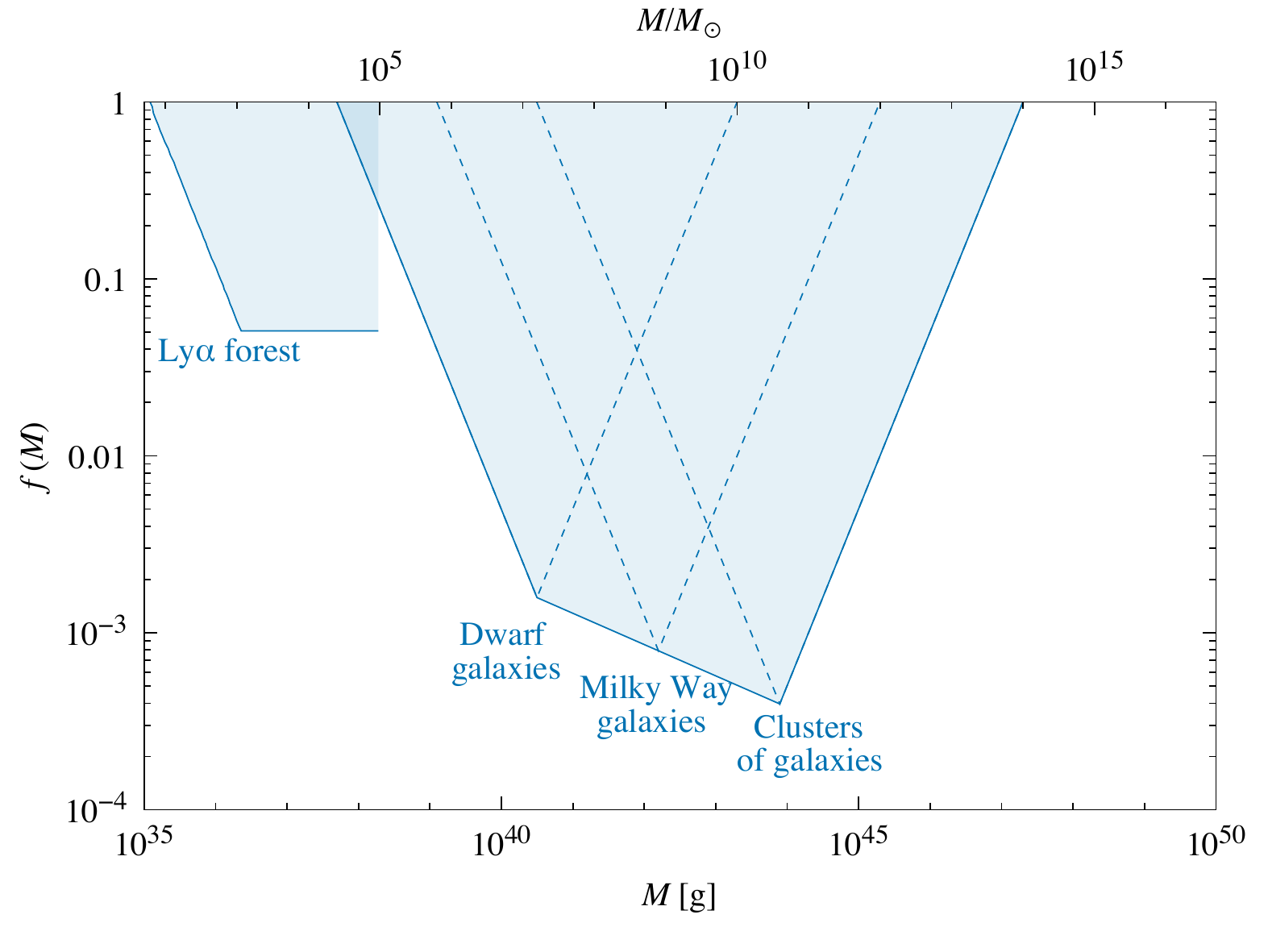}
\end{center}
\caption{Large-scale structure constraints from the requirement that various types of cosmic structures do not form too early~\cite{Carr:2018rid} and Lyman-$\alpha$ forest observations~\cite{Murgia:2019duy}.}
\label{fig:LSS}
\end{figure}

Even if PBHs play no role in generating cosmic structures, one can still place interesting upper limits on the fraction of DM in them by requiring that various types of structure do not form too early.
Carr and Silk find the constraints associated with dwarf galaxies, Milky-Way-type galaxies and clusters of galaxies, these being shown in Fig.~\ref{fig:LSS} and discussed below.
There is also a Poisson constraint associated with observations of the Lyman-$\alpha$ forest~\cite{Afshordi:2003zb,Murgia:2019duy}, which was somewhat misrepresented in our earlier work~\cite{Carr:2009jm}.

The effect of PBH Poisson fluctuations - also termed PBH shot noise - on cosmic structure is also of interest because it enhances the matter power spectrum on small scales.
In particular, several authors have suggested that this would allow the first bound halos to form earlier and more abundantly than in the standard scenario~\cite{Carr:2018rid,2005Natur.438...45K}.
Most recently, Oguri and Takahashi~\cite{Oguri:2020ldf} have pointed out that PBHs larger than $0.1\,M_{\odot}$ would significantly enhance the power spectrum above $10^5\,h\,\mathrm{Mpc}^{-1}$, thereby increasing the gravitational lensing dispersions at $10\text{--}100\,\mathrm{Hz}$ by more than an order of magnitude.

\subsubsection{Lyman-$\alpha$ systems}

Afshordi \textit{et al.}~\cite{Afshordi:2003zb} used observations of the Lyman-$\alpha$ forest to obtain an upper limit of about $10^4\,M_{\odot}$ on the mass of PBHs which provide the DM.
CKSY extended this result to the case in which the PBHs provide a fraction $f(M)$ of the DM on the assumption that the Lyman-$\alpha$ objects are associated with a single mass $M_{\mathrm{Ly}\alpha} \sim 10^{10}\,M_{\odot}$\,.
Since the Poisson fluctuation in the number of PBHs on this mass scale grows between $z_\mathrm{eq} \approx 4000$ and the redshift up to which the Lyman-$\alpha$ forest is observed ($z_{\mathrm{Ly}\alpha} \approx 4$), corresponding to a growth factor of $z_\mathrm{eq}/z_{\mathrm{Ly}\alpha} \approx 10^3$, CKSY claimed that the forest will form earlier than observed unless $f(M) < (M/10^4\,M_{\odot})^{-1}$.
However, there was some confusion here because the Afshordi \textit{et al.} limit is independent of the nature of the Lyman-$\alpha$ objects themselves;
it just uses their power spectrum as a tracer of the density fluctuations.
So the CKSY limit just corresponds to the requirement that $10^{10}\,M_{\odot}$ objects (i.e.\ dwarf galaxies) must not bind too early, as discussed below.

Murgia \textit{et al.} \cite{Murgia:2019duy} have recently improved the Lyman-$\alpha$ limit, using a grid of hydrodynamic simulations to explore different values of astrophysical parameters.
They obtain a ($2 \sigma$) upper limit of the form $f(M) < (M/60\,M_{\odot})^{-1}$ when a Gaussian prior on the reionisation redshift is imposed, weakening to $f(M) < (M/170\,M_{\odot})^{-1}$ when a conservative flat prior is assumed.
Both limits are significantly stronger than the Afshordi \textit{et al.} one and the first is included in Fig.~\ref{fig:LSS}.
The limit flattens off at $10^{3}\, M_\odot$ with $f \sim 0.05$ but they do not discuss how to extend the limit above $10^5 M_{\odot}$\,.
This is problematic because the seed effect then dominates but only influences a small fraction of regions, so the interpretation of the observations is unclear.

\subsubsection{Galaxies and clusters of galaxies}

Galaxies clearly span a range of masses and form over a range of redshifts.
However, if we apply the above argument to Milky-Way-type galaxies, assuming these have a typical mass of $10^{12}\, M_\odot$ and must not bind before a redshift $z_\mathrm B \sim 3$, we obtain
\begin{equation}
f(M)
<
\begin{cases}
(M/10^6\,M_{\odot})^{-1}
& (10^6\,M_\odot < M \lesssim 10^9\,M_\odot) \\
M/10^{12}\,M_{\odot}
& (10^9\,M_\odot \lesssim M < 10^{12}\,M_\odot)\,.
\end{cases}
\label{eq:galaxy}
\end{equation}
This limit is shown in Fig.~\ref{fig:LSS} and bottoms out at $M \sim 10^9\,M_{\odot}$ with a value $f \sim 0.001$.
The first expression can be obtained by putting $\mathcal M \sim 10^{12}\,M_{\odot}$ and $z_\mathrm B \sim 3$ in Eq.~\eqref{eq:bind}.
The second expression corresponds to having just one PBH per galaxy and is also the line above which the seed effect dominates the Poisson effect.
Indeed, since the initial seed fluctuation is $M/\mathcal M$, the seed mass required for a Milky-Way-type galaxy to bind at $z \approx 3$ is immediately seen to be $10^9\,M_{\odot}$\,.
There is no constraint on PBHs below this line because the fraction of the Universe going into galaxies would be small, with most of the baryons presumably going into the intergalactic medium.

If we apply the same argument to dwarf galaxies, assuming these have a mass of $10^{10}\,M_\odot$ and must not bind before $z_\mathrm B \sim 10$, we obtain
\begin{equation}
f(M)
<
\begin{cases}
(M/10^5\,M_{\odot})^{-1}
& (10^5\,M_\odot < M \lesssim 3 \times10^{7}\,M_\odot) \\
M/10^{10}\,M_{\odot}
& (3 \times 10^{7}\,M_\odot \lesssim M < 10^{10}\,M_\odot)\,,
\end{cases}
\end{equation}
this bottoming out at $M \sim 3 \times 10^{7}\,M_{\odot}$ with a value $f \sim 0.003$.
On the other hand, if we apply the argument to clusters of galaxies, assuming these have a mass of $10^{14}\,M_\odot$ and must not bind before $z_\mathrm B \sim 1$, we obtain
\begin{equation}
f(M)
<
\begin{cases}
(M/10^7\,M_{\odot})^{-1}
& (10^7\,M_\odot < M \lesssim 3 \times 10^{10}\,M_\odot ) \\
M/10^{14}\,M_{\odot}
& (3 \times 10^{10}\,M_\odot \lesssim M < 10^{14}\,M_\odot)\,,
\end{cases}
\end{equation}
this bottoming out at $M \sim 3 \times 10^{10}\,M_{\odot}$ with a value $f \sim 0.0003$.
Clearly all these limits are very approximate and the division into different types of bound structure is just made for simplification.
In practice, one should combine all the limits shown in Fig.~\ref{fig:LSS} to obtain the boundary indicated by the solid line.

\subsubsection{First baryonic clouds}
\label{sec:clouds}

The first bound baryonic clouds would be expected to have a mass of $10^6 M_{\odot}$ and be surrounded by minihalos of CDM in the usual picture.
We cannot apply the above argument to constrain PBHs directly because there is no observational evidence for the formation redshift of such clouds.
However, they would form at $z_\mathrm B \sim 100$ in the CDM picture, so we can still derive a limit corresponding to the requirement that the standard picture is not perturbed.
In this case, Eq.~\eqref{eq:galaxy} is replaced by 
\begin{equation}
f(M)
<
\begin{cases}
(M/10^3\,M_{\odot})^{-1}
& (10^3\,M_\odot < M \lesssim 3 \times 10^4\,M_\odot) \\
M/10^{6}\,M_{\odot}
& (3 \times 10^4\,M_\odot \lesssim M < 10^{6}\,M_\odot)\,,
\end{cases}
\end{equation}
this bottoming out at $M \sim 3 \times 10^4\,M_{\odot}$ with a value $f \sim 0.03$.
This is not shown in Fig.~\ref{fig:LSS} since it is not an actual constraint.
This scenario has been studied in more detail by Kadota and Silk~\cite{Kadota:2020ahr}, who argue that a large fraction of the DM can reside in minihalos which formed at $z >100$.
Since they would have a higher density than in the conventional scenario (without PBHs), they would be less prone to tidal disruption.

Kashlinsky has been prompted by the LIGO/Virgo observations to consider the formation of bound clouds by the Poisson fluctuations associated with a population of $30\,M_{\odot}$ PBHs \cite{Kashlinsky:2016sdv}.
However, he adds an interesting new feature to the scenario by suggesting that this could explain the spatial coherence of the fluctuations in the source-subtracted near-IR cosmic infrared background (CIB) detected by the \textit{Spitzer/Akari} satellites and the X-ray background~\cite{2007ApJ...654L...5K,2013ApJ...769...68C,2012ApJ...753...63K,Kashlinsky:2018mnu}.
The idea is that the infrared radiation is generated by the stars which form in the first halos and the X-rays by the PBHs.
It has long been appreciated that the CIB and its fluctuations would be a crucial test of scenarios in which the DM comprises black holes \cite{Bond:1985pc} but in this case the PBHs are merely triggering high-redshift star formation and not generating the CIB directly.

In a related scenario, the formation of bound clusters of PBHs through the Poisson effect is invoked to generate the supermassive black holes (SMBHs) thought to reside in galactic nuclei \cite{Duechting:2004dk,Khlopov:2004sc};
if one replaces $ \mathcal M$ with $ 10^8\,M_\odot $ and $ z_{\mathrm B}$ with $ 10 $, the limiting mass implied by Eq.~\eqref{eq:bind} is reduced to $10^3\,M_\odot $\,.
The process whereby a cluster of PBHs can evolve into the single SMBH associated with a quasar has been studied by Dokuchaev \textit{et al.}~\cite{Dokuchaev:2004kr}.
Recently, Kashlinksky~\cite{Kashlinsky:2020ial} has linked these scenarios by arguing that advection will lead to a common motion of DM and baryons on the scales relevant for the formation of the first luminous sources.
The advection rate reaches a minimum at around $10^9 M_{\odot}$\,, associated with the formation of quasars, and then rises to a maximum near $10^{12} M_{\odot}$\,, associated with galaxy formation.
This scenario was also discussed by Serpico \textit{et al.}~\cite{Serpico:2020ehh}, who pointed out that it is constrained by the CMB polarisation measurements because the advection-dominated accretion inevitably emit extra photons at high redshifts.

\subsubsection{Clusters of PBHs}

Clesse and Garc\'ia-Bellido stress that PBHs could be spatially clustered into subhaloes if they are part of a larger-scale overdense region \cite{Clesse:2016vqa}.
Indeed, this idea goes back to the early work of Chisholm \cite{Chisholm:2005vm,Chisholm:2011kn}.
They infer that the lensing constraints in the mass range required to explain the LIGO/Virgo detections can be relaxed and this also has important implications for the efficacy of PBH merging in the early Universe.
However, this argument depends on details of small-scale structure formation which are not fully understood.
Indeed, Bringmann \textit{et al.} \cite{Bringmann:2018mxj} argue that the PBH constraints become even stronger if there is large initial clustering.
Ballesteros \textit{et al.}~\cite{Ballesteros:2018swv} argue that this only applies if the PBHs have a very wide mass function but they agree that PBHs with a narrow mass function cannot be numerous enough to explain the LIGO/Virgo detections.

Ali-Ha\"imoud \cite{Ali-Haimoud:2018dau} has studied the initial clustering of PBHs formed from the gravitational collapse of large density fluctuations and does not find small-scale clustering beyond that expected from Poisson effects.
He claims that Chisholm's analysis is flawed due to an improper series expansion.
He also derives an analytic expression for the two-point correlation function of large-threshold fluctuations, generalizing previous results to arbitrary separation.
This conclusion is supported by Moradinezhad Dizgah \textit{et al.}~\cite{MoradinezhadDizgah:2019wjf}.
Desjacques and Riotto \cite{Desjacques:2018wuu} have discussed the impact of PBH spatial clustering on the large-scale power spectrum.
While a Poissonian term is always present in the zero-lag correlation, this need not imply that PBHs are Poisson distributed because their initial clustering depends on the shape of the small-scale power spectrum.
However, this is not relevant for a narrow spectral feature and for the PBH masses still allowed by observations.

Young and Byrnes \cite{Young:2019gfc} consider the impact of local non-Gaussianity on the initial PBH clustering and mass function due to mode coupling between long and short wavelength modes.
They show that even a small amount of non-Gaussianity results in a significant enhancement of clustering and merging, shifting the mass function to higher masses.
However, as the clustering becomes strong, the increased local PBH number density leads to a large theoretical uncertainty in the merger rate.
Using a functional integration approach, Suyama and Yokoyama \cite{Suyama:2019cst} find that PBH clustering on super-Hubble scales can never be induced if the initial primordial fluctuations are Gaussian but that it can be enhanced by the local trispectrum of the primordial curvature perturbations.
Franciolini \textit{et al.}~\cite{Franciolini:2018vbk} have used a path integral approach to study the effects of non-Gaussianity on both the formation and clustering of PBHs.
Matsubara \textit{et al.} have considered the clustering of PBHs if they form in an early matter-dominated epoch from non-Gaussian fluctuations \cite{Matsubara:2019qzv}.
Clustering is less significant in this case but can still be much larger than Poisson shot noise.
The situation is complicated and a clear picture of the effect on non-Gaussianities on PBH abundance and distribution has not yet emerged.

\subsection{Accretion and $\mu$-distortion constraints}

\begin{figure}[ht]
\begin{center}
\includegraphics[width=.60\textwidth]{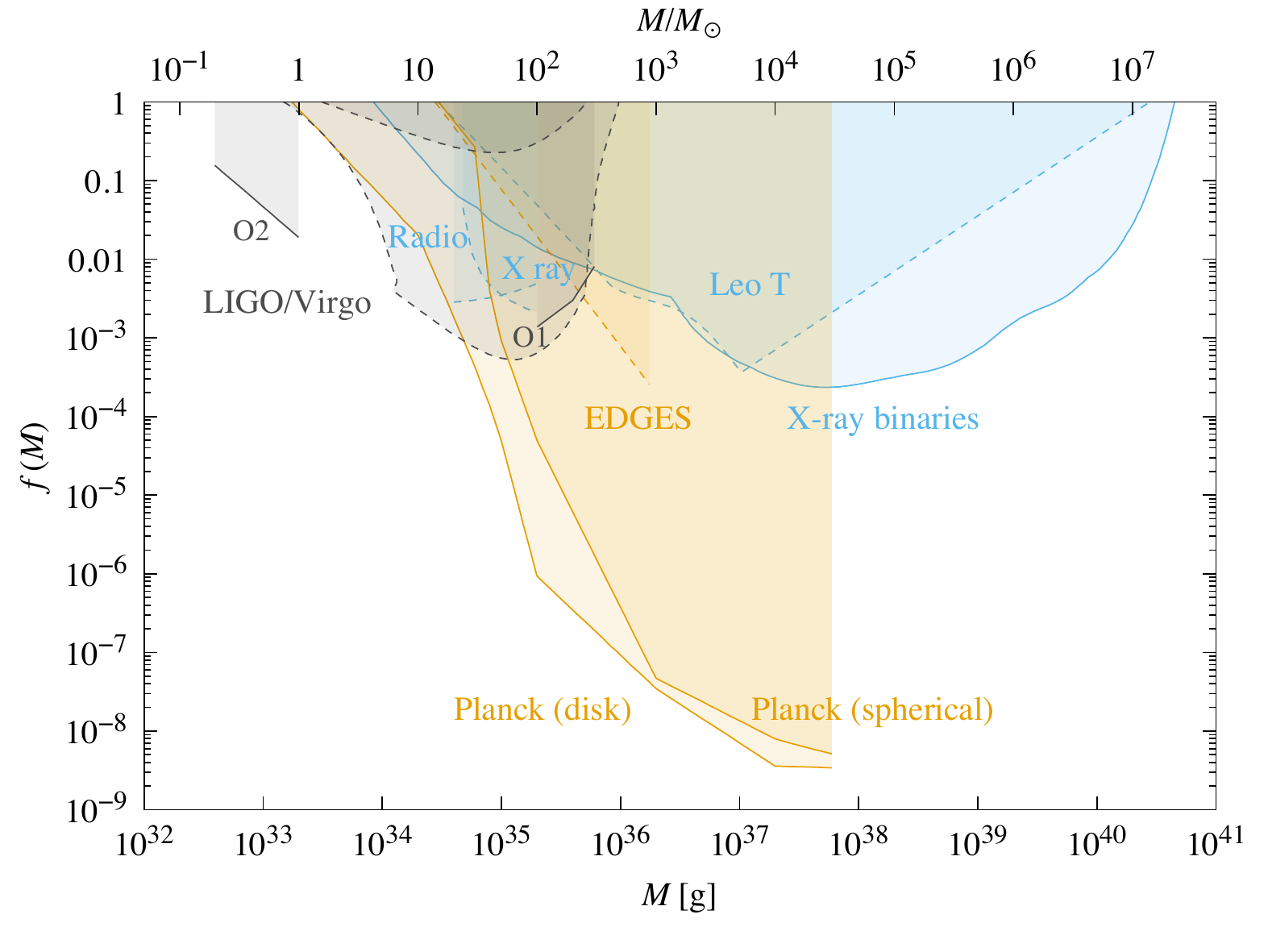}
\end{center}
\caption{Accretion (cyan and yellow) and gravitational wave (grey) constraints.
The Planck limits \cite{Serpico:2020ehh} assume that dark halos form in the late matter-dominated epoch.
The EDGES limit~\cite{Hektor:2018qqw} from the 21\,cm absorption signal is shown by a broken line because it depends on several ill-determined parameters.
The X-ray binaries limit \cite{Inoue:2017csr} and X-ray/radio limits \cite{Manshanden:2018tze} come from source counts; the latter are shown by a broken line because of the criticism of Ref.~\cite{Hektor:2018rul}.
The limit from the heating of Leo T~\cite{Lu:2020bmd} is shown by a broken line because it depends on several astrophysical assumptions.
The LIGO/Virgo limits come from the merger rates in O1 \cite{Ali-Haimoud:2017rtz} and O2 \cite{Authors:2019qbw} (solid line) and from the small number of burst detections (lower broken line) and non-detection of a stochastic GWB (upper broken line) in O2 \cite{Vaskonen:2019jpv}.}
\label{fig:accretion}
\end{figure}

There are a large number of constraints associated with the accretion of PBHs and the strongest of these are summarised in Fig.~\ref{fig:accretion}.
However, it should be stressed that all these limits depend on various astrophysical parameters and qualitative features, such as whether one has disc or spherical accretion.
Also most studies assume Bondi accretion, even though there is no direct evidence that this applies.
Therefore these limits may be less secure than the dynamical ones.
Since accretion can modify both the mass and mass function of PBHs, this also changes one's interpretation of any limits on $f(M)$, considerably weakening the constraints for PBHs larger than a few solar masses~\cite{DeLuca:2020fpg}.

\subsubsection{Accretion during radiation-dominated era}

There are good reasons for believing that PBHs cannot grow much during the radiation-dominated era.
Although a simple Bondi argument suggests that they could grow as fast as the horizon \cite{1967SvA....10..602Z}, this does not account for the background cosmological expansion and a fully relativistic calculation shows that such self-similar growth is impossible \cite{Carr:1974nx,1978ApJ...219.1043B,1978ApJ...225..237B}.
Consequently there is very little growth during the radiation era.
However, self-similar black hole solutions in an expanding background do exist for cosmological models dominated by a ``dark energy'' fluid with $p < -\rho/3$, as in the quintessence scenario \cite{Harada:2007tj,Maeda:2007tk,Carr:2010wk}.
This may support the suggestion of Bean and Magueijo \cite{Bean:2002kx} that intermediate-mass PBHs might accrete quintessence efficiently enough to evolve into the SMBHs in galactic nuclei.

\subsubsection{Limit from ionisation and thermal history}

Even if PBHs cannot accrete appreciably in the radiation-dominated era, they might still do so in the period after decoupling and the Bondi-type analysis \emph{should} then apply.
The associated accretion and emission of radiation could have a profound effect on the thermal history of the Universe.
There are two types of constraints: one associated with the background radiation density generated (usually X-rays) and the other with the thermal history.
This problem was first analysed by Carr \cite{1981MNRAS.194..639C} on the assumption that the PBHs accrete at the Bondi rate with a luminosity bounded by the Eddington limit.
However, as discussed in a recent reassessment of this analysis \cite{Carr:2020erq}, some of the assumptions made there were unrealistic and the Bondi formula itself fails above $10^4\,M_{\odot}$ because the accretion timescale exceeds the cosmic expansion timescale.
We therefore do not discuss the analysis of Ref.~\cite{1981MNRAS.194..639C} further here.

The problem was investigated more carefully by Ricotti \textit{et al.} \cite{Ricotti:2007au}, who studied the effects of accreting PBHs on the ionisation and temperature evolution of the Universe.
The emitted X-rays would produce measurable anisotropies and spectral distortions in the CMB.
Using WMAP data to constrain the first, they obtained the limit
\begin{equation}
f(M)
<
\begin{cases}
( M / 30\,M_{\odot} )^{-2}
& ( 30\,M_{\odot}< M \lesssim 10^{4}\,M_{\odot} ) \\
10^{-5}
& ( 10^{4}\,M_{\odot} \lesssim M < 10^{11}\,M_{\odot} ) \\
M / M_{\ell = 100}
& ( M > 10^{11}\,M_{\odot} )\,,
\end{cases}
\label{eq:wmap}
\end{equation}
where the last expression is not included in Ref.~\cite{Ricotti:2007au} but corresponds to having one PBH on the scale associated with the CMB anisotropies;
for $\ell = 100$, this is $M_{\ell = 100} \approx 10^{16}\,M_{\odot}$\,.
Using FIRAS data to constrain the second, they obtained the limit
\begin{equation}
f(M)
<
\begin{cases}
( M / 1\,M_{\odot} )^{-2}
& (1\,M_{\odot} < M \ll 10^{3}\,M_{\odot} ) \\
0.015
& ( 10^{3}\,M_{\odot}\lesssim M < 10^{14}\,M_{\odot} ) \\
M / M_{\ell = 100}
& ( M > 10^{14}\,M_{\odot} )\,.
\end{cases}
\end{equation}
Both limit flatten off above $10^3\text{--}10^4\,M_{\odot}$ because the luminosity tends to the Eddington limit but we note that the Bondi formula also fails in this regime.
These limits are stronger than the ones inferred from the X-ray background itself~\cite{1979MNRAS.189..123C} and appear to exclude $f = 1$ down to masses as low as $1\,M_{\odot}$\,.
However, they are very model-dependent and there was also a technical error (an incorrect power of redshift in one of the integrals), so they are not shown in Fig.~\ref{fig:accretion}.
Nevertheless, the qualitative form of the above limits is correct.

This problem has subsequently been reconsidered by various groups.
Ali-Ha\"imoud and Kamionkowski \cite{Ali-Haimoud:2016mbv} include the suppression of accretion by Compton drag and Compton cooling from CMB photons, and allow for the PBH velocity relative to the background gas.
They find that the spectral distortions are too small to be detected, while the anisotropy constraints have a similar form to Eq.~\eqref{eq:wmap} but only exclude $f=1$ above $10^2\,M_{\odot}$\,.
Horowitz \cite{Horowitz:2016lib} performs a similar analysis and obtains an upper limit of $30\,M_{\odot}$\,.
However, neither of these analyses extends to sufficiently high mass to exhibit the flattening of the limit.
Aloni \textit{et al.} \cite{Blum:2016cjs} also calculate the effect of PBH accretion on the ionisation history but do not give an explicit constraint on the DM fraction.
Chen \textit{et al.} \cite{Chen:2016pud} use Planck data to constrain the DM contribution from PBHs for two different reionisation models and improve the upper limit \eqref{eq:wmap} by two orders of magnitude.
However, their result depends on the Ricotti \textit{et al.} analysis and is affected by the same error.
Although all these analyses exclude PBHs comprising the DM above some critical mass, they do not exclude the small fraction required to seed galaxies.
The fraction of the DM in SMBHs in galactic nuclei is $\Omega_\mathrm{SMBH}/\Omega_\mathrm{CDM} \sim 2 \times 10^{-5}$, whereas the accretion limit generally flattens off about this value.

All the above arguments assume spherical accretion.
However, Poulin \textit{et al.} \cite{Poulin:2017bwe} argue that this approximation breaks down, with an accretion disk forming instead, and this affects the statistical properties of the CMB anisotropies.
Even under conservative accretion assumptions, their constraints exclude a monochromatic distribution of PBHs with mass above $2\,M_{\odot}$ as the dominant form of DM.
However, the bound on the median PBH mass gets more stringent if a broad (lognormal) mass function is considered.
In follow-up work, Serpico \textit{et al.} \cite{Serpico:2020ehh} study disk and spherical accretion onto a PBH plus halo system, this being strongly constrained by the latest observations of CMB polarisation \cite{Aghanim:2018eyx,Aghanim:2019ame}, and these are the accretion limits shown in Fig.~\ref{fig:accretion}.

Ewall-Wice \textit{et al.} \cite{Ewall-Wice:2018bzf} estimate the 21\,cm radio background from accretion onto intermediate-mass BHs between $z = 30$ and $z = 16$.
Combining plausible scenarios for black hole formation and growth with empirical correlations between luminosity and radio emission observed in low-redshift AGN, they find that the black hole remnants of Population III stars are able to produce a 21\,cm background.
In particular, this could explain the surprisingly large amplitude of the 21\,cm absorption feature recently reported by the EDGES collaboration \cite{Bowman:2018yin}.
The black holes could also contribute to the $0.5\text{--}2\,\mathrm{keV}$ soft X-ray background at a level consistent with existing constraints.
Although the Ewall-Wice \textit{et al.} analysis may also apply for PBHs, they do not explicitly constrain the function $f(M)$.

Hektor \textit{et al.} \cite{Hektor:2018qqw} have used the EDGES results to place bounds on energy injection from the accretion of $\mathcal O(1\text{--}100)\,M_{\odot}$ PBHs.
By requiring that the gas temperature at $z \sim 17$ does not exceed the EDGES value, they infer 
\begin{equation}
f(M)
< C(\beta)\,\left(\frac{0.15}{f_E}\right)\,\left(\frac{\lambda}{0.01}\,\frac{M}{10\,M_{\odot}}\right)^{-1-\beta}\,,
\end{equation}
where $\lambda$, $\beta$ and $f_E$ are accretion parameters.
We include this as a broken line in Fig.~\ref{fig:accretion} as an example of a constraint which can be expressed analytically but depends on various ill-determined parameters.
Mena \textit{et al.} \cite{Mena:2019nhm} show that observations of the 21\,cm transition in neutral hydrogen during the epoch of reionisation provide stringent constraints on solar-mass PBHs.
There are three distinct effects: (i) the modification to the primordial power spectrum induced by Poisson noise, (ii) a uniform heating and ionisation of the intergalactic medium via X-rays produced during accretion, and (iii) a local modification to the temperature and density of the ambient medium surrounding isolated PBHs.
Using a $4$-parameter astrophysical model, they show that experiments like SKA and HERA (Hydrogen Epoch of Reionization Array) could potentially improve upon existing constraints by more than an order of magnitude.
However, H\"utsi \textit{et al.}~\cite{Hutsi:2019hlw} show that PBH accretion could be suppressed due to enhanced substructure in the PBH distribution, softening the bounds from the 21\,cm signature.

Abe \textit{et al.} \cite{Abe:2019kon} have investigated the Sunyaev--Zel'dovich (SZ) effect on the CMB caused by the emission of UV and X-ray photons by accreting PBHs, these photons ionizing and heating the intergalactic medium (IGM).
They compute the ionisation and temperature profiles around each PBH and infer the $y$-parameter created by the IGM gas.
They find that the angular power spectrum of the SZ temperature anisotropy is a flat up to the scale of the ionised region ($l \leq 2000$) and could dominate the primordial temperature spectrum below the Silk scale.

Bosch-Ramon and Bellomo~\cite{Bosch-Ramon:2020pcz} argue that outflows can reduce the accretion rate of PBHs with masses in the LIGO/Virgo range by at least one order of magnitude if aligned with the PBH motion.
If the outflow is perpendicular to the motion, the effect is less important.
Hasinger~\cite{Hasinger:2020ptw} estimates the contribution of baryon accretion onto a population of $10^{-8}\text{--}10^{-10}\,M_{\odot}$ PBHs to the infrared and X-ray cosmic backgrounds, as indicated by the cross-correlation between the fluctuations in these backgrounds~\cite{Kashlinsky:2018mnu}.
Assuming Bondi capture and advection-dominated disk accretion, he claims the model fits the X-ray data, with the predicted flux peaking at redshifts $z \approx 17\text{--}30$, although the PBH contribution to the infrared background fluctuations is only about 1\,\%.

\subsubsection{Limit from accretion at present epoch}

Limits associated with accretion at the present epoch avoid the uncertainties associated with early cosmological evolution but depend upon the environment and are not specific to black holes of primordial origin.
An early attempt to constrain $f(M)$ for black holes in galactic discs, galactic halos and clusters of galaxies by requiring that the background radiation generated not exceed the observed values in the appropriate waveband was made in Ref.~\cite{1979MNRAS.189..123C}.
Because the Bondi and Eddington accretion rates scale as $M^2$ and $M$, respectively, the limit on $f(M)$ was found to always have the same form, scaling as $M^{-1}$ at low $M$ and flattening off when $M$ gets large enough for the Eddington limit to apply.
This constraint depends on the luminosity efficiency (taken to be $0.1$) and the waveband in which radiation is emitted, with the upper limit on the mass of black holes providing the DM varying from $10\,M_{\odot}$ for X-ray emission to $10^2\,M_{\odot}$ for UV-ray emission.
However, these limits are not included in Fig.~\ref{fig:accretion} because they are very old and dependent on uncertain astrophysical parameters.

Another type of limit is derived from discrete source counts.
PBH accretion within galaxies should result in a significant X-ray flux, thereby contributing to the observed number of compact X-ray objects in galaxies.
For example, Mack \textit{et al.} \cite{Mack:2006gz} considered the growth of large PBHs through the capture of DM halos and suggested that the accretion of such enhanced holes when passing through molecular clouds could give rise to the ultra-luminous X-ray sources observed in nearby galactic discs.
They concluded that such PBHs could provide seeds for SMBHs in galactic nuclei but not all the DM.
Kawaguchi \textit{et al.} \cite{Kawaguchi:2007fz} considered the accretion of the interstellar medium by the intermediate mass PBHs formed in a double-inflation scenario.
More recently, Inoue and Kusenko \cite{Inoue:2017csr} have used existing X-ray data to constrain the PBH number density in the mass range from a few to $2 \times 10^7\,M_{\odot}$\,.
Their limit is shown in Fig.~\ref{fig:accretion} and marginally allows the black holes detected by LIGO/Virgo.

Lu \textit{et al.}~\cite{Lu:2020bmd} use data from the Leo T dwarf galaxy to set a limit on $f(M)$ in the mass range $1\text{--}10^7\,M_{\odot}$ by considering their heating of the interstellar medium by accretion, dynamical friction and disk outflows.
However, they make the strong assumption that dissipation by gas turbulence in Leo T induces a radiatively inefficient-accretion flow around a PBH, so their limit is only shown by a broken line in Fig.~\ref{fig:accretion}.

Gaggero \textit{et al.} \cite{Gaggero:2016dpq} model the accretion of gas onto a population of massive PBHs in the Milky Way and compare the predicted radio and X-ray emission with observational data.
With conservative assumptions about the accretion process, they use the VLA radio catalog at 1.4 GHz and the Chandra X-ray catalog to exclude $\mathcal O(10)\,M_{\odot}$ PBHs from providing all of the DM at the $5 \sigma$ level.
More sensitive future radio and X-ray surveys could identify a PBH population.
Similar arguments have been made by Manshanden \textit{et al.} \cite{Manshanden:2018tze}, whose limit is also shown in Fig.~\ref{fig:accretion}.
Hektor \textit{et al.} \cite{Hektor:2018rul} have considered the PBH constraints based on X-ray data from the Galactic centre and point out that all these accretion limits could be substantially weakened due to gas turbulence inside the interstellar medium and molecular clouds, this modifying the relative velocity between the PBHs and gas.

\subsubsection{$\mu$ distortion}

If PBHs form from the high-$\sigma$ tail of Gaussian density fluctuations, as in the simplest scenario \cite{Carr:1975qj}, then another interesting limit comes from the dissipation of those density fluctuations by Silk damping at a much later time.
This merely decreases the baryon-to-photon ratio for $t < 7 \times 10^6\,\mathrm s$, with the success of the BBN scenario then giving a weak limit.
However, it leads to a $\mu$ distortion in the CMB spectrum \cite{Chluba:2012we} for $7 \times 10^6\,\mathrm s < t < 3 \times 10^9\,\mathrm s$, giving an upper limit $\delta (M) < \sqrt{\mu} \sim 10^{-2}$ over the mass range $10^3 < M/M_{\odot} < 10^{12}$.
This limit was first given in Ref.~\cite{Carr:1993aq}, based on a result in Ref.~\cite{1991MNRAS.248...52B}, but the limit on $\mu$ is now stronger and will become still stronger in the future~\cite{Chluba:2019nxa,Delabrouille:2019thj}.
There is also a $y$ distortion for $3 \times 10^9\,\mathrm s < t < 3 \times 10^{12}\,\mathrm s$ (the time of decoupling) but this only applies for much larger masses and is not discussed here.

In principle, limits on the $\mu$ distortion can be translated into a very strong constraint on $f(M)$ in the range $10^4 < M/M_{\odot} < 10^{12}$ \cite{Kohri:2014lza}.
However, the assumption that the fluctuations are Gaussian may be incorrect.
For example, Nakama \textit{et al.} \cite{Nakama:2016kfq} have proposed a ``patch'' model, in which the relationship between the background inhomogeneities and the overdensity in the tiny fraction of volumes collapsing to PBHs is modified.
Kawasaki and Murai \cite{Kawasaki:2019iis} have another non-Gaussian model, in which PBH seeds for SMBHs are produced by inhomogeneous baryogenesis in a modified Affleck--Dine mechanism.
The $\mu$ distortion constraint could thus be much weaker and one needs to consider its dependence on the possible non-Gaussianity of primordial fluctuations.
A phenomenological description of such non-Gaussianity was introduced in Ref.~\cite{Nakama:2016kfq} and involves a parameter $p$, such that the dispersion of the primordial fluctuations becomes smaller as $p$ is reduced, thereby reducing the $\mu$ distortion.

Recently, Nakama \textit{et al.} \cite{Nakama:2017xvq} have calculated the $\mu$-distortion constraints on $f(M)$ for the $p$ model, using both the current FIRAS limit and the projected upper limit from PIXIE~\cite{Abitbol:2017vwa}.
However, one must be careful in the interpretation of $M$ here.
The diffusion mass is given by
\begin{equation}
M_\mathrm{D} \sim \sqrt{M_{\tau}\,M_\mathrm{H}}
\sim
\begin{cases}
10^{10}\,(t/t_\mathrm{eq})^{7/4}\,M_{\odot}
& (t < t_\mathrm{eq}) \\
10^{13}\,(t/t_\mathrm{dec})^{11/6}\,M_{\odot}
& (t_\mathrm{eq} < t < t_\mathrm{dec})\,,
\end{cases}
\end{equation}
where $M_\mathrm{H}$ is the horizon mass and $M_{\tau}$ is the mass of unit optical depth to Thomson scattering.
A PBH has the horizon mass at formation and it subsequently remains constant, whereas the radiation mass is reduced by redshift effects.
One can show that the PBH and diffusion masses on a given scale are related by
\begin{equation}
M
\sim
\begin{cases}
10^{2}\,(M_\mathrm{D}/M_{\odot})^{6/7}\,M_{\odot}
& (t < t_\mathrm{eq}) \\
10\,(M_\mathrm{D}/M_{\odot})^{10/11}\,M_{\odot}
& (t_\mathrm{eq} < t < t_\mathrm{dec})\,.
\end{cases}
\end{equation}
The probability distribution for the curvature perturbation $\zeta$ is taken to have the non-Gaussian form,
\begin{equation}
P(\zeta)
= \frac{1}{2\sqrt{2}\,\tilde{\sigma}\,\Gamma\left(1+1/p\right)}\,
  \exp\left[-\left(\frac{|\zeta |}{\sqrt{2}\,\tilde{\sigma}}\right)^p\right]\,,
\end{equation}
where $\tilde{\sigma}$ is the root-mean-square curvature fluctuation, this becoming Gaussian for $p=2$.
The collapse fraction is therefore
\begin{equation}
\beta
= \int_{\zeta_c}^\infty P(\zeta) \mathrm d\zeta
= \frac{\Gamma(1/p, 2^{-p/2}(\zeta_c/\tilde{\sigma})^p)}{2p\,\Gamma(1+1/p)}\,,
\label{eq:betamu}
\end{equation}
where $\zeta_c$ is the threshold for PBH formation and $\Gamma(a,z)$ is the incomplete gamma function.
The $\mu$ distortion is
\begin{equation}
\mu
\simeq
  2.2\,\sigma^2\,
  \biggl[
   \exp\biggl(-\frac{\hat{k}_*}{5400}\biggr)
   - \exp\biggl(-\biggl[\frac{\hat{k}_*}{31.6}\biggr]^2\biggr)
  \biggr]\,,
\label{eq:mumu}
\end{equation}
where $\hat{k}_*$ is the wave-number in units of $\mathrm{Mpc}^{-1}$, this peaking at $k=80\,\mathrm{Mpc}^{-1}$ or $3 \times 10^9\,M_{\odot}$\,.
This is required to be less than $9 \times 10^{-5}$ from FIRAS and $10^{-9}$ from HYPERPIXIE.
Equations~\eqref{eq:betamu} and \eqref{eq:mumu} and the relation
\begin{equation}
k
\simeq
  7.5 \times 10^5\,\gamma^{1/2}\,
  \left(\frac{g_*}{10.75}\right)^{-1/12}\,
  \left(\frac{M}{30\,M_\odot}\right)^{-1/2}\, \mathrm{Mpc}^{-1}\,,
\end{equation}
where the parameters $\gamma$ and $g_*$ are defined in Sec.~\ref{sec:bg}, then lead to constraints on $\beta(M)$ and $f(M)$.

\begin{figure}[ht]
\begin{center}
\includegraphics[width=.50\textwidth]{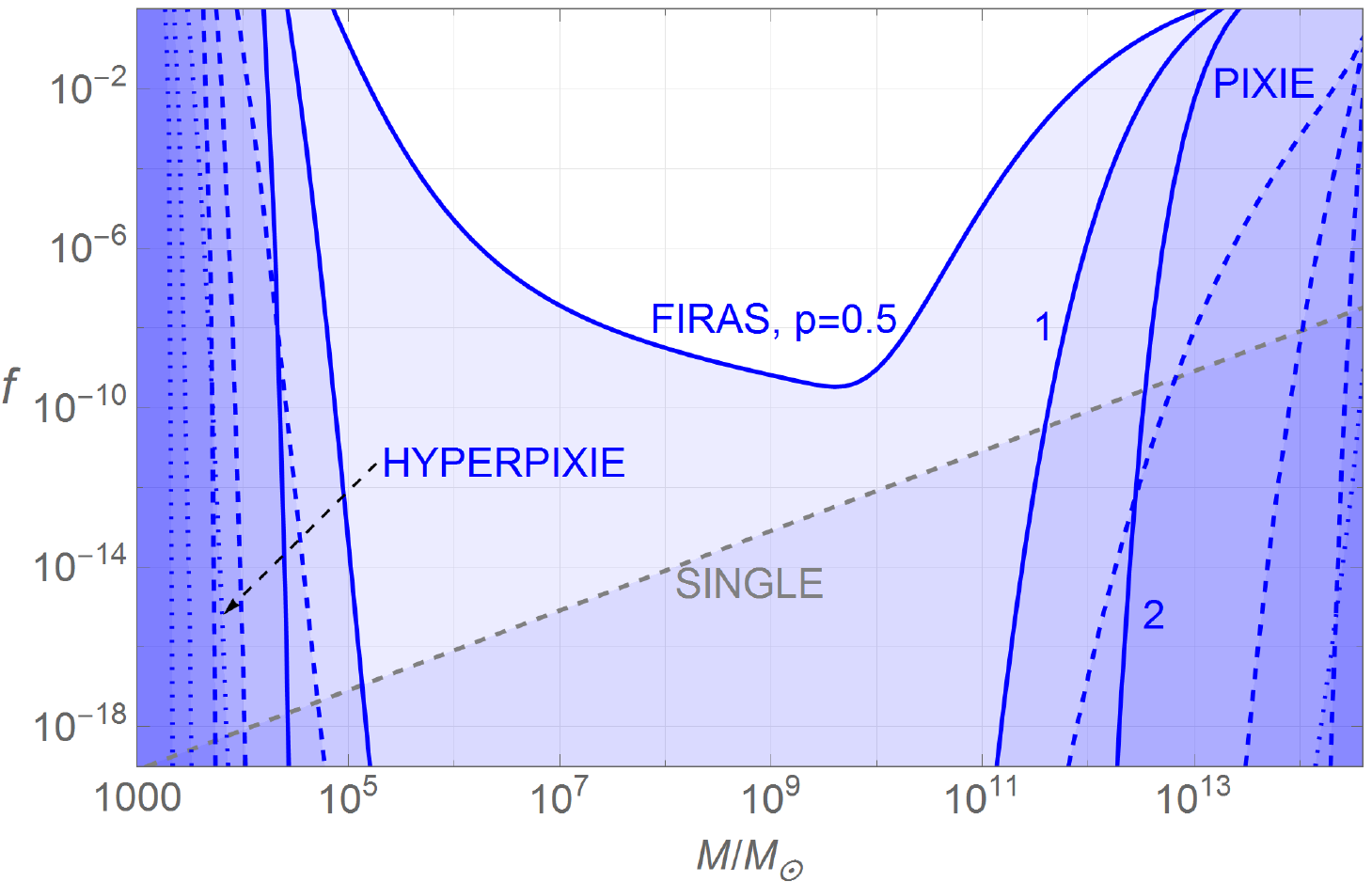}
\end{center}
\caption{Upper limits on $f$ from $\mu$ constraints for different values of the non-Gaussianity parameter $p$, from Ref.~\cite{Nakama:2017xvq}.
The solid curves are for FIRAS ($\mu=9\times 10^{-5}$) and correspond to $p=0.5,1,2$ (from top to bottom).
The dashed and dotted curves are for PIXIE ($\mu=3.6\times 10^{-7}$) and HYPERPIXIE ($\mu=10^{-9}$), respectively, with the same values of $p$ from top to bottom.
The SINGLE line corresponds to the incredulity limit of one PBH per Hubble volume.}
\label{fig:mu}
\end{figure}

The limit on $f(M)$ is shown in Fig.~\ref{fig:mu}, and this indicates that the PBH density is severely constrained in the range $10^5\,M_\odot < M < 10^{11}\,M_\odot$\,, with the mass range around $10^9\,M_\odot$ being most restricted.
We need a low value of $p$ (i.e.\ large non-Gaussianity) if such massive PBHs are to evade the $\mu$ distortion constraints (cf.\ \cite{Garcia-Bellido:2017aan}).
It would therefore be more plausible to invoke PBHs with initial masses below $10^5\,M_\odot$ which undergo substantial accretion between the $\mu$-distortion era and the time of matter-radiation equality.
In particular, PBHs with $10^6 < M/M_{\odot} < 10^{10}$ and $f \sim 10^{-4}$, such as are required to power quasars, require either $p < 0.5$ or an accretion factor of $(M/10^5\,M_{\odot})^{-1}$.

\subsection{Gravitational wave constraints}

In recent years interest in PBHs has intensified because of the detection of gravitational waves (GWs) from coalescing binary black holes by LIGO/Virgo \cite{Abbott:2016blz,Abbott:2016nmj,Abbott:2016nhf,LIGOScientific:2018mvr} and the possibility that these might be of primordial rather than stellar origin.
Even if this is not the case, the observations place important constraints on the number of PBHs.
Indeed, as illustrated in Fig.~\ref{fig:accretion}, the combination of the accretion and GW limits already makes it unlikely that the LIGO/Virgo black holes (even if primordial) could explain \emph{all} the DM.
On the hand, if the PBHs have an extended mass function, the mass at which the GW signal peaks is expected to be larger than the mass at which most of the density resides, so this need not preclude PBHs providing the dark matter at a somewhat lower mass.
In this section, we discuss three different ways in which GW searches can constrain the PBH proposal.

\subsubsection{Gravitational wave background}

A population of black holes would be expected to generate a stochastic gravitational wave background (GWB) \cite{1980A&A....89....6C}.
This would be especially interesting if there were a population of binary black holes, coalescing at the present epoch due to gravitational-radiation losses.
This was first discussed by Bond and Carr \cite{1984MNRAS.207..585B} in the context of Population III black holes and by Nakamura \textit{et al.}~\cite{Nakamura:1997sm} and Ioka \textit{et al.}~\cite{Ioka:1998gf} in the context of PBHs.
However, the precise formation epoch of the holes is not crucial since the coalescences occur much later.
In either case, the black holes would be expected to cluster inside galactic halos (along with other forms of DM) and so the detection of the associated GWB would provide a probe of the halo distribution \cite{Inoue:2003di}.
Dalianis and Kouvaris~\cite{Dalianis:2020gup} have pointed out that regions forming PBHs in an early matter-dominated era should collapse asymmetrically to form pancakes and the modified form of GWB generated would be probed by future interferometers.

The LIGO data had already placed weak constraints on the PBH scenario over a decade ago \cite{Abbott:2006zx} but the constraints are now much stronger.
A few years ago Raidal \textit{et al.} \cite{Raidal:2017mfl} showed that the expected GWB could be detected by the LIGO/Virgo runs, with the data at that time already giving the strongest available constraint on $f(M)$ in the mass range $0.5\text{--}30\,M_{\odot}$\,.
Indeed, Wang \textit{et al.} \cite{Wang:2016ana} showed that Advanced LIGO's first run gave the best PBH constraint in the range $1\text{--}100\,M_{\odot}$\,.
However, as we will discuss shortly, Vaskonen and Veerm\"ae \cite{Vaskonen:2019jpv} showed that the merger rate of binaries would be significantly reduced for $ f \sim 1 $ due to perturbations in small halos.
They found a more robust limit from a potential non-detection of stochastic GWB in the LIGO/Virgo O2 data and this is shown by the upper broken line in Fig.~\ref{fig:accretion}.
More recently, Kapadia \textit{et al.}~\cite{Kapadia:2020pir,Kapadia:2020pnr} have searched for a stochastic GWB in the LIGO's second observing run and inferred upper limits on $f(M)$ in the mass range $10^{-20}\text{--}10^{-19}\,M_{\odot}$\,.
Such PBHs should have evaporated by now but this implies a tight limit even if Hawking evaporation is somehow avoided.
Bartolo \textit{et al.} \cite{Bartolo:2019zvb} have calculated the anisotropies and non-Gaussianity in a GWB generated by PBHs on account of propagation effects and conclude that PBHs cannot provide all the DM if these effects are large.
The non-Gaussianity required to avoid the $\mu$ limit (discussed above) would result in more mergers and therefore increase the GWB~\cite{Young:2019gfc}.

Hooper \textit{et al.}~\cite{Hooper:2020evu} consider a more exotic scenario in which the early Universe was dominated by PBHs with $M < 5 \times 10^8\,\mathrm g$ which form binaries and undergo merging before evaporation and prior to BBN.
The merged PBHs will possess substantial angular momentum, causing their evaporation to produce a significant background of high energy gravitons and their mergers to produce a high-frequency GWB.
Inomata \textit{et al.}~\cite{Inomata:2020lmk} consider a similar scenario, with reheating through PBH evaporations changing the equation of state of the Universe much more suddenly than conventional particle decays.
This enhances the GWB and future observations could give a constraint $\beta < 10^{-5}\text{--}10^{-8}$ for $10^3\text{--}10^5\,\mathrm g$ if the PBH mass function is very narrow.

If PBHs form in an early matter-dominated era and eventually reheat the Universe by Hawking evaporation, then Dom\`enech \textit{et al.}~\cite{Domenech:2020ssp} and Papanikolaou \textit{et al.}~\cite{Papanikolaou:2020qtd} argue that the density fluctuations associated with their inhomogeneous distribution will generate a GWB at the time of reheating.
BBN constraints on the GWB imply that the initial collapse fraction is below $10^{-4}$ but this limit decreases as $M$ increases and Papanikolaou \textit{et al.} find it has the form $10^{-4}\,(M/10^9\,\mathrm g)^{-1/4}$.
Although this constraint is outside the mass range shown in Fig.~\ref{fig:combined}, Dom\`enech \textit{et al.} point out that the GWB is in the LIGO and DECIGO windows for $M \sim 10^4\text{--}10^8\,\mathrm g$, so this scenario could soon be testable.

Looking further to the future, Wang \textit{et al.} \cite{Wang:2019kzb} point out that some PBHs may accumulate at the center of a galaxy and follow a prograde or retrograde orbit due to the gravity of the central SMBH.
If the PBH mass is $\mathcal{O}(1)\,M_{\odot}$ or smaller, the incoherent superposition of all the extreme-mass-ratio inspirals in the Universe could then generate a GWB detectable by LISA.
In an extension of this argument, K{\"u}hnel \textit{et al.}~\cite{Kuhnel:2018mlr} investigate GW production by PBHs in the mass range $10^{-13}\text{--}1\,M_{\odot}$ orbiting a SMBH.
While the individual objects would be undetectable, the extended stochastic emission from a large number of them might be.

\subsubsection{PBHs and the LIGO/Virgo events}

During the first two LIGO/Virgo runs, $10$ events were observed with component masses in the range $8\text{--}51\,M_{\odot}$\,, which is larger than originally predicted for black holes of stellar origin.
After the first detection, Bird \textit{et al.}~\cite{Bird:2016dcv} claimed that the expected merger rate for PBHs providing the DM was compatible with the range $9\text{--}240\,\mathrm{Gpc}^{-3}\,\mathrm{yr}^{-1}$ obtained by the LIGO analysis and this was supported by other studies~\cite{Clesse:2016vqa,Blinnikov:2016bxu}.
However, Sasaki \textit{et al.}~\cite{Sasaki:2016jop} argued that the merger rate would be in tension with the CMB distortion constraints unless the PBHs provided only a small fraction of the DM.
This conclusion could be avoided if the LIGO/Virgo back holes derived from the accretion and merger of smaller PBHs \cite{Clesse:2016vqa}.
However, Eroshenko~\cite{Eroshenko:2016hmn} argued that the PBH merger rates would be highly suppressed by tidal forces and inferred that the LIGO/Virgo results would allow only 0.1\,\% of the DM to be in PBHs.

A cleaner PBH signature would be a mass below $1\,M_{\odot}$ since such black holes definitely cannot result from stellar evolution.
However, the LIGO/Virgo runs have found no compact binary systems with component masses in the range $0.2\text{--}1.0\,M_{\odot}$ \cite{Abbott:2018oah}.
Indeed, for monochromatic non-spinning PBHs, the merger rate of $0.2\,M_{\odot}$ and $1.0\,M_{\odot}$ binaries is constrained to be less than $3.7 \times 10^5\,\mathrm{Gpc}^{-3}\,\mathrm{yr}^{-1}$ and $5.2 \times 10^3\,\mathrm{Gpc}^{-3}\,\mathrm{yr}^{-1}$, respectively.
This corresponds to at most 16\,\% or 2\,\% of the DM, respectively~\cite{Authors:2019qbw}, this limit being shown in Fig.~\ref{fig:accretion}.
Future sensitivities for mergers below $\mathcal{O}(1)\,M_{\odot}$ are studied in Ref.~\cite{Wang:2019kaf}.

A crucial issue for the PBH model is whether the binaries form in the early Universe or after galaxy formation.
Sasaki \textit{et al.}~\cite{Sasaki:2016jop} argued that the first is more important unless the formation of binaries in the radiation-dominated era is significantly reduced for some reason.
Bird \textit{et al.}~\cite{Bird:2016dcv} invoked the second mechanism and this is the basis of their claim that PBHs could provide all the DM.
However, Ali-Ha\"imoud \textit{et al.} \cite{Ali-Haimoud:2017rtz}, a subset of the same authors, calculated the binary merger rate resulting from gravitational capture in halos and concluded that this should be subdominant if binaries formed in the early Universe survive until the present, as suggested by their analytic estimates.
They found that tidal torquing by other PBHs and the halo field, which might reduce the surviving fraction, is much weaker than previously calculated.
Their limit from the LIGO/Virgo O1 data is $f(M) < 0.01$ for $10\text{--}300\,M_{\odot}$ PBHs but only the part above $100\,M_{\odot}$ is shown in Fig.~\ref{fig:accretion} because the lower part is just a potential constraint.

Raidal \textit{et al.} \cite{Raidal:2017mfl} studied the production and merging of PBH binaries for an extended PBH mass function and showed that it is possible to satisfy all current PBH constraints if this has a lognormal form.
However, this ignored the suppression of mergers due to the binaries being perturbed and they later considered the possibility of a clustered spatial distribution.
In this case, the probability distribution of orbital parameters is modified and binaries are disrupted when they get too close to other PBHs~\cite{Raidal:2018bbj}.
The merger estimates are not reliable when PBHs provide more than 10\,\% of the DM, in which case the constraint from the observed LIGO merger rate is strongest in the mass range $2\text{--}160\,M_{\odot}$\,.
Their conclusion is that the LIGO/Virgo events can result from PBH mergers but that the PBHs can only provide the DM if they have an extended mass function.

The simulations of Vaskonen and Veerm\"ae~\cite{Vaskonen:2019jpv} show that the fraction of disrupted initial binaries can be much larger than estimated in Ref.~\cite{Ali-Haimoud:2017rtz} if PBHs make up a large fraction of DM.
Nevertheless, even for scenarios in which early substructure survives and PBH binary disruption is certain, the merger rate is still large enough to rule out PBH DM in some mass range.
This puts the LIGO/Virgo merger constraints shown by the lower broken line in Fig.~\ref{fig:accretion} on a more solid footing.
However, this is only a potential limit since it assumes a small number of burst detections in the O2 survey.
Jedamzik~\cite{Jedamzik:2020omx} uses full numerical integration to study the evolution of binaries in clusters due to interactions with a third passing PBH.
His Monte-Carlo analysis shows that the formerly predicted merger rates are reduced by orders of magnitude due to such interactions.

The robustness of the LIGO/Virgo bounds on PBH of $\mathcal O(10)\,M_{\odot}$ depends on the accuracy with which the formation of PBH binaries in the early Universe can be described.
Ballesteros \textit{et al.} \cite{Ballesteros:2018swv} revisit the standard estimate of the merger rate, focusing on the spatial distribution of nearest neighbours and the initial clustering of PBHs associated with the primordial power spectrum.
They confirm the robustness of previous results for a narrow mass function and infer $ f \sim 0.001\text{--}0.01 $.
PBH clustering might tighten this constraint but only for very broad mass functions, corresponding to bumps in the primordial power spectra extending over several decades.
Korol \textit{et al.} \cite{Korol:2019jud} investigate the formation of binaries in small clusters of $30$ PBHs in the absence of initial binaries.
They find that the binaries act as heat sources for the cluster, increasing its velocity dispersion and inhibiting mergers through two-body captures.

Several authors have considered the PBH merger rate as a function of the binary mass ratio.
In particular, Gow \textit{et al.} \cite{Gow:2019pok} consider the sensitivity of LIGO for five different PBH mass functions, showing that the empirical preference for nearly equal-mass binaries in current data is consistent with the PBH hypothesis once observational selection effects are taken into account.
Young and Byrnes~\cite{Young:2019gfc} find that the initial PBH clustering due to primordial non-Gaussianity increases the present merger rate, although there is a large theoretical uncertainty when the clustering is large.
Diego \cite{Diego:2019rzc} has constrained PBHs of $ 5\text{--}50\,M_{\odot}$ from the gravitational lensing of GWs at the LIGO frequency.
For the magnifications expected for observed GW events, he finds that $f(M)$ can be constrained at the percent level.
Indeed, if a small fraction of the DM is in the form of microlenses of $\mathcal{O}(10)\,M_{\odot}$\,, all GWs will show interference effects at sufficiently large magnifications.

Most of the observed coalesced black holes have effective spins compatible with zero.
Although the statistical significance of this result is low, it goes against a stellar binary origin \cite{Gerosa:2018wbw} but is a prediction of the PBH scenario \cite{Garcia-Bellido:2017fdg}.
Fernandez \textit{et al.} \cite{Fernandez:2019kyb} also argue that the spin information favours a primordial rather than astrophysical origin for the LIGO/Virgo events and they discuss the number of events necessary to acquire information on the relative fraction of primordial and non-primordial binary black holes.
De Luca \textit{et al.}~\cite{DeLuca:2020bjf,DeLuca:2020qqa} study the evolution of the PBH spin and the corresponding probabilities for GW events.
In order to realize efficient PBH accretion, they take the lowest relative velocity between the PBH and baryon fluid.
They compute the best-fit PBH mass distribution at formation compatible with current GW data and infer that the maximum DM fraction is around $10^{-3}$.
However, the fact that at least one component of GW190412 is moderately spinning is incompatible with a primordial origin unless accretion or hierarchical mergers are significant.
With accretion, the current PBH constraints are significantly relaxed and the GW bounds is the most stringent ones in the relevant mass range.

Carr \textit{et al.}~\cite{Carr:2019kxo} argue that three recent LIGO/Virgo events - GW190425, GW190814 and GW190521~\cite{Abbott:2020uma,Abbott:2020khf,Abbott:2020tfl,Abbott:2020mjq} - may support a scenario in which the PBHs form at the QCD epoch.
On the other hand, Takhistov \textit{et al.}~\cite{Takhistov:2020vxs} argue that the black holes in the range $1.5\text{--}2.6\,M_{\odot}$ required to explain GW190425 and GW190814 could be produced via neutron star implosions induced by capture of small PBHs, with the mass distribution of such ``transmuted'' BHs following that of the neutron stars.
De Luca \textit{et al.} \cite{DeLuca:2020sae} argue that GW190521 can only be explained within the PBH scenario without violating other constraints if the PBHs accrete efficiently before the reionisation epoch.
Inomata~\cite{Inomata:2020cck} argues that the power spectrum of scalar perturbations required to produce such PBHs through the non-linear interaction appearing in perturbations could avoid the BBN and CMB $\mu$-distortion constraints.

Miller \textit{et al.}~\cite{Miller:2020kmv} argue that current searches for continuous waves from PBH binaries imply $f < 1$ for chirp masses in the range $4 \times 10^{-5} \text{--}10^{-3}\,M_{\odot}$ but that one could push this limit down to $f < 10^{-2}$ with the Einstein Telescope.
Herman \textit{et al.}~\cite{Herman:2020wao} have discussed the detection of high-frequency GWs from planetary-mass PBH binaries with electromagnetic detectors.
This would allow the detection of PBH binary mergers with mass around $10^{-5}\,M_{\odot}$ if $f \sim 10^{-4}$.
Of course, the LIGO/Virgo events could be a mixture of PBHs and stellar black holes and H\"utsi \textit{et al.}~\cite{Hutsi:2020sol} argue that PBHs alone are unlikely to explain the data.

\subsubsection{Constraints from second order tensor perturbations}

\begin{figure}[ht]
\begin{center}
\includegraphics[width=.60\textwidth]{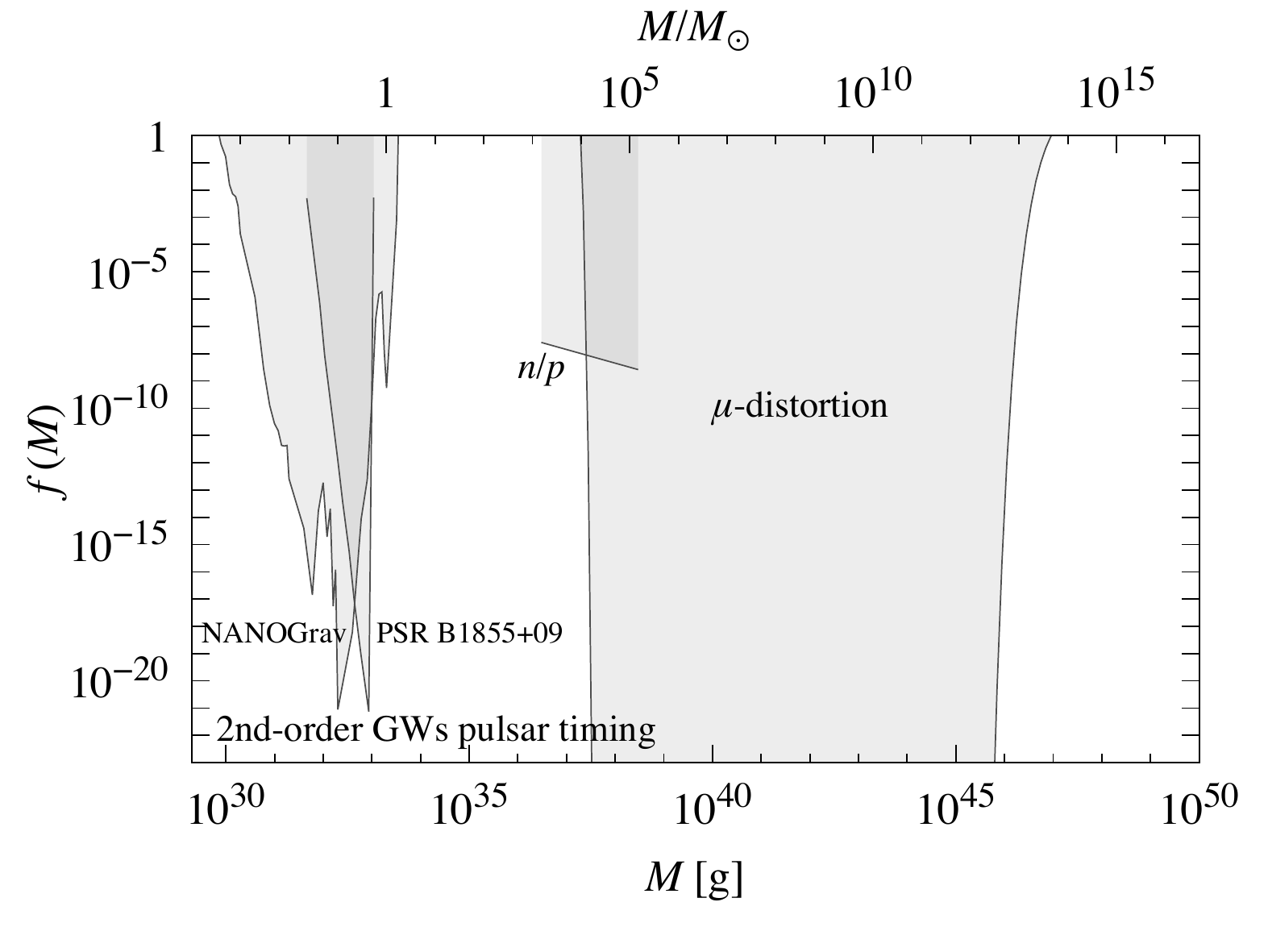}
\end{center}
\caption{Constraints on $f(M)$ from observations of various kinds of background radiation: 
second-order gravitational waves from pulsar timing~\cite{Saito:2008jc} and NANOGrav~\cite{Chen:2019xse}, the CMB $\mu$ distortion \cite{Nakama:2017xvq} and the $n/p$ ratio \cite{Inomata:2016uip}.}
\label{fig:background}
\end{figure}

A different type of GW constraint on $f(M)$, first pointed out by Saito and Yokoyama \cite{Saito:2008jc,Saito:2009jt}, arises because of the second order tensor perturbations generated by the scalar perturbations which produce the PBHs, the associated frequency being $10^{-10}\,(M/10^{3}M_{\odot})^{-1/2}\,\mathrm{Hz}$.
Such a background could be detected by space-based interferometers in the planetary mass range and by pulsar timing in the intermediate mass range, these being the two ranges in which PBHs might provide the DM.
The limit on $f(M)$ relates to the amplitude of the density fluctuations at the horizon epoch and the associated constraints are shown in Fig.~\ref{fig:background}.
Conversely, one can use PBH limits to constrain the primordial GWB \cite{Nakama:2015nea,Nakama:2016enz}.

These effects have subsequently been studied by many other authors \cite{Bugaev:2009zh,Assadullahi:2009jc,Bugaev:2010bb,Alabidi:2012ex}.
In particular, Orlofsky \textit{et al.} \cite{Orlofsky:2016vbd} claim that PTA measurements can be used to probe whether the LIGO/Virgo mergers were from inflationary PBHs, while Ballesteros \textit{et al.}~\cite{Ballesteros:2020qam} claim that PBHs in the asteroid mass range might be detected by Earth-based interferometers.
The modifications required if the PBHs form in a matter-dominated rather than radiaton-dominated era have been discussed in Refs.~\cite{Alabidi:2013lya,Kohri:2018awv,Inomata:2019zqy,Inomata:2019ivs}.
The spectrum of the induced GWB is also important.
For a wide class of perturbations giving rise to PBHs, the infrared behavior of the power spectrum of the GWB may have the general form $\Omega_\mathrm{GW} \propto k^3$ for a wide peak and $\Omega_\mathrm{GW} \propto k^2$ for a narrow peak \cite{Cai:2019cdl}.
However, Yuan \textit{et al.}~\cite{Yuan:2019wwo} find a log-dependent slope, with the $k^3$ or $k^2$ parts being highly suppressed.

These analyses apply only if the PBHs are generated by Gaussian fluctuations.
Although this is questionable for the large-amplitude fluctuations relevant to PBH formation~\cite{Hidalgo:2007vk,Hidalgo:2009fp}, the non-Gaussian effects are not expected to be large~\cite{Saito:2008em}.
However, even if the scalar curvature perturbations have a subdominant non-Gaussian component, this can still be the dominant source of induced GWs~\cite{Unal:2018yaa}, so this has motivated studies of the detectability of the GWB non-Gaussianity.
For example, if PBHs provide the DM in the window around $10^{21}\,\mathrm g$, Bartolo \textit{et al.} \cite{Bartolo:2018rku,Bartolo:2018evs} show that LISA will be able to detect the GW power spectrum but not the bispectrum.
A related point has been considered by Cai \textit{et al.} \cite{Cai:2018dig}, who find that GWB induced by non-Gaussian curvature perturbations can exceed that induced by the Gaussian part.
If PBHs with masses of $10^{20}\text{--}10^{22}\,\mathrm g$ provide the DM, they find that the corresponding GWs will be detectable by LISA, irrespective of the value of $f_\mathrm{NL}$\,.
However, Cai \textit{et al.} \cite{Cai:2019elf} argue that the PTA constraints can be relieved if the perturbations are locally non-Gaussian with $f_\mathrm{NL} \gtrsim \mathcal{O}(10)$.
They conclude that the rate of PBH mergers is at least an order of magnitude smaller than previously suggested, so that GW constraints on the PBH contribution to the DM are relieved.

Chen \textit{et al.} \cite{Chen:2019xse} have obtained PBH limits by searching for induced GWs in the 11-year PTA data set of the North American Nanohertz Observatory for Gravitational waves (NANOGrav).
They infer $f < 1$ in the range $0.0004\text{--}2\,M_{\odot}$ and $f < 10^{-6}$ in the range $0.002\text{--}0.4\,M_{\odot}$ and this constraint is shown in Fig.~\ref{fig:background}.
However, the NANOGrav collaboration have recently claimed a detection.
Six groups have suggested that this could be the second order background associated with PBHs but they advocate very different mass-scales:
Kohri and Terada~\cite{Kohri:2020qqd} advocate $\mathcal O(1)\,M_{\odot}$ PBHs with $f = 10^{-3}$;
De Luca \textit{et al.}~\cite{DeLuca:2020agl} argue that $f \sim 1$ can be realised in the mass range $M \sim 10^{-15}\text{--}10^{-11}\,M_{\odot}$ if the PBHs have a broad mass spectrum;
Vaskonen and Veerm\"ae~\cite{Vaskonen:2020lbd} argue that $10^{3}\text{--}10^{6}\,M_{\odot}$ PBHs can seed SMBHs and fit the GW signal but that a wider mass range of $10^{-1}\text{--}10^{6}\,M_{\odot}$ is also consistent with the data at the $2 \sigma$ level for $f < 0.01$;
Dom\`{e}nech \textit{et al.}~\cite{Domenech:2020ers} argue that a flat secondary GW spectrum fits the NANOGrav data and that this is expected if the PBHs form in a matter-dominated early Universe with mass in the range $10^{-5}\text{--}10^{-1}\,M_{\odot}$;
Atal \textit{et al.}~\cite{Atal:2020yic} argue that PBHs in the range $10^{11}\text{--}10^{12}\,M_{\odot}$ with $f \sim 10^{-3}$ could explain the signal, with the $\mu$ distortion limits being avoided if they arise from non-Gaussian fluctuations;
Sugiyama \textit{et al.}~\cite{Sugiyama:2020roc} argue that sublunar PBHs arising from inflation could generate the NANOGrav signal, pointing out that the PBHs could also be probed by the Subaru observations discussed in Sec.~\ref{sec:ml}.
Even if the NANOGrav detection is real, there is clearly considerable ambiguity in its interpretation.

\section{Constraints for extended PBH mass functions and mixed dark matter}

\begin{figure}[ht]
\begin{center}
\includegraphics[width=.60\textwidth]{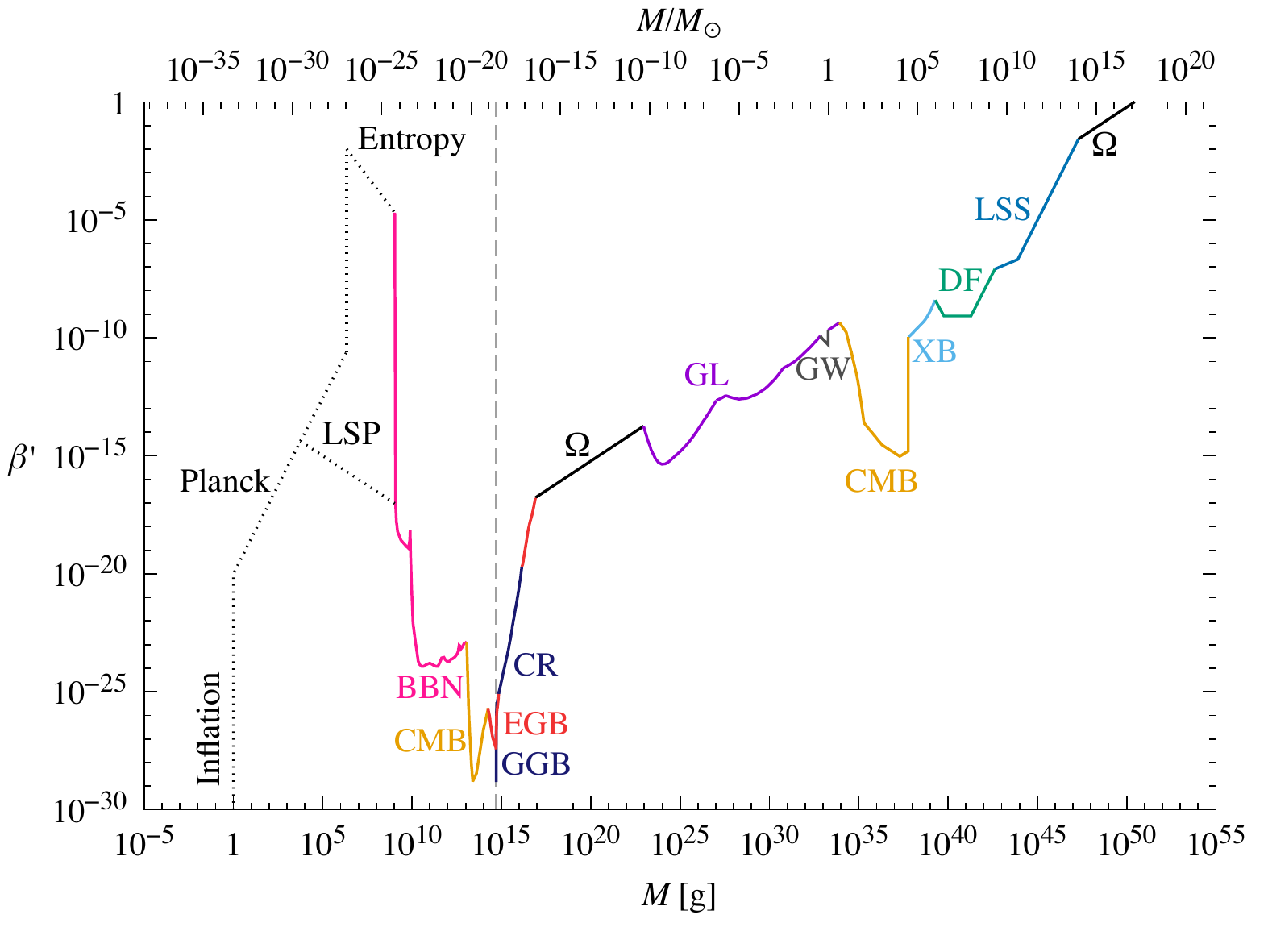}
\end{center}
\caption{Combined constraints on $\beta(M)$ for a monochromatic PBH mass function.
For each mass range, this gives the strongest limit for either evaporating and non-evaporating PBHs.
As discussed in the text, the dotted lines below $10^9\,\mathrm g$ involve less secure assumptions.
No PBHs form to the left of the `inflation' line, so this is a theoretical rather than observational constraint.}
\label{fig:master}
\end{figure}

Our master diagram of all the $\beta'(M)$ constraints for a monochromatic PBH mass function is shown in Fig.~\ref{fig:master}.
For each mass range, this gives the strongest limit shown in Fig.~\ref{fig:combined} for evaporating PBHs and in Fig.~\ref{fig:latest} for non-evaporating ones, where we have used relation \eqref{eq:f} in the latter case to convert from $f(M)$ to $\beta'(M)$.
The lines below $10^9\,\mathrm g$ are shown dotted since they involve less secure assumptions (Planck relics, LSP particles, inflation) than the ones at higher mass.
The relic and entropy constraints come from Eqs.~\eqref{relics} and \eqref{eq:entropy}, respectively, while the LSP constraint comes from Eq.~\eqref{eq:stable} and is shown in Fig.~\ref{fig:combined}.
The vertical `inflation' line at $M \sim 1\,\mathrm g$ corresponds to the maximum reheat temperature of $10^{16}\,\mathrm{GeV}$ implied by the CMB quadrupole anisotropy, no PBHs being expected below this mass.
The vertical line at $10^9\,\mathrm g$ corresponds to the lower bound on the reheat temperature, $T_\mathrm R>5\,\mathrm{MeV}$, obtained from BBN~\cite{Hasegawa:2019jsa}, although this analysis applies only for $\beta'(M) > 10^{-14}$.

\begin{figure}[ht]
\begin{center}
\includegraphics[width=.60\textwidth]{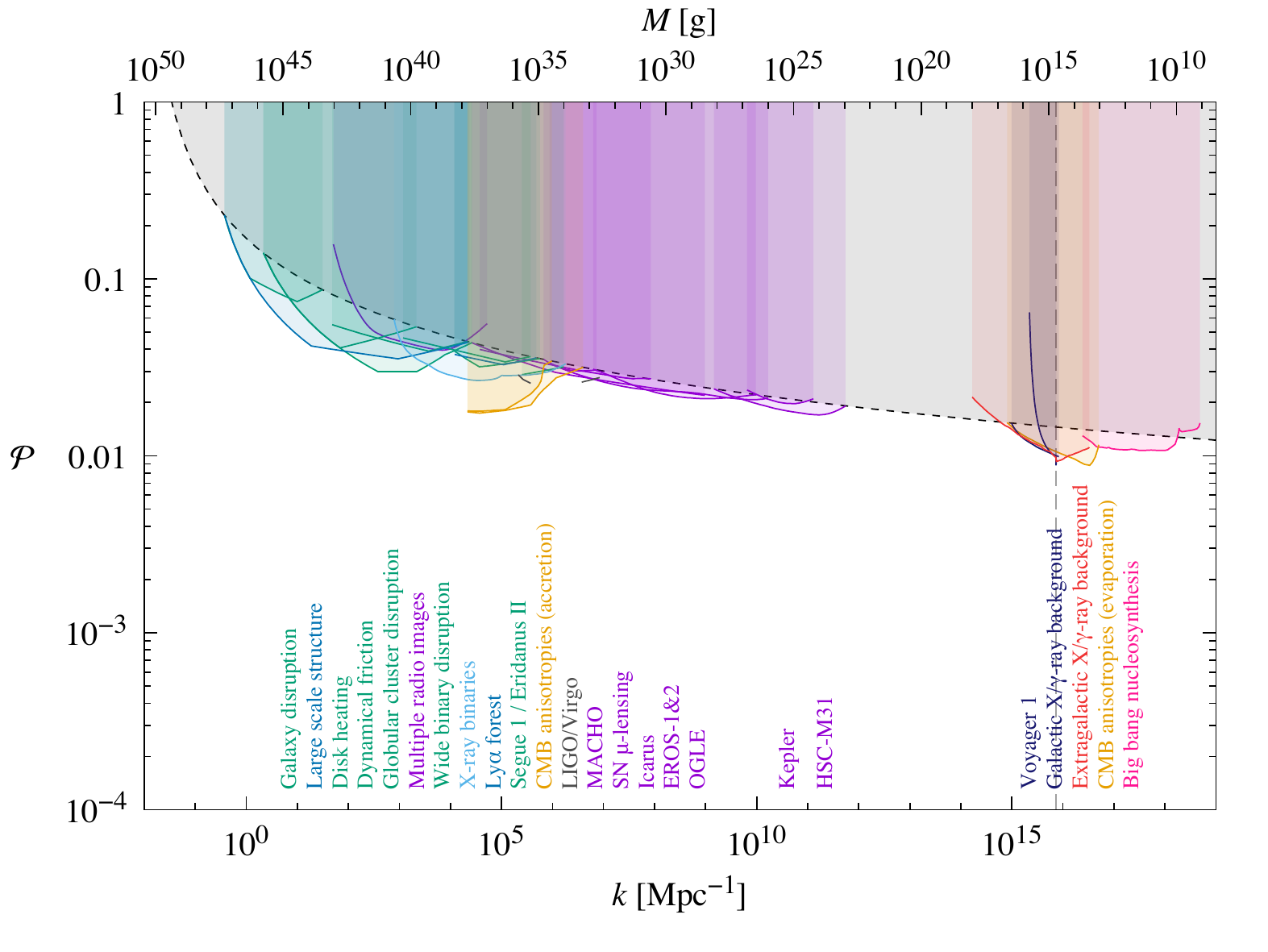}
\end{center}
\caption{Constraints on power spectrum $\mathcal P(k)$ implied by Figs.~\ref{fig:combined} and \ref{fig:latest} and Eq.~\eqref{eq:betasigma} for Gaussian fluctuations.}
\label{fig:power}
\end{figure}

If the PBHs form from primordial inhomogeneities and the horizon-scale fluctuations have a Gaussian distribution with dispersion $\sigma$, then one expects the fraction of horizon patches collapsing to a black hole to be \cite{Carr:1975qj}
\begin{equation}
\beta
\approx
  \mathrm{Erfc}\!
  \left[
   \frac{ \delta_\mathrm c }{ \sqrt{2}\,\sigma }
  \right]\,.
\label{eq:betasigma}
\end{equation}
Here `Erfc' is the complementary error function and $\delta_\mathrm c$ is the density contrast required for PBH formation.
This implies the limits on the primordial power spectrum $\mathcal P(k)$ shown in Fig.~\ref{fig:power}.
However, this does not include the constraints shown in Fig.~\ref{fig:background} because these are not \emph{direct} PBH limits.
For example, the $\mu$-distortion constraints imply limits on both PBHs and the power spectrum, whereas observational limits on PBHs directly constrain the power spectrum in the other cases.
There are many other constraints on $\mathcal P(k)$ but the PBH limits provide the most stringent ones on small scales and indeed cover a range of $k$ not probed by any other type of observation.

If the primordial fluctuations can be described by a scalar spectral index and its running at the CMB scale~\cite{Li:2018iwg}, then Fig.~\ref{fig:power} implies constraints on these two parameters.
However, the situation may be more complicated than this since most models for PBH formation from inflation do not predict a simple power law with running and some require a spike in the power-spectrum.
Also Eq.~\eqref{eq:betasigma} may not apply in a more detailed analysis~\cite{Akrami:2016vrq}.
In particular, Kalaja \textit{et al.} \cite{Kalaja:2019uju} have used results from numerical simulations and peak theory to study PBH formation for a range of perturbation profiles, including a careful treatment of non-linearities and smoothing and filtering scales.
Sato-Polito \textit{et al.} \cite{Sato-Polito:2019hws} incorporate uncertainties in the critical overdensity $\delta_\mathrm c$ associated with ellipsoidal collapse.
Other relevant results can be found in Refs.~\cite{Yoo:2018esr,Young:2020xmk} for peak theory, in Refs.~\cite{Germani:2019zez,Suyama:2019npc,Yoo:2019pma} for some extensions of peak theory, in Ref.~\cite{Ando:2018nge,Young:2019osy} for different choices of the window function and in Refs.~\cite{Yoo:2018esr,Germani:2018jgr,Kawasaki:2019mbl} for various non-linear relations between the curvature and density perturbations.
Finally, Eq.~\eqref{eq:betasigma} does not apply for PBHs which form in a matter-dominated era or through some other mechanism unrelated to primordial inhomogeneities.

The constraints shown in Fig.~\ref{fig:master} assume that the PBH mass function is quasi-monochromatic (i.e.\ with a width $\Delta M \sim M$).
However, in many scenarios one would expect the mass function to be extended, specific examples being discussed below.
In the context of the DM problem, this is a two-edged sword \cite{Carr:2016drx}.
On the one hand, it means that the \emph{total} PBH density may suffice to explain the DM, even if the density in any particular mass band is small and within the observational bounds discussed in Sec.~\ref{sec:Constraints}.
On the other hand, even if PBHs can provide all the DM at some mass scale without violating the constraints there, the extended mass function may still violate the constraints at some other scale.
This problem is particularly pertinent if the mass function extends over many decades.

A detailed assessment of this problem requires a knowledge of the expected PBH mass fraction, $f_\mathrm{exp}(M)$, and the maximum allowed fraction, $f_\mathrm{max}(M)$, in the monochromatic case.
However, the procedure is non-trivial even when these functions are known.
In particular, one cannot just plot $f_\mathrm{exp}(M)$ for a given model in Fig.~\ref{fig:latest} and infer that the model is allowed because it does not intersect $f_\mathrm{max}(M)$, even though this procedure is sometimes used in the literature.
If the PBHs span an extended range of masses, the differential mass function is written as $\mathrm dn/\mathrm dM$ where $\mathrm dn$ is the number density of PBHs in the mass range $(M,M+\mathrm d M)$.
One can then define the mass density and DM fraction of PBHs with mass `around' $M$ as
\begin{equation}
\rho(M)
\equiv
  M^{2}\,\frac{\mathrm dn}{\mathrm dM}\,,
\quad
f(M)
\equiv
  \frac{\rho(M)}{\rho_{\mathrm{CDM}}}\,.
\label{extended}
\end{equation}
Alternatively, one can define $n(M)$, $\rho(M)$ and $f(M)$ as integrated values for PBHs with mass larger or smaller than $M$.
However, these are only simply related to the functions in Eq.~\eqref{extended} for a power-law spectrum.

One approach, adopted by Carr \textit{et al.} in Ref.~\cite{Carr:2016drx}, is to study the possible DM windows by breaking up each constraint into narrow mass bins and treating it as a sequence of flat constraints.
However, this is quite a complicated procedure and it has attracted some criticism from Green \cite{Green:2016xgy}.
A more elegant methodology has been adopted by Carr \textit{et al.} in Ref.~\cite{Carr:2017jsz}.
In this, one introduces the distribution function for $\log{M}$,
\begin{equation}
\psi(M)
\propto
  M\,\frac{\mathrm dn}{\mathrm dM}\,,
\end{equation}
normalised so that the \emph{total} fraction of the DM in PBHs is
\begin{equation}
f_\mathrm{PBH}
\equiv
  \frac{\Omega_\mathrm{PBH}}{\Omega_\mathrm{CDM}}
= \int_{M_\mathrm{min}}^{M_\mathrm{max}} \mathrm dM\,\psi(M)\,.
\end{equation}
The lower cut-off in the mass integral necessarily exceeds $M_* \approx 4 \times 10^{14}\,\mathrm g$, the mass of the PBHs evaporating at the present epoch \cite{Carr:2009jm}.
We follow the discussion of Ref.~\cite{Carr:2017jsz} below, although other approaches have been used by Bellomo \textit{et al.}~\cite{Bellomo:2017zsr} and De Luca \textit{et al.}~\cite{DeLuca:2020ioi}.

\subsection{Possible PBH mass functions}

Reference~\cite{Carr:2017jsz} considers four types of mass function, which we summarise briefly here.

\begin{enumerate}
\item A lognormal mass function of the form
\begin{equation}
\psi(M)
= \frac{f_\mathrm{PBH}}{\sqrt{2\pi}\,\sigma M}\,
  \exp\left(-\frac{\log^2(M/M_c)}{2\sigma^2}\right)\,,
\label{eq:dist}
\end{equation}
where $M_c$ is the mass at which the function $M\psi(M)$ peaks and $\sigma$ is its width.
This is a good approximation if the PBHs result from a smooth symmetric peak in the inflationary power spectrum.
It was first suggested in Ref.~\cite{Dolgov:1992pu} and then demonstrated numerically in Ref.~\cite{Green:2016xgy} and analytically in Ref.~\cite{Kannike:2017bxn} for the slow-roll case.
For a given value of $f_\mathrm{PBH}$\,, the mass function is therefore described by two parameters.
Note that the lognormal mass function used in Refs.~\cite{Green:2016xgy,Horowitz:2016lib,Kuhnel:2017pwq} omits the $M^{-1}$ term in Eq.~\eqref{eq:dist}, in which case the function $M\psi(M)$ peaks at $e^{\sigma^2}M_c$ rather than $M_c$\,.
The form \eqref{eq:dist} is more useful for our purposes.
However, Gow \textit{et al.}~\cite{Gow:2020cou} point out that a narrow peak in the power spectrum is more likely to produce a skew-lognormal mass function.

\item A power-law mass function of the form
\begin{equation}
\psi(M)
\propto
  M^{\gamma -1}
\quad (M_\mathrm{min} < M < M_\mathrm{max})\,.
\label{eq:power}
\end{equation}
For $\gamma\neq 0$, either the lower or upper cut-off can be neglected if $M_\mathrm{min} \ll M_\mathrm{max}$\,, so this scenario is effectively described by two parameters.
Only in the $\gamma = 0$ case are both cut-offs important.
A mass function of this form arises naturally if the PBHs form from scale-invariant density fluctuations or from the collapse of cosmic strings.
In both cases, $\gamma = -2w/(1+w)$, where $w$ specifies the equation of state, $p = w\,\rho\,c^2$, when the PBHs form \cite{Carr:1975qj}.
In a non-inflationary model, $w \in (-1/3,1)$ and $\gamma \in (-1,1)$, so the exponent in Eq.~\eqref{eq:power} is always negative.
Inflation itself corresponds to $w \in (-1,-1/3)$ and $\gamma \in (1,\infty)$ but Eq.~\eqref{eq:power} is inapplicable in this case since PBHs do not form \textit{during} inflation but only afterwards from the inflationary fluctuations.

\item A critical collapse mass function of the form
\begin{equation}
\psi(M)
\propto
  M^{2.85}\,\exp(-(M/M_f)^{2.85})\,,
\label{eq:crit}
\end{equation}
where the exponent 2.85 assumes radiation-domination~\cite{Yokoyama:1998xd,Niemeyer:1999ak,Musco:2012au,Carr:2016hva}.
This applies generically if the PBHs form from density fluctuations with a $\delta$-function power spectrum.
In this case, the mass spectrum extends down to arbitrarily low masses but there is an exponential upper cut-off at the mass-scale $M_f$\,, which is roughly to the horizon mass at the collapse epoch.
Although the mass function is described by a single parameter, Eq.~\eqref{eq:crit} must be modified if the density fluctuations are themselves extended~\cite{Carr:2016drx}, as expected in the inflationary scenario.
Indeed, the lognormal distribution may then be appropriate, so two parameters may be required in the more realistic critical collapse situation.

\item Special consideration is required in the $w=0$ (matter-dominated) case~\cite{Polnarev:1986bi}, because both cut-offs in \eqref{eq:power} can then be relevant.
For example, this may arise due to some form of phase transition in which the mass is channeled into non-relativistic particles \cite{Khlopov:1980mg,1981SvA....25..406P} or due to slow reheating after inflation \cite{Carr:1994ar}.
Since Eq.~\eqref{eq:power} predicts $\gamma = 0$ in this case, $\rho(M)$ should increase logarithmically with $M$.
However, the analysis breaks down for this scenario because the Jeans length is much smaller than the particle horizon, so collapse is prevented by deviations from spherical symmetry rather than pressure and the probability of PBH formation becomes
\begin{equation}
\beta(M)
\approx
  0.056\,\delta_H(M)^{5}\,.
\end{equation}
This is small for $\delta_H(M) \ll 1$ but much larger than the exponentially suppressed fraction in the radiation-dominated case.
The factor $\delta_H(M)^{5}$ was first derived in Refs.~\cite{Khlopov:1980mg,1981SvA....25..406P} and is in agreement with the more recent analysis of Ref.~\cite{Harada:2016mhb}.
As discussed in Sec.~\ref{sec:collapse}, there may be a correction factor which scales as $\delta_H(M)^{3/2}$ but we will neglect that here.
We also need to consider the effect of spin~\cite{Kohri:2018qtx} because this is inevitably produced during collapse in a matter-dominated period and can exponentially suppress PBH production for smaller density perturbations~\cite{Harada:2017fjm}.
If the matter-dominated phase extends from $t_1$ to $t_2$\,, PBH formation is enhanced over the mass range
\begin{equation}
M_\mathrm{min} \sim M_H(t_1)
< M
< M_\mathrm{max} \sim M_H(t_2)\,\delta_H(M_\mathrm{max})^{3/2}\,.
\end{equation}
The lower limit is the horizon mass at the start of matter-dominance and the upper limit is the horizon mass at the epoch when the regions which bind at the end of matter-dominance enter the horizon.
For nearly scale-invariant primordial fluctuations, $\beta(M)$ is constant and $f(M)$ is logarithmic, but the mass function is still determined by two parameters, which can be taken to be $t_1$ and $t_2$ \cite{Carr:2017edp}.
\end{enumerate}

To compare with the lognormal case, we describe the mass function in the first three cases by the mean and variance of the $\log M$ distribution:
\begin{equation}
\log M_{c}
\equiv
  \langle \log M \rangle_{\psi}\,,
\quad
\sigma^{2}
\equiv
  \langle \log^{2} M \rangle_{\psi} - \langle \log M \rangle_{\psi}^{2}\,,
\end{equation}
where
\begin{equation}
\langle X \rangle_{\psi}
\equiv
  f_\mathrm{PBH}^{-1}\,\int \mathrm dM\,\psi(M)\,X(M)\,.
\end{equation}
For a power-law distribution these are
\begin{equation}
M_{c}
= M_\mathrm{cut}\,e^{-1/\gamma},
\quad
\sigma
= 1/|\gamma|\,,
\label{eq:par}
\end{equation}
where $M_\mathrm{cut} \equiv \mathrm{max}(M_\mathrm{min}, M_*)$ if $\gamma < 0$ or $M_\mathrm{max}$ if $\gamma > 0$.
For the critical-collapse distribution, the exponential cut-off is very sharp, so the mass function is well approximated by a power-law distribution with $\gamma = 3.85$ and $M_\mathrm{max} \approx M_f$\,.
As it is relatively narrow, Eq.~\eqref{eq:par} implying $\sigma = 0.26$, even the monochromatic mass function provides a good fit.
Since critical collapse should be a fairly generic feature of PBH formation, $\sigma = 0.26$ always provides a lower limit to the width.
However, critical collapse may not be relevant in the cosmic string or matter-dominated scenarios.

One might expect two parameters to suffice to describe the PBH mass function \emph{locally} (i.e.\ close to a peak) since this just corresponds to the first two terms in a Taylor expansion.
However, the mass function could be more complicated than this and might have several peaks.
For example, Hasegawa \textit{et al.} \cite{Hasegawa:2017jtk} have proposed a scenario in the minimally supersymmetric standard model which generates both intermediate-mass PBHs to explain the LIGO detections and lunar-mass PBHs to explain the DM.
Kusenko \textit{et al.} \cite{Kusenko:2020pcg} envisage PBH production through the nucleation of false vacuum bubbles during inflation, this simultaneously accounting for the DM, the HSC lensing candidate, the LIGO/Virgo black holes and the seeds for SMBHs.
Carr \textit{et al.}~\cite{Carr:2019kxo} have a scenario in which dips in the pressure during the thermal history of the Universe naturally generate four peaks in the PBH mass function, this explaining an even greater variety of cosmological conundra.

The following analysis does not apply in these situations.

\subsection{Constraints for extended PBH mass function}

Reference~\cite{Carr:2017jsz} introduces a general method for dealing with extended mass functions.
Any astrophysical observable $A[\psi(M)]$ depending on the PBH abundance (e.g.\ the number of ML events of given duration in a given time interval) can be expanded as
\begin{equation}
A[\psi(M)]
= A_0
  + \int \mathrm dM\,\psi(M)\,K_{1}(M)
  + \int \mathrm dM_{1} \mathrm dM_{2}\,\psi(M_{1})\,\psi(M_{2})\,K_{2}(M_{1},M_{2})
  + \ldots \,,
\label{eq:A_observable}
\end{equation}
where $A_0$ is the background contribution and the functions $K_j$ depend on the details of the underlying physics and the nature of the observation.
If PBHs with different mass contribute independently to the observable, only the first two terms in Eq.~\eqref{eq:A_observable} need to be considered.
Explicit expressions are given for lensing and survival of stars in Ref.~\cite{Green:2016xgy}, for evaporation in Ref.~\cite{Carr:2016drx} and for neutron star capture and accretion in Ref.~\cite{Kuhnel:2017pwq}.

If a measurement puts an upper bound on the observable,
\begin{equation}
A[\psi(M)]
\leq
  A_\mathrm{exp}\,,
\end{equation}
then for a monochromatic mass function with $M=M_c$ we have
\begin{equation}
\psi_\mathrm{mon}(M)
\equiv
  f_\mathrm{PBH}(M_{c})\,\delta(M - M_{c})\,.
\end{equation}
The maximum allowed fraction of DM in PBHs of mass $M_{c}$ is then
\begin{equation}
f_\mathrm{PBH}(M_{c})
\leq
  \frac{A_\mathrm{exp} - A_0}{K_{1}(M_{c})}
\equiv
  f_\mathrm{max}(M_{c})\,.
\label{eq:f_max}
\end{equation}
Combining Eqs.~\eqref{eq:A_observable}--\eqref{eq:f_max} then yields
\begin{equation}
\int \mathrm dM\,\frac{\psi(M)}{f_\mathrm{max}(M)}
\leq
  1\,.
\label{eq:general_constraint}
\end{equation}
Once $f_\mathrm{max}$ is known, we can apply Eq.~\eqref{eq:general_constraint} to obtain the constraint for an arbitrary mass function $\psi (M)$.
We first integrate Eq.~\eqref{eq:general_constraint} over the mass range ($M_1,M_2$) for which the constraint applies, assuming a particular function $\psi (M;f_\mathrm{PBH},M_c,\sigma)$.
After specifying $M_1$ and $M_2$\,, this constrains $f_\mathrm{PBH}$ as a function of $M_c$ and $\sigma$.
The procedure must be implemented separately for each observable.
Different constraints can be combined by using the relation
\begin{equation}
\sum_{j=1}^N \left(\int \mathrm dM\,\frac{\psi(M)}{f_{\mathrm{max},j}(M)} \right)^2
\leq
  1\,,
\label{eq:combined_constraint}
\end{equation}
where $f_{\mathrm{max},j}(M)$ correspond to the different bounds for a monochromatic mass function, as defined by Eq.~\eqref{eq:f_max}.
Most constraints rely on a single observable, such as the number of lensing events or the abundance of wide binaries.

\begin{figure}[ht]
\begin{center}
\includegraphics[width=.4\textwidth]{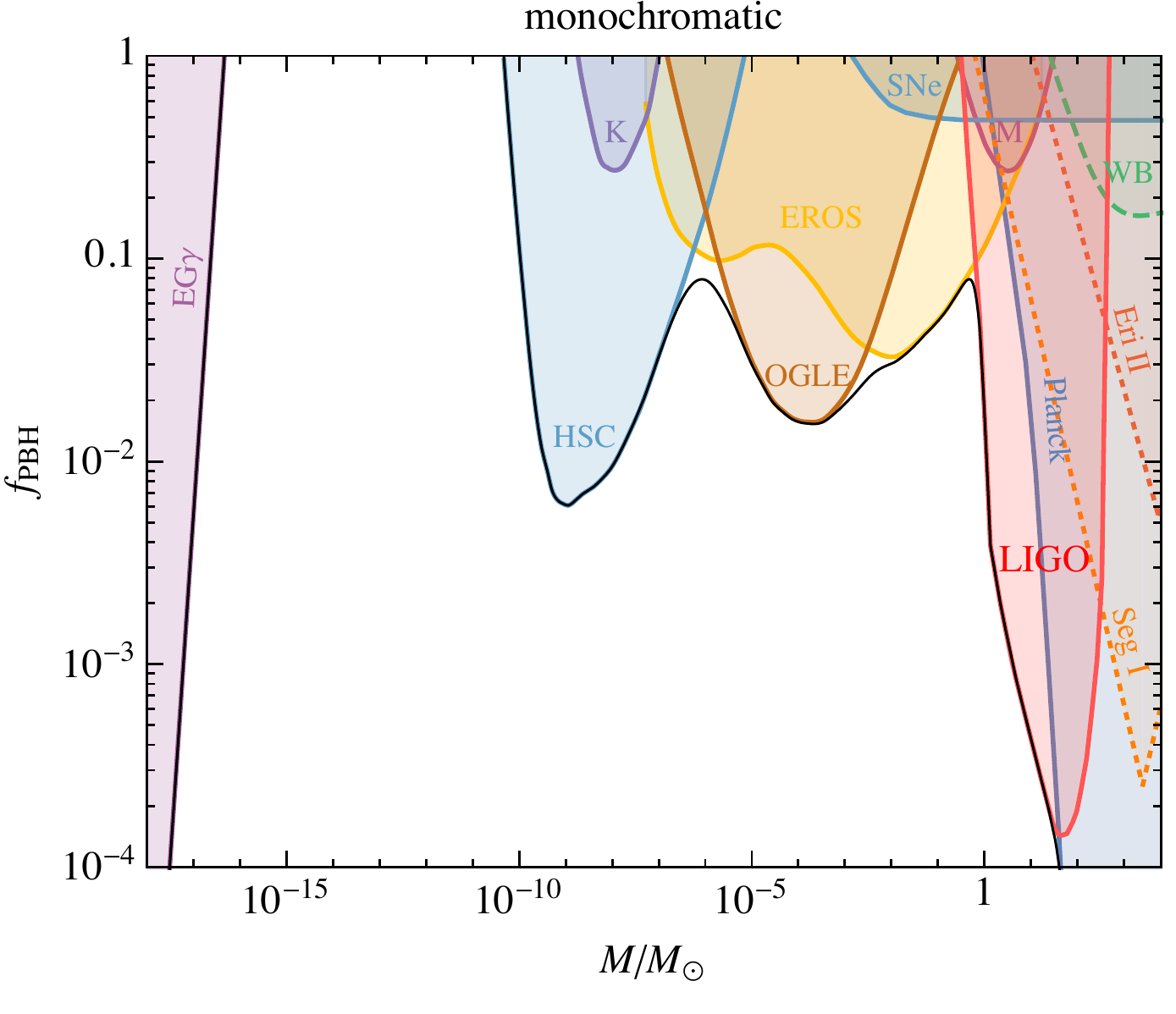} 
\includegraphics[width=.4\textwidth]{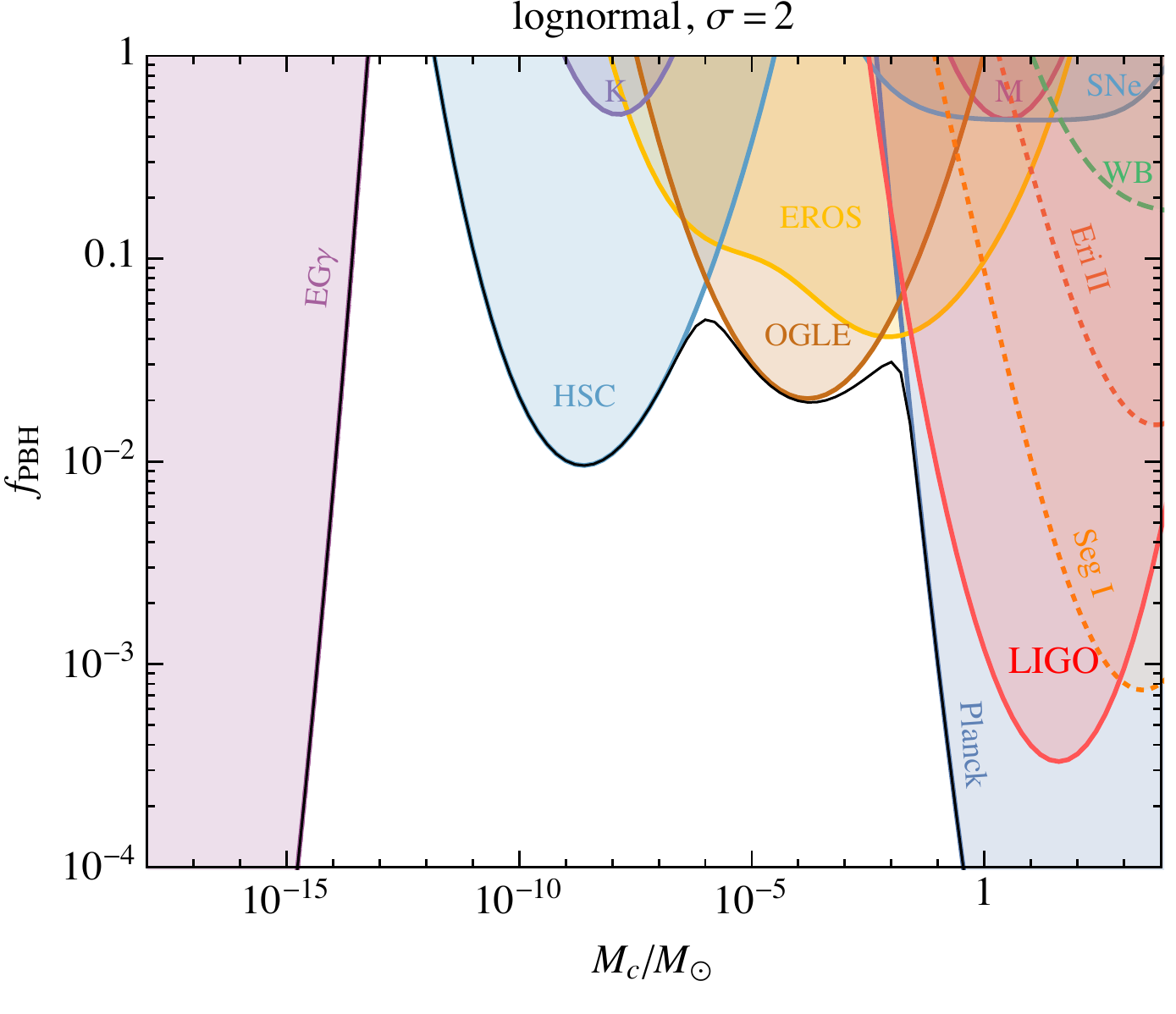}
\end{center}
\caption{Constraints on the fraction of PBH dark matter, $f_\mathrm{PBH} \equiv \Omega_\mathrm{PBH}/\Omega_\mathrm{DM}$\,, as a function of the PBH mass $M_c$; this is an updated version of Fig.~1 in Ref.~\cite{Carr:2017jsz} (Raidal, Vaskonen \& Veermae, private communication).
\textit{Left panel}: This assumes a monochromatic mass function and corresponds to a subset of the constraints in Fig.~\ref{fig:latest}.
The purple region is excluded by evaporations \cite{Carr:2009jm}, the blue, violet, yellow, brown, purple and dark blue regions by the microlensing results from HSC~\cite{Niikura:2017zjd}, Kepler (K) \cite{Griest:2013aaa}, EROS \cite{Tisserand:2006zx}, OGLE \cite{Niikura:2019kqi}, MACHO (M) \cite{Allsman:2000kg} and supernovae (SNe) \cite{Zumalacarregui:2017qqd}, respectively.
The red and blue regions on the right are excluded by LIGO/Virgo \cite{Vaskonen:2019jpv} and Planck data \cite{Ali-Haimoud:2016mbv}.
Various dynamical exclusions are indicated by broken lines, in orange for survival of stars in Segue I (Seg I) \cite{Koushiappas:2017chw} and Eridanus II (Eri II) \cite{Brandt:2016aco} and in green for survival of wide binaries (WB) \cite{Monroy-Rodriguez:2014ula}.
The black solid lines show the combined constraint.
\textit{Right panel:} Equivalent constraints for a lognormal PBH mass function with $\sigma=2$ as a function of $M_c$\,.}
\label{fig:mono}
\end{figure}

The constraints on the allowed DM fraction in PBHs are shown in Fig.~\ref{fig:mono}, which is an updated version of Fig.~1 in Ref.~\cite{Carr:2017jsz}.
The left panel shows the constraint for a monochromatic mass function and corresponds to a subset of the constraints in Fig.~\ref{fig:latest} for the mass range $10^{-18}\text{--}10^5\,M_{\odot}$\,.
The most restrictive forms for the constraints are used and some are depicted by the dotted lines.
The right panel shows the equivalent constraints for a lognormal mass function with $\sigma=2$ and a general value of $M_c$\,.
We do not show the limits for a power-law mass function, since we only wish to demonstrate the methodology, but these are given in Ref.~\cite{Carr:2017jsz} for an earlier form of the monochromatic constraints.
The important qualitative point is that the form of the constraints in this case is itself dependent on the PBH mass function.
One cannot just compare a predicted extended mass function with the monochromatic form of the constraints.

\begin{figure}[ht]
\begin{center}
\includegraphics[width=.4\textwidth]{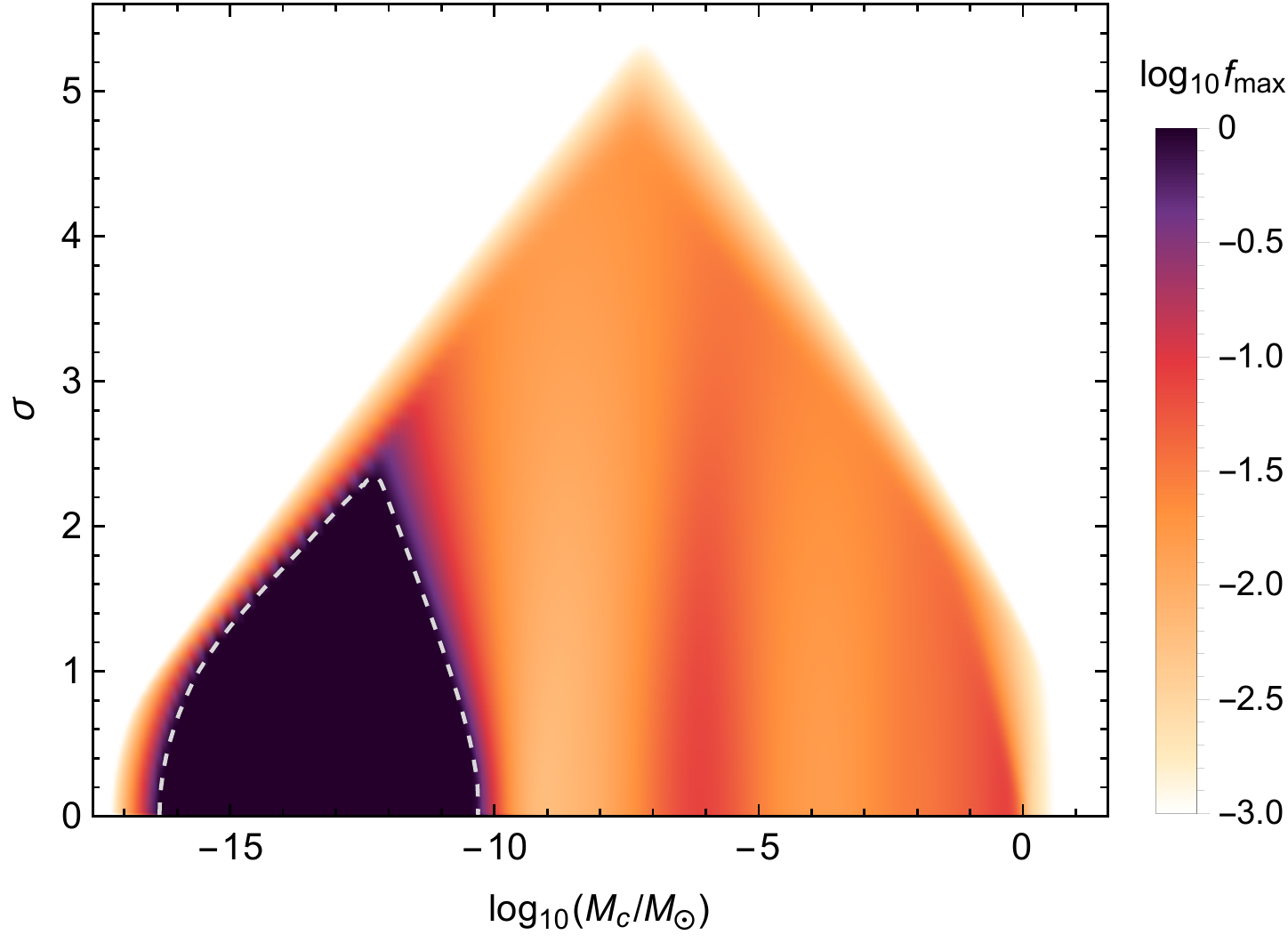}
\end{center}
\caption{Observational constraints on $M_c$ and $\sigma$ for a lognormal PBH mass function; this is an updated version of Fig.~2 in Ref.~\cite{Carr:2017jsz} (Raidal, Vaskonen \& Veermae, private communication).
The color coding shows the maximum allowed fraction of PBH dark matter.
In the white region $f_\mathrm{max}< 10^{-3}$, while the dashed white contour correspond to $f_\mathrm{max}=1$.}
\label{fig:densityplot}
\end{figure}

Figure~\ref{fig:densityplot} is an update of Fig.~2 in Ref.~\cite{Carr:2017jsz} and shows the maximum allowed DM fraction, $f_\mathrm{max}$\,, in the ($M_c\,,\sigma$) plane for lognormal PBH mass function.
The constraints have been combined using Eq.~\eqref{eq:combined_constraint}.
The region where 100\,\% of DM can consist of PBHs is indicated by the dashed white contour.
The shape of the constraints in Fig.~\ref{fig:densityplot} makes it clear that the allowed mass range for fixed $f_\mathrm{PBH}$ decreases with increasing width $\sigma$, thus ruling out the possibility of evading the constraints by simply extending the mass function.
Moreover, Fig.~\ref{fig:densityplot} gives an upper bound $\sigma \lesssim 2.4$ if all the DM is in the form of PBHs.
For a power-law mass function, Ref.~\cite{Carr:2017jsz} conclude $|\gamma| \gtrsim 1$, which effectively rules out PBH DM from the collapse of cosmic strings or scale-invariant density fluctuations.
This agrees with the conclusions of Refs.~\cite{Green:2016xgy,Horowitz:2016lib,Kuhnel:2017pwq}.
The last paper provides a more comprehensive analysis, covering the mass range $10^{-18}\text{--}10^4\,M_\odot$\,, but it does not include the recent HSC constraint \cite{Niikura:2017zjd} and it uses the unnecessarily stringent Planck constraint from Ref.~\cite{Chen:2016pud}.

\subsection{Combined PBH and particle dark matter}

If most of the DM is in the form of elementary particles, these will be accreted around any PBHs to form minihalos.
For weakly interacting massive particles (WIMPs) and light enough PBHs, this can even happen during the radiation-dominated era, since Eroshenko \cite{Eroshenko:2016yve} has shown that a low-velocity subset of particles will accumulate around PBHs as density spikes shortly after the WIMPs kinetically decouple from the background plasma.
For somewhat heavier PBHs, WIMP accretion is not important until after the time of matter-radiation equality ($t_\mathrm{eq}$) and it then arises due to secondary infall.
Both cases lead to a similar halo profile and the subsequent annihilation of the WIMPs within the halos will give rise to bright $\gamma$-ray sources.
Comparison of the expected signal with Fermi-LAT data then severely constrains $f_\mathrm{PBH}$ for PBHs larger than about $10^{-12}\,M_{\odot}$\,.
Indeed, this leads to much stronger limits than any other constraint discussed in this review.

\begin{figure}[ht]
\begin{center}
\includegraphics[width = 0.36\textwidth]{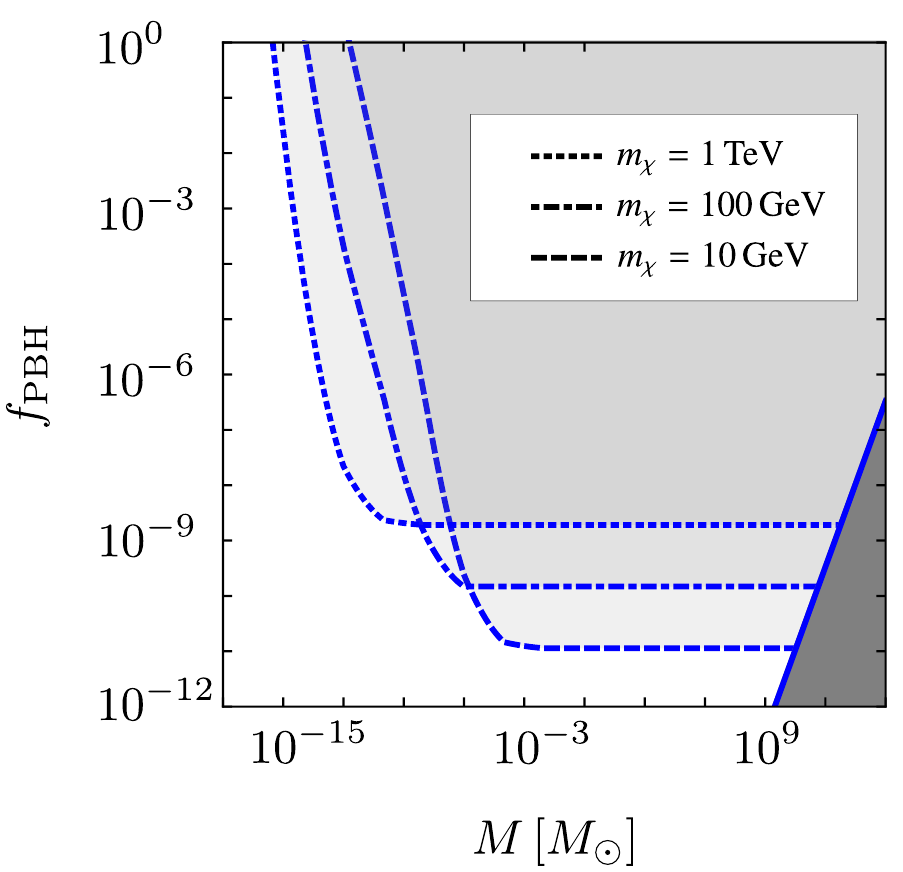}
\end{center}
\caption{Constraints on $f_\mathrm{PBH}$ as a function of PBH mass, adapted from Ref.~\cite{Carr:2020mqm}, for a WIMP mass of $10\,\mathrm{GeV}$ (dashed line), $100\,\mathrm{GeV}$ (dashed-dotted line) and $1\,\mathrm{TeV}$ (dotted line), with $\langle \sigma v \rangle = 3 \times 10^{-26}\,\mathrm{cm}^{3}\,\mathrm s^{-1}$.
Also shown is the incredulity limit (solid line), corresponding to one PBH per horizon.}
\label{fig:fPBHWIMP}
\end{figure}

Figure~\ref{fig:fPBHWIMP} shows constraints on $f_\mathrm{PBH}$ for WIMP masses of $10\,\mathrm{GeV}$, $100\,\mathrm{GeV}$ and $1\,\mathrm{TeV}$ and the standard value for the velocity-averaged annihilation cross-section, $\langle \sigma v \rangle_\mathrm F = 3 \times 10^{-26}\,\mathrm{cm}\,\mathrm s^{-1}$.
The falling limit at low $M$ is associated with halos formed after kinetic decoupling and is taken from Ref.~\cite{Carr:2020xqk}.
Boucenna \textit{et al.}~\cite{Boucenna:2017ghj} have also investigated this constraint but for a larger range of values for $\langle \sigma v \rangle$ and $m_{\chi}$.
The flat limit at larger $M$ is associated with halos formed by secondary infall and comes from the background generated by the integrated emission of many WIMP halos.
It corresponds to the ($M$-independent) constraint $f_\mathrm{PBH} \lesssim 10^{-10}$ for the assumed values of $\langle \sigma v \rangle$ and $m_{\chi}$\,.
Adamek \textit{et al.}~\cite{Adamek:2019gns} first derived this limit for solar-mass PBHs but the argument can be extended to much larger masses, as discussed in Ref.~\cite{Carr:2020erq}.

More precisely, the flat limit on the DM fraction can be expressed as:
\begin{equation}
f_\mathrm{PBH}
< \frac{ 16 }{ 3 }\,
  \frac{ \Phi^\mathrm{Fermi}_{100\,\mathrm{MeV}}\,H_{0} }
       { \rho_\mathrm{DM}\,\tilde{N}_{\gamma}( m_{\chi} ) }\,
  \left(\frac{ 2 m_{\chi}^{4}\,t_{0}^{2} }
             { \langle\sigma v \rangle\,\rho_\mathrm{eq} }\right)^{1 / 3}\,,
\end{equation}
where $\Phi^\mathrm{Fermi}_{100\,\mathrm{MeV}} = 6 \times 10^{-9}\,\mathrm{cm}^{-2}\,\mathrm s^{-1}$ is the Fermi point-source sensitivity above the threshold energy $E_\mathrm{th} = 100\,\mathrm{MeV}$ and $\rho_\mathrm{eq} = 2.1 \times 10^{-19}\,\mathrm g\,\mathrm{cm}^{-3}$ is the density at $t_\mathrm{eq}$\,.
$\tilde{N}_{\gamma}$ is the average number of photons produced,
\begin{equation}
\tilde{N}_{\gamma}( m_{\chi} )
\equiv
  \int_{E_\mathrm{th}}^{m_{\chi}} \mathrm d E\,
  \frac{ dN_{\gamma} }{ \mathrm d E }\,
  \int_{0}^{\infty}\!\mathrm d z\,
  \frac{ H_{0} }{ H( z ) }\,
  \exp\left[-\tau(z,E)\right]\,,
\end{equation}
where $\mathrm d N_{\gamma} / \mathrm d E$ is the number of $\gamma$-rays emitted from the annihilations per unit energy and $\tau$ is the optical depth for photon-matter pair production, photon-photon scattering and photon-photon pair production~\cite{Cirelli:2009dv}.
No bound can be placed above the mass where the lines intersects the cosmological incredulity limit
\begin{equation}
M_\mathrm{IL}
= 2.5 \times 10^{10}\,
  \left(\frac{ m_{\chi} }{ 100\,\mathrm{GeV} }\right)^{1.11}\,
  \left(\frac{ \langle \sigma v \rangle}{ \langle \sigma v \rangle_\mathrm F }\right)^{-1 / 3}\,
  M_{\odot}\,.
\end{equation}
These constraints are not included in Fig.~\ref{fig:accretion}, since they depend on the assumption that the DM is dominated by WIMPs, but they have the important implication that there cannot be an appreciable amount of DM in \emph{both} WIMPs and PBHs.
As discussed by Kadota and Silk~\cite{Kadota:2020ahr}, an interesting application of this limit arises if the first bound cosmic structures are minihalos formed from PBH Poisson fluctuations (cf.\ Sec.~\ref{sec:clouds}).

\section{Conclusions}

Although this review has focused on PBH constraints, our motivation has not been to exclude their existence, so the large number of limits discussed should not merely be regarded as `nails in the coffin' of the PBH scenario.
All the constraints in Fig.~\ref{fig:latest} come with caveats, either because of uncertainties in the observations or because of flexibility in the models.
Some limits can be relaxed in various ways (e.g.\ by invoking PBH clustering for the lensing constraints or non-Gaussian fluctuations for the $\mu$ distortion constraints), although these `let-outs' may themselves be controversial.
Ideally, all the constraints should have 95\,\% confidence level but calculating a confidence level is difficult in some cases.
While we have tried to cover all the literature for historical completeness, we have represented constraints which are questionable or no longer believed by broken lines in our figures.

The constraints also have a positive aspect in that they could each in principle provide \emph{evidence} for PBHs.
Indeed, historically some of them have already been thought to do so.
For example, microlensing of stars in the LMC once suggested that PBHs with $M \sim 1\,M_{\odot}$ could provide the dark matter in the Galactic halo~\cite{Alcock:2000ph};
later observations seemed to exclude this but the source of the extra events remains unclear, so there could still be some PBHs.
There were also claims to have detected PBHs with $M \sim 10^{-3}\,M_{\odot}$ from microlensing of quasars~\cite{1993Natur.366..242H}, although this mass estimate was later revised upwards.
At one stage it was argued that PBHs with $M \sim 10^6\,M_{\odot}$ could generate the observed heating of the Galactic disc~\cite{1985ApJ...299..633L}, although this now seems unlikely.
It was once proposed that accretion by massive PBHs could generate the extragalactic X-ray background~\cite{1980Natur.284..326C} and that exploding PBHs with $M \sim 10^{15}\,\mathrm g$ could explain some short-period gamma-ray bursts~\cite{Cline:1996zg}.

Most of these claims have now been superseded but there is great current interest in whether PBHs can explain other cosmological conundra.
In particular, many authors have discussed the possibility that PBHs could provide the dark matter~\cite{Carr:2020xqk,Green:2020jor}.
In this context, Fig.~\ref{fig:latest} might suggest that the asteroid range $10^{17}\,\mathrm g < M < 10^{23}\,\mathrm g$ between the EGB and HSC microlensing limits is most plausible.
However, the intermediate range $10\,M_{\odot} < M < 10^{2}\,M_{\odot}$ between the LMC microlensing and wide binary limits is still favored by some theorists, even though this appears to violate other constraints, because PBHs may naturally form in this range.
In principle, Fig.~\ref{fig:latest} suggests that a lot of dark matter could reside in ``stupendously large black holes'' in the range $10^{14}\,M_{\odot} < M < 10^{17}\,M_{\odot}$\,, although the lack of constraints there probably just reflects the fact that little attention has been paid to this possibility and obviously such SLABs could not provide dark matter inside galactic halos~\cite{Carr:2020erq}.
A final possibility is the Planck-mass relics of Hawking evaporation in the range around $M \sim 10^{-5}\,\mathrm g$, although this is probably untestable unless the relics are electrically charged~\cite{Lehmann:2019zgt}.

PBHs could have important cosmological consequences even if $f$ is small.
For example, we have emphasised the possible association of PBHs with the LIGO/Virgo events, the merging black holes having masses in the range $8\text{--}50\,M_{\odot}$\,.
While the mainstream view may be that these are stellar remnants \cite{2016Natur.534..512B}, this range is certainly larger than was originally expected for such remnants, so PBHs offer a plausible alternative.
On the other hand, it is unlikely that the LIGO/Virgo black holes themselves can provide all the dark matter, although PBHs with an extended mass function still might do so because one would then expect the PBH density to peak at a lower mass than the LIGO/Virgo events.
The viability of this proposal depends on whether the binaries form in the early Universe or much later in galactic halos.
In any case, it is clear that LIGO/Virgo/KAGRA observations will soon confirm or exclude this proposal.
As another example, we have seen that intermediate-mass PBHs could influence the development of large-scale cosmic structure for $f \sim 10^{-3}$ and they could seed the SMBHs in galactic nuclei for $f \sim 10^{-5}$.

The three proposed roles for PBHs discussed above - accounting for dark matter, LIGO/Virgo events, seeds for large-scale structure - are essentially independent if the PBHs have a monochromatic mass function.
However, if they have a very extended mass function, they could in principle play all three roles.
Indeed, Ref.~\cite{Carr:2019kxo} suggests that the thermal history of the Universe could naturally produce a PBH mass spectrum with peaks at $10^{-6}$, $1$, $30$ and $10^{6}\,M_{\odot}$ and that this might provide a unified solution of a multitude of observational conundra:
(1) microlensing events towards the Galactic bulge generated by planetary-mass objects, these being much more frequent than expected for free-floating planets~\cite{Niikura:2019kqi};
(2) microlensing of quasars \cite{Mediavilla:2017bok}, including ones that are so misaligned with the lensing galaxy that the probability of lensing by a star is very low;
(3) the unexpectedly high number of microlensing events towards the Galactic bulge by objects between $2$ and $5\,M_{\odot}$ \cite{Wyrzykowski:2019jyg}, where stellar evolution models fail to form black holes \cite{Brown:1999ax};
(4) unexplained correlations in the source-subtracted X-ray and cosmic infrared background fluctuations \cite{2013ApJ...769...68C};
(5) the non-observation of ultra-faint dwarf galaxies below the critical radius associated with dynamical disruption by PBHs \cite{Clesse:2017bsw};
(6) the unexplained correlation between the masses of galaxies and their central SMBHs;
(7) the observed mass and spin distributions for the coalescing black holes found by LIGO/Virgo \cite{LIGOScientific:2018mvr}.
There are additional observational problems which Silk has argued may be solved by PBHs in the intermediate mass range \cite{Silk:2017yai}.

Even if PBHs turn out to play none of these roles, their non-detection over various mass ranges still provides important constraints on the early Universe.
In particular, the limits on $ \beta(M) $ can be used to constrain models involving inflation, an early matter-dominated phase or the collapse of cosmic strings.
They also restrict the form of the primordial inhomogeneities and their statistical distribution.
Although we have not discussed them here, the PBH limits could also constrain more exotic models of the early Universe, such as those involving a variable gravitational constant~\cite{Harada:2001kc} or a different number of spatial dimensions~\cite{Guedens:2002km,Majumdar:2002mra,Guedens:2002sd,Clancy:2003zd,Sendouda:2003dc,Sendouda:2004hz}.
In the latter context, it is well known that the existence of extra dimensions which are large compared to the Planck length could permit the formation of black holes in accelerators~\cite{Giddings:2001bu}.
Even though such black holes are not themselves primordial, the existence of the extra dimensions in the early Universe would have important implications for PBH formation.

\begin{acknowledgments}
For many years of fruitful PBH collaboration we thank Laila Alabidi, Dick Bond, Francesca Calore, Sebastien Clesse, Juan Garc\'ia-Bellido, Jaume Garriga, Jonathan Gilbert, Tomohiro Harada, Stephen Hawking, Kimitake Hayasaki, Sanjay Jhingan, Takafumi Kokubu, Florian K\"uhnel, Koutarou Kyutoku, Matt Lake, Jim Lidsey, Hideki Maeda, Jane MacGibbon, Tomohiro Matsuda, Hiroki Matsui, Jonas Mureika, Shigehiro Nagataki, Tomohiro Nakama, Ken-ichi Nakao, Piero Nicolini, Vivian Poulin, Marti Raidal, Martin Rees, Mari Sakellariadou, Ryo Saito, Marit Sandstad, Misao Sasaki, Katsuhiko Sato, Pasquale Serpico, Joe Silk, Teruaki Suyama, Keitaro Takahashi, Tommi Tenkanen, Takahiro Terada, Ville Vaskonen, Hardi Veerm\"ae and Chul-Moon Yoo.
For useful input we also thank Anne Green, Carlos Hidalgo, Maxim Khlopov, Alex Kusenko, Karim Malik, Takahiko Matsubara, Shi Pi and Sasha Polnarev.
We are also indebted to the two referees, who suggested numerous improvements to an earlier version of this review.
B.J.C.\ thanks the Research Center for the Early Universe (RESCEU), University of Tokyo, for hospitality received during this work.
K.K.\ is grateful for the hospitality of The University of Oxford where a part of his work was done.
The work of K.K.\ is supported in part by JSPS KAKENHI grant No.~JP17H01131 and MEXT KAKENHI Grant Nos.~JP15H05889, JP18H04594, JP19H05114, JP20H04750.
Y.S.\ is supported in part by JSPS KAKENHI Grant Number 16K17675.
J.Y.\ is supported in part by JSPS KAKENHI Grant JP15H02082, 20H00151, 20H05639 and Grant on Innovative Areas JP15H05888.
\end{acknowledgments}

\bibliographystyle{apsrev4-1}
\bibliography{refs}

\end{document}